\documentclass[letterpaper,twocolumn,10pt]{article}
\usepackage{usenix-2020-09}

\usepackage[utf8]{inputenc}

\usepackage[english]{babel}
\usepackage{cite}
\usepackage{bbm}
\usepackage{url}
\usepackage{graphicx}

\usepackage{caption}
\usepackage[caption=false]{subfig}

\usepackage{etex}

\usepackage{color}
\usepackage{framed}
\usepackage{gensymb}
\usepackage{textcomp}
\usepackage{longtable}
\usepackage{array}
\usepackage{psfrag}
\usepackage{multirow}

\usepackage{diagbox}

\usepackage{adjustbox}

\usepackage{cleveref} 

\usepackage{footnote}

\usepackage{rotating}

\usepackage{xcolor,colortbl}

\usepackage{booktabs}

\usepackage{tikz-qtree}
\usepackage{etex}
\usepackage{pgf}

\usepackage[
  separate-uncertainty = true,
  multi-part-units = repeat
]{siunitx}

\makesavenoteenv{tabular}

\graphicspath{{figures/}}

\definecolor{Gray1}{gray}{0.75}
\definecolor{Gray2}{gray}{0.9}

\newcommand{\jm}[1]{\textcolor{blue}{}}

\pgfmathsetmacro{\widthfate}{0.18}

\begin{document}

\title{ML-based tunnel detection and tunneled application classification}

\author{
  {\rm Johan Mazel}\\
  ANSSI\\
  \and
  {\rm Matthieu Saudrais}\\
  \and
  {\rm Antoine Hervieu}\\
}

\maketitle

\begin{table*}[h]
  \centering
  \begin{tabular}{llll|ccccc|cccccccc}
    \toprule
    &\textbf{Authors} & \textbf{Refs.} & \multicolumn{1}{l}{\textbf{Year}} & 
    \multicolumn{5}{c}{\textbf{Tunnels}} & 
    \multicolumn{8}{c}{\textbf{MM methods}}\\
    \cmidrule(lr){5-9}\cmidrule(lr){10-17}\\
    
    & & & &
    
    \multicolumn{1}{c}{\rotatebox{90}{\textbf{SSL/TLS}}} & 
    \multicolumn{1}{c}{\rotatebox{90}{\textbf{Tor}}} & 
    \multicolumn{1}{c}{\rotatebox{90}{\textbf{SSH}}} & 
    \multicolumn{1}{c}{\rotatebox{90}{\textbf{IPSec}}} & 
    \multicolumn{1}{c}{\rotatebox{90}{\textbf{PPTP}}} & 
    \multicolumn{1}{c}{\rotatebox{90}{\textbf{Decision Tree}}} & 
    \multicolumn{1}{c}{\rotatebox{90}{\textbf{K-Nearest Neighbor}}} & 
    \multicolumn{1}{c}{\rotatebox{90}{\textbf{Naive Bayes}}} & 
    \multicolumn{1}{c}{\rotatebox{90}{\textbf{Support Vector}}} & 
    \multicolumn{1}{c}{\rotatebox{90}{\textbf{Gaussian Mixture Model}}} & 
    \multicolumn{1}{c}{\rotatebox{90}{\textbf{K-Means}}} & 
    \multicolumn{1}{c}{\rotatebox{90}{\textbf{Markov Chains}}} & 
    \multicolumn{1}{c}{\rotatebox{90}{\textbf{Neural Networks}}}\\
    
    \midrule
    
    \multirow{4}{*}{\rotatebox{90}{\parbox{5em}{\centering Protocol 
         classification}}}
    & Bernaille et al.   & \cite{Bernaille2006Traffic} & 2006 &
    - & - & - & - & - &   &   & x &   &   &   &   &   \\
    & Bernaille et al.   & \cite{Bernaille2006Early}   & 2006 &
    - & - & - & - & - &   & x &   &   & x &   & x &   \\
    & Kim et al.         & \cite{Kim2008Internet}      & 2008 &   
    - & - & - & - & - & x & x & x & x &   &   &   & x \\
    & Jaber et al.      & \cite{Jaber2011Can}         & 2011 &   
    - & - & - & - & - &   &   &   &   &   & x &   &   \\
    
    \midrule
    
    \multirow{8}{*}{\rotatebox{90}{\parbox{9em}{\centering Protocol 
          classification inside tunnel}}}
    
    &Wright et al.      & \cite{Wright2006Inferring}  & 2006 & 
    x &   & x &   &   &   & x &   &   &   &   & x &\\
    &Bernaille et al.   & \cite{Bernaille2007Early}   & 2007 & 
    x &   &   &   &   &   &   &   &   & x &   &   &\\
    &Maiolini et al.    & \cite{Maiolini2009Real}     & 2009 &   
    &   & x &   &   &   &   &   &   &   & x &   &\\
    &Sun et al.         & \cite{Sun2010Novel}         & 2010 & 
    x & x &   &   &   &   &   & x &   &   &   &   &\\
    &Okada et al.       & \cite{Okada2011Application} & 2011 &   
    &   &   & x & x &   &   & x &   &   &   &   &\\
    &Okada et al.       & \cite{Okada2011Comparisons} & 2011 &   
    &   &   & x & x & x &   & x & x &   &   &   &\\
    & Korczyński et al. & \cite{Korczynski2014Markov} & 2014 & 
    x &   &   &   &   &   &   &   &   &   &   & x &\\
    &Kumano et al.      & \cite{Kumano2014Towards}    & 2014 &   
    &   &   & x & x & x &   &   & x &   &   &   &\\
    \bottomrule
  \end{tabular}
  \caption{Tunnelling procotocols and machine learning techniques used in 
    previous work.}
  \label{table:ml}
\end{table*}

\newcommand{\fnl}{1.8cm}

\newcommand{\ad}{$\mathcal{A}$}
\newcommand{\dd}{$\mathcal{D}$}
\newcommand{\bd}{$\mathcal{B}$}
\newcommand{\id}{$\mathcal{I}$}
\newcommand{\pe}{$\mathcal{P}$}

\begin{table*}[t!]
  \setlength\tabcolsep{2.5pt} 
  \begin{tabular}{@{}lllllcccccccccccc@{}}
    \toprule
    
    \textbf{Type} & \textbf{Authors} & \textbf{Ref.} & \textbf{Year} & 
    
    \multicolumn{4}{c}{\textbf{Flow}} & & & 
    \multicolumn{4}{c}{\parbox{1.4cm}{\textbf{N first}}}
    \\
    
    \cmidrule(lr){5-8} \cmidrule(lr){11-14} \\
    
    & & & & & & \multicolumn{2}{c}{\textbf{Time}} \\
    
    \cmidrule(lr){7-8} \\
    
    & & & &
    
    \rotatebox{90}{\parbox{\fnl}{\textbf{\# packet}}} & 
    
    \rotatebox{90}{\parbox{\fnl}{\textbf{Packet size}}} & 
    
    \rotatebox{90}{\parbox{\fnl}{\textbf{Duration}}} & 
    \rotatebox{90}{\parbox{\fnl}{\textbf{IAT}}} & 
    \rotatebox{90}{\textbf{TCP-based (eg. flag)}} & 
    \rotatebox{90}{\parbox{\fnl}{\textbf{TLS type}}} &     
    
    \rotatebox{90}{\parbox{\fnl}{\textbf{Direction}}} & 
    \rotatebox{90}{\parbox{\fnl}{\textbf{Packet size}}} & 
    \rotatebox{90}{\parbox{\fnl}{\textbf{Elapsed time}}} & 
    \rotatebox{90}{\parbox{\fnl}{\textbf{IAT}}} & 
    \rotatebox{90}{\parbox{\fnl}{\textbf{Packet/byte burst}}} & 
    \\    
    
    \midrule
    
    \multirow{4}{*}{\rotatebox{90}{\parbox{4em}{\centering Protocol 
          classification}}}
    
    & Bernaille et al.                         & \cite{Bernaille2006Traffic} & 
    2006   &   
    &                                          &           
    &                          &   &   &   & x     \\
    & Bernaille et al.                         & \cite{Bernaille2006Early}   & 
    2006   &   
    &                                          &           
    &                          &   &   &   & x/\dd \\
    
    & Kim et al.                               & \cite{Kim2008Internet}      & 
    2008   &   
    x     & \{T,m,M,$\mu$,$\sigma$\}/\bd       & x & 
    \{T,m,M,$\mu$,$\sigma$\}/\bd     & x &   & & x &       \\
    
    & Jabber et al.                            & \cite{Jaber2011Can}         & 
    2011   &   
    &                                    &   &                                  
    &   &   &   &       &  & x  \\
    
    \midrule
    
    \multirow{8}{*}{\rotatebox{90}{\parbox{3.0cm}{\centering Protocol 
          classification inside tunnel}}}
    
    & Wrigh et al.                             & \cite{Wright2006Inferring}  & 
    2006   &   
    &                                          &   
    &                                  &   &   &   & x/\dd &  & x \\
    
    & Bernaille et al.                         & \cite{Bernaille2007Early}   & 
    2007   &   
    &                                          &   
    &                                  &   &   &   & x/\dd &  \\
    
    & Maiolini et al.                          & \cite{Maiolini2009Real}     & 
    2009   &   
    &                                          &   
    &                                  &   &   & x & x & x & \\
    
    & Sun et al.                               & \cite{Sun2010Novel}         & 
    2010   &   
    x     & $\mu$,m,M                          &   & 
    $\mu$,m,M                        &   &   & \\
    
    & Okada et al.                             & \cite{Okada2011Application} & 
    2011   &   
    x/\ad & \{T,m,M,\pe,$\mu$,$\sigma^2$\}/\ad & x & 
    \{m,M,\pe,$\mu$,$\sigma^2$\}/\ad &   &   & 
    \\
    
    & Okada et al.                             & \cite{Okada2011Comparisons} & 
    2011   &   
    x/\ad & \{T,m,M,\pe,$\mu$,$\sigma^2$\}/\ad & x & 
    \{m,M,\pe,$\mu$,$\sigma^2$\}/\ad &   &   & \\ 
    
    & Korczyński et al.                        & \cite{Korczynski2014Markov} & 
    2014   &   
    &                                          &   
    &                                  &   & x & &  \\
    
    & Kumano et al.                            & \cite{Kumano2014Towards}    & 
    2014   &   
    x/\ad & \{T,m,M,\pe,$\mu$,$\sigma^2$\}/\ad & x & 
    \{m,M,\pe,$\mu$,$\sigma^2$\}/\ad &   &   & \\

    \midrule
    
    & Netflow v5                               & \cite{Netflow_v1578}        
    &        &   
    x     & T,$\mu$                            & x & 
    ($\mu$)                          &   &   & \\
    & Netflow v9                               & \cite{Netflow_v9}           
    &        &   
    x/\ad & T/\ad,\{m,M\}/\id                  & x 
    &                                  &   &   & \\
    
    \midrule

    & Argus                                    & \cite{Argus}                
    &        &   
    x/\ad  & T/\ad                             & x & 
    m                                &   &   & &   &   &  \\
    
    & Zeek (conn.log)                           & \cite{zeek}                  
    &        &   
    x/\ad  & T/\ad                             & x & 
    m                                &   &   & &   &   &  \\
    
    \midrule
    
    & Our work  &                               
    &        &   
    x/\ad  & \{T,m,M,$\mu$,$\sigma$\}/\ad          & x & 
    \{m,M,$\mu$,$\sigma$\}/\ad             &   &   & x & x/\dd & x & x & x/\dd \\
    
    \bottomrule        
    
  \end{tabular}
  
  \caption{Network traffic features used for protocol classification and tunnel 
    protocol identification. 
    Abbreviations: x: used, \dd: sign encodes packet direction 
    ; $\mu$: mean, $\sigma$: standard deviation, $\sigma^2$: variance ; \bd: 
    uni-directional 
    ; \id: incoming only ; \ad: all direction
    ; T: total ; m: minimum ; M: maximum ; \pe: $25^{th}$ centile, $50^{th}$ 
    centile and $75^{th}$ centile.
  }
  \label{table:feature}
\end{table*}

\begin{abstract}

Encrypted tunneling protocols are widely used.
Beyond business and personal uses, malicious actors also deploy 
tunneling to hinder the detection of Command and Control and data exfiltration.
A common approach to maintain visibility on tunneling is to rely on network 
traffic metadata and machine learning to analyze tunnel occurrence without 
actually decrypting data.
Existing work that address tunneling protocols however exhibit several weaknesses:
their goal is to detect application inside tunnels and not tunnel identification,
they exhibit limited protocol coverage (e.g. OpenVPN and Wireguard are not 
addressed), and both inconsistent features and diverse machine learning techniques 
which makes performance comparison difficult.

Our work makes four contributions that address these limitations and provide 
further analysis.
First, we address OpenVPN and Wireguard.
Second, we propose a complete pipeline to detect and classify tunneling 
protocols and tunneled applications.
Third, we present a thorough analysis of the performance of both network traffic 
metadata features and machine learning techniques.
Fourth, we provide a novel analysis of domain generalization regarding background 
untunneled traffic, and, both domain generalization and adversarial learning regarding
Maximum Transmission Unit (MTU).

\end{abstract}

\section{Introduction}

Off-the-shelf encrypted tunneling protocols such as SSH or OpenVPN provide 
a reliable way for users to protect their network traffic from passive analysis.
Companies often deploy such tools between locations or to access remote 
infrastructure.
General public also use tunneling to avoid passive monitoring or tampering 
\cite{verizon}.
These same tools are also used by attackers to hide C2 (Command and Control)
and data exfiltration.
Although many tools already exists, such as IPsec, OpenVPN or SSH, new ones
also recently appeared (e.g. Wireguard).
The encrypted tunneling protocol landscape thus keep getting more complex
and makes malicious actors monitoring more difficult.

Port-based heuristic relying on tunnel default configuration are easy to 
circumvent by attackers.
Supervised machine learning (ML) applied to network traffic metadata (usually 
byte- and time-based) is thus a natural port-agnostic approach to identify some 
protocols without actually inspecting encrypted data.
Several work use this approach to identify applications inside
tunneling tools such as SSL/TLS \cite{Wright2006Inferring,Bernaille2007Early,Sun2010Novel,Korczynski2014Markov},
both IPsec and PPTP \cite{Okada2011Application,Okada2011Comparisons,Kumano2014Towards},
and SSH \cite{Wright2006Inferring,Maiolini2009Real}.
Their tunneling coverage is however limited: they do not address
common protocols such as OpenVPN or Wireguard.
These work also use non-overlapping network traffic metadata feature sets.
Furthermore, they use different machine learning techniques.
These last two aspects makes it difficult to compare their results.
Although machine learning is now easier to use than ever thanks to libraries 
such as Scikit-learn \cite{scikit-learn}, clear guidelines on how to use 
machine learning to analyze tunneling protocols are missing.
Beyond application classification inside tunnel, up to our knowledge, 
the preliminary task of identifying tunnels remains unaddressed.

Our goal is to address these limitations and design an approach to detect and classify
existing tunneling protocols.
We also want to provide a detailed analysis of network traffic metadata features
and machine learning techniques that provides clear and actionable guidelines for 
real-world deployment.
Finally, we want to address domain generalization as it is paramount to assess 
the whether a previously learned model will be usable beyond its initial training
context.
We here address two aspects: background untunneled traffic and Maximum 
Transmission Unit (MTU).

Our work provide four main contributions.
First, we include the two tunneling protocols that are not addressed in 
existing work: OpenVPN and Wireguard.
Second, we design a complete pipeline to detect and classify tunneling 
protocols and tunneled applications.
Third, we provide a thorough comparisons of network traffic metadata features 
and machine learning techniques in the context of tunneling traffic detection 
and classification.
Fourth, we address domain generalization regarding background untunneled 
traffic, and then, both domain generalization and adversarial learning regarding MTU.
Up to our knowledge, both of these aspects have never been addressed before.

Our paper is structured as follows.
\Cref{sec:related_work} present existing work on tunneling protocols analysis.
\Cref{sec:methodology} details our methodology.
\Cref{sec:results} presents our detection and classification results.

\vspace{3em}

\section{Related work}
\label{sec:related_work}

The idea of using network traffic metadata can be traced back to the original 
work of Wright et al. \cite{Wright2006Inferring} who applied it to application 
classification inside SSL/TLS and SSH tunnels.
Several works extend this initial contribution regarding other 
tunneling tools such as Tor \cite{Sun2010Novel} or IPsec and PPTP
\cite{Okada2011Application,Okada2011Comparisons,Kumano2014Towards}.
Some works revisit previously addressed protocols but are relevant 
considering the quick evolution of the tunneling and encryption landscapes
\cite{Bernaille2007Early,Maiolini2009Real,Korczynski2014Markov}.

Another use case of network traffic metadata that generated many contributions 
is protocol classification.
Bernaille et al. \cite{Bernaille2006Traffic} is the seminal work for this use 
case.
Several other works then improve on this initial contribution 
\cite{Bernaille2006Early,Kim2008Internet,Jaber2011Can}.

\Cref{table:ml} describes the targeted tunneling tools and ML techniques used.
ML techniques and targeted tunneling protocols widely 
vary across studies which make their results extremely difficult to compare.
These work also do not address tunneling tools that are common such OpenVPN or 
Wireguard.
\Cref{table:feature} describes features used in existing work and tools.
They are very diverse.
The first papers on protocol classification inside tunnel used N first packet size 
or IAT, while later work usually leverage five tuple flow-related feature such as 
total number of packet or byte exchanged.
Features that were more recently proposed such as bursts \cite{wang2014effective}
have also not been evaluated in the context of tunneling protocol analysis.

Up to our knowledge, there is no existing work on the preliminary steps to 
application classification inside tunnels which are tunnelling protocol 
detection and/or classification.

Our goal is thus to 1) provide a complete pipeline for the analysis of tunelling 
protocol (detection, classification, and application classification within tunnels),
and 2) address the diversity of existing work by providing a unified 
point of view on performance of both network traffic features, and machine 
learning techniques.

\jm{TODO: add velan paper}

\begin{figure*}[t!]
  \centering
  \includegraphics[width=0.99\textwidth]{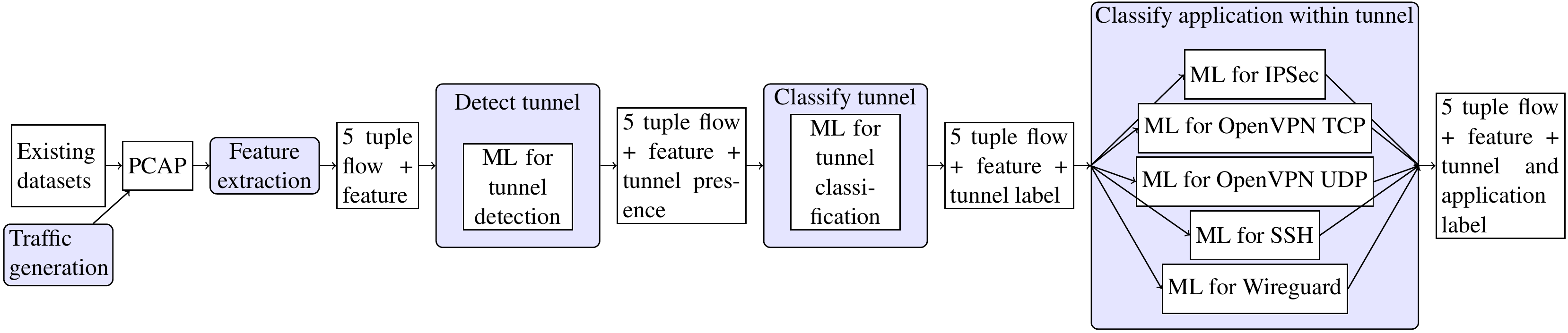}
  \caption{Processing pipeline.
  Functional boxes have a purple background and rounded corners.
  Data boxes have a white background and square corners.
  }
  \label{fig:pipeline}
\end{figure*}

\begin{figure*}[t!]
  \centering
  \includegraphics[width=0.9\textwidth]{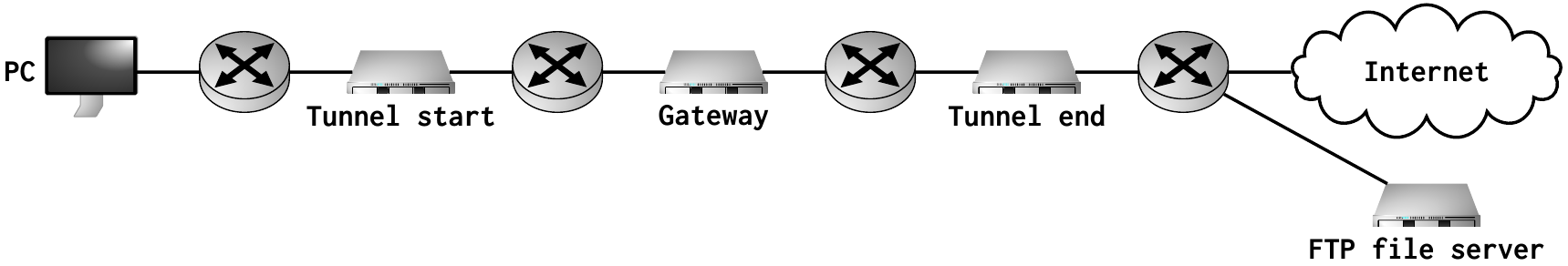}
  \caption{Traffic generation topology.}
  \label{fig:topology}
\end{figure*}

\begin{table*}[t]
  \centering
  \begin{tabular}{lllrrrrrr}
    \toprule
    \multicolumn{1}{l}{\textbf{Application}} & 
    \multicolumn{1}{l}{\textbf{Tool}} &
    \multicolumn{1}{l}{\textbf{Target}} &
    \multicolumn{1}{r}{\textbf{Target nb}} &
    \multicolumn{1}{r}{\textbf{Max application run}} & 
    \multicolumn{1}{r}{\textbf{Max target nb}} & \\
    
    \midrule
    Web     & Firefox/browsh & Alexa Top              &    1M & 20 &  1 \\
    wget    & wget           & Debian packages subset & 42669 &  4 &  1 \\
    FTP get & lftp/vsftp     & Debian packages subset &  1000 & 20 & 10 \\
    FTP put & lftp/vsftp     & Debian packages subset &  1000 & 20 & 10 \\
    \bottomrule
    
  \end{tabular}
  \caption{Deployed applications.
  Max application run is the maximum number of times an application is run inside a tunnel.
  Max target nb is the maximum number of target (website or debian package) retrieved by each application run.
  In both cases, the actual number is randomly picked between 1 and this maximum value.
  Timewise, browsing a website lasts between 30 and 45s and file downloads are consecutive.
  }
  \label{table:application}
\end{table*}

\begin{figure}[t!]
  \centering
  \includegraphics[width=\columnwidth]{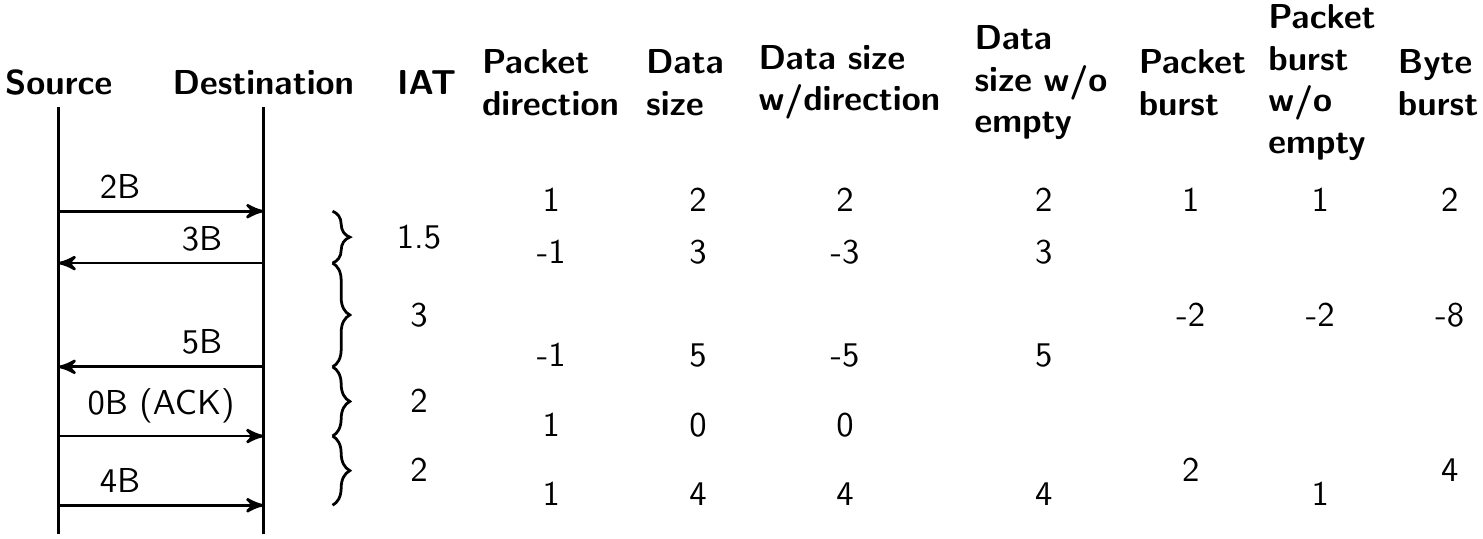}
  \caption{Example of used N first feature.}
  \label{fig:feature}
\end{figure}

\begin{table}[h]
  \setlength\tabcolsep{1pt}
  \centering
  \begin{tabular}{lrrrrrrrr}
    \toprule
    & & \multicolumn{6}{c}{\textbf{Tunnels}}\\    
    \cmidrule(lr){3-8}\\
    \textbf{Source} & 
    \rotatebox{90}{\textbf{Untunneled}} &
    \rotatebox{90}{\textbf{SSH}} & 
    \rotatebox{90}{\textbf{OpenVPN-TCP}} &
    \rotatebox{90}{\textbf{OpenVPN-UDP}} & 
    \rotatebox{90}{\textbf{IPSec}} &
    \rotatebox{90}{\textbf{Wireguard}} & 
    \rotatebox{90}{\textbf{Total}} \\
    
    \midrule
    UN-B \cite{unibs} &  66,585 &     27 &   - &   - &   - &   - &  66,612 \\
    UN-S \cite{unibs} &       0 & 27,978 &   - &   - &   - &   - &  27,978 \\
    UPC \cite{bujlow2015independent,upc} 
                      & 525,537 &  4,093 &   - &   - &   - &   - & 529,630 \\
    \midrule
    This work         & 241,941 &   2,400 & 2,400 & 2,400 & 2,400 & 2,400 & 253,941 \\
    \bottomrule    
  \end{tabular}
  \caption{Data used in our work.
  UN-B (resp. UN-S) is the UNIBS dataset with untunneled (resp. SSH) traffic.
  }
  \label{table:datasets}
\end{table}

\begin{table*}[t!]
  \setlength\tabcolsep{1pt}

  \centering
  \begin{tabular}{llllllll}
    \toprule
    \textbf{Algorithms} & 
    \parbox{4em}{\textbf{Scikit learn function}} & 
    \parbox{4em}{\textbf{Short name}} & 
    \textbf{Parameter name} & 
    \parbox{5em}{\textbf{Parameter default value}} & 
    \textbf{Parameter values for grid search} & \\
    
    \midrule
    AdaBoost             & \cite{sklearn_ada_boost} & AB 
                         & learning rate    & 1.0 & $10^{-3,-2,-1,0,1}$ \\
    \midrule

    Gradient boosting    & \cite{sklearn_gradient_boosting} & GB
                         & max depth        &   3 & 1, 2, 3, 4, 5, 6, 7, 10, 20, 50 \\
    \midrule
    
    \multirow{2}{*}{\parbox{7em}{Gaussian naive Bayes}}
                         & \cite{sklearn_gaussian_naive_bayes} & GNB 
                         & -                &    - & - & \\ 
    \\
    \midrule

    K nearest neighbors  & \cite{sklearn_knn} & KNN & n neighbors      &    5 & 1, 5, 10, 50 \\
    \midrule

    \multirow{2}{*}{\parbox{7em}{Support vector with liblinear}}
                         & \cite{sklearn_linear_svc} & SV LL 
                         & C                &  1.0 & $10^{-1,0,1,2,3,4,5,6,7}$ \\
    \\
    \midrule
                         
    Logistic regression  & \cite{sklearn_logistic_regression} & LR
                         & solver           & saga & saga \\
                     & & & C                &  1.0 & $10^{-1,0,1,2,3,4,5,6,7}$ \\
    \midrule 
    
    \multirow{2}{*}{\parbox{7em}{Logistic regression with SGD}}
                         & \cite{sklearn_sgd_classifier} & LR SGD
                         & loss             &  log & log \\
                     & & & alpha            & $10^{-4}$ & $10^{-8,-7,-6,-5,-4,-3,-2,-1,0,1}$ \\
    \midrule

    \multirow{2}{*}{\parbox{7em}{Support vector with SGD}}
                         & \cite{sklearn_sgd_classifier} & SV SGD 
                         & loss             & hinge & hinge \\
                     & & & alpha            & $10^{-4}$ & $10^{-8,-7,-6,-5,-4,-3,-2,-1,0,1}$ \\

    \midrule

    Random forest and    & \cite{sklearn_random_forest} & RF 
                         & criterion        & gini & gini, entropy \\

    decision tree        & \cite{sklearn_decision_tree} & DT 
                         & min samples split &   2 & 2,3,4,5,10,50,100 \\
    \bottomrule
  \end{tabular}
  \caption{Machine learning algorithms used and parameters explored.
  Random forest and decision tree both use the criterion and min samples split parameters.
  Short names are used to identify algorithms in the rest of the paper.
  }
  \label{table:ml_parameters}
\end{table*}

\section{Methodology}
\label{sec:methodology}

In this section, we present our approach.
First, we address the general machine learning-based pipeline in \Cref{sec:pipeline}, 
and then, we detail the main steps: traffic generation in \Cref{sec:traffic_generation},
used datasets in \Cref{sec:datasets},
feature extraction in \Cref{sec:feature_extraction}, and, machine learning in
\Cref{sec:machine_learning}.

\subsection{General pipeline}
\label{sec:pipeline}

\Cref{fig:pipeline} picture the whole pipeline that we use to analyze tunneling 
protocols.
First, we generate network traffic (see \Cref{sec:traffic_generation}) to 
complete the external datasets that we obtained (see \Cref{sec:datasets}).
Then, using generated PCAP files, we extract network traffic metadata feature 
(see \Cref{sec:feature_extraction}).
The next step is the use of ML to determine if a given flow is a tunnel.
This task is named \emph{tunnel detection} in the remainder of the paper.
If it is the case, we then identify which tunnel is present.
This step is called \emph{tunnel classification}.
If a tunnel is present, we classify the application used inside the tunnel.
We call this phase \emph{application classification inside tunnel}.

\subsection{Traffic generation}
\label{sec:traffic_generation}

Although some existing dataset provide some tunneling protocol network traffic 
(eg. SSH in UNIBS \cite{unibs}), we could not find any data for IPsec, OpenVPN 
or Wireguard.
We thus design a generic topology to collect tunnel traffic.
Here, a host uses a tunnel to access some resources on Internet or on an 
FTP file server.
This topology is presented on \Cref{fig:topology}.
The tunnel is setup between two hosts (tunnel start and tunnel end) and is 
routed through a machine called gateway.
This topology is setup using Vagrant with the libvirt provider.
All hosts use Debian Testing with the version 20210228.1 
\footnote{https://app.vagrantup.com/debian/boxes/testing64} 
(this is equivalent to Debian Bullseye).
We deactivate NIC offloading on all machines in order to avoid packet size 
alteration during capture.

We deploy each tunnel between the tunnel start and tunnel end machines.
We setup the following tunnels: IPsec, OpenVPN TCP/UDP, SSH and Wireguard.

We use several application inside each tunnel: web browsing, FTP get/put and wget.
We actually use a single application inside each tunnel instance.
\Cref{table:application} provides additional details regarding applications used
inside tunnels.

\subsection{Datasets}
\label{sec:datasets}

\Cref{table:datasets} present the data we use in this work.
The UNIBS datasets \cite{unibs} were collected in 2009 and contain both SSH 
tunneled and untunneled traffic.
The UPC dataset \cite{upc} was gathered in 2014 and contains mostly untunneled 
traffic and SSH and Tor tunnels.
In this work, we only use the untunneled and SSH tunneled traffic.
The network traffic generation in \Cref{sec:traffic_generation} create 100 flows 
for six MTU values (see \Cref{sec:dg_al_mtu} for additional details on MTU  
choice) and four applications used with 5 tunnels and without tunnel.
We thus obtain 2400 flows for each tunneling protocol.
Untunneled traffic does not yield the same number of flows for each MTU due 
to varying interactions with third-party during website browsing.

\subsection{Feature extraction}
\label{sec:feature_extraction}

Once network traffic has been generated, we extract five tuple flow (source and 
destination IP, transport protocol, and source and destination port) and their 
associated features.
We extract the following features about each five tuple flow: transport protocol 
(e.g. TCP, UDP), packet number, duration, and, 
mean, minimum, maximum, standard deviation in both direction and the total of 
both packet size and Inter-Arrival Times (IAT).
We also extract the following N first feature (see example in \Cref{fig:feature}): 
elapsed time since the start of the flow, IAT, packet direction, packet size,
packet size with direction in sign (called packet size with direction in the 
remainder of the paper), packet burst and byte burst.
Burst are build by grouping consecutive packet in the same direction.
Packet bursts are the grouping sizes and byte bursts are the grouping total byte 
number.
We also build packet size from source and destination only, and byte burst from 
source and destination only.
The \Cref{table:feature} presents our features and those of previous work.

\jm{TODO: byte number is total of payload $\rightarrow$ replace by unique bytes, ie drop retransmission}

\subsection{Machine learning}
\label{sec:machine_learning}

\jm{add XGBoost}

We use algorithms available in scikit-learn \cite{scikit-learn} for multiclass
 classification: random forest, decision tree, AdaBoost, logistic regression
and support vector with 
SGD, logistic regression with the saga solver, Gaussian naive Bayes, and K 
nearest neighbors.
Some algorithms are omitted in the following experiments due to excessive 
running time (e.g. K nearest neighbors for tunnel detection).

In terms of performance metric, we use the F1 score for tunnel detection and 
the average F1 score across classes (macro in scikit learn) for tunnel 
classification and application classification inside tunnels.

Performance comparison between algorithms is performed using a nested 
cross-validation \cite{raschka2018model}.
The inner loop uses a grid search to find the best parameters for each 
algorithm (see \Cref{table:ml_parameters}).
The best parameters are determined using F1 score as specified above.
The outer loop then reuse these optimal parameters to provide a performance lower 
bound for comparison with other algorithms.

\begin{figure*}[t!]
  \centering
  \begin{tabular}{ccccc}

    50 first packet sizes & 50 first packet directions & 50 first packet sizes with direction \\
    \includegraphics[width=0.32\textwidth]{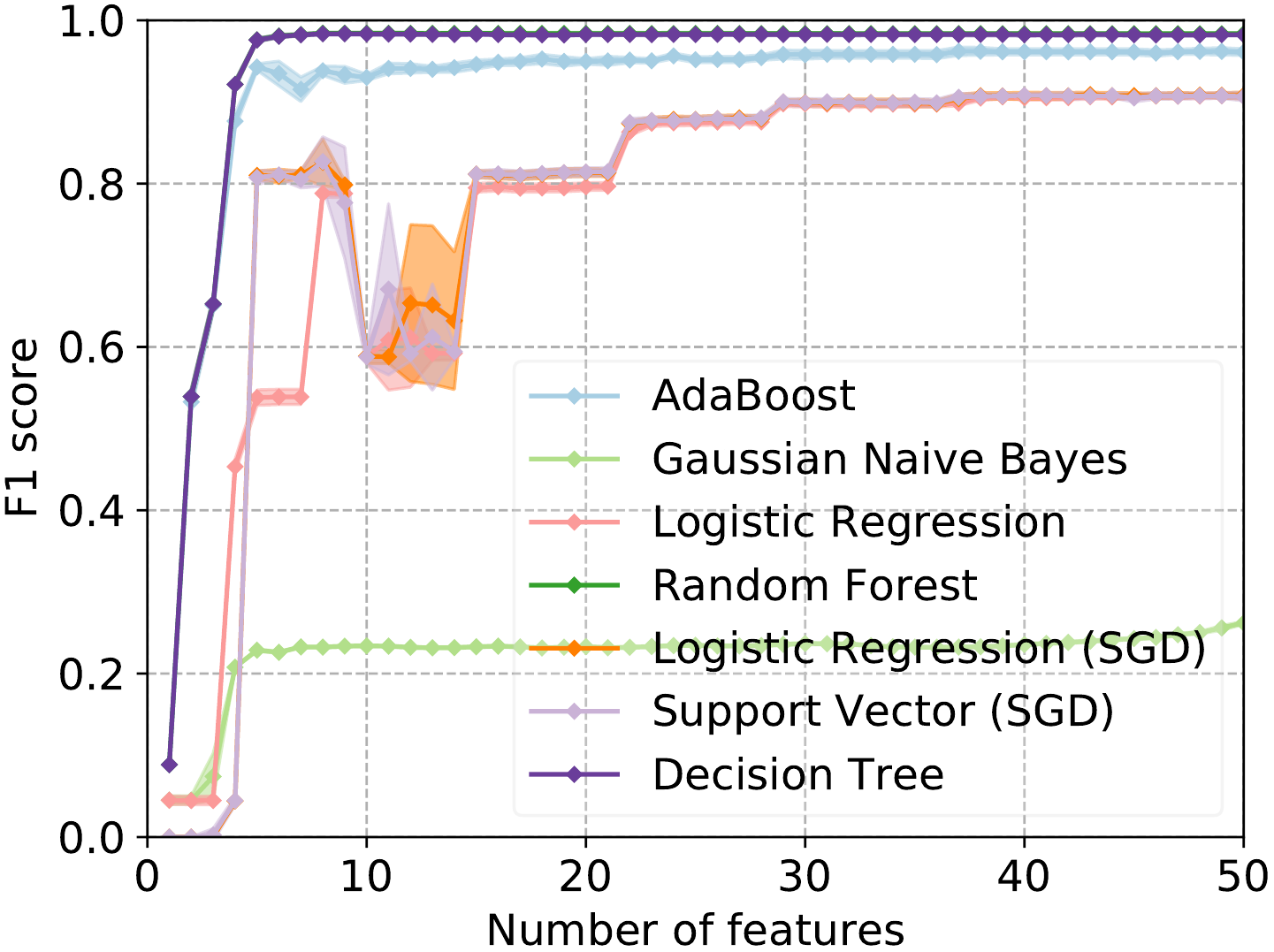} &
    \includegraphics[width=0.32\textwidth]{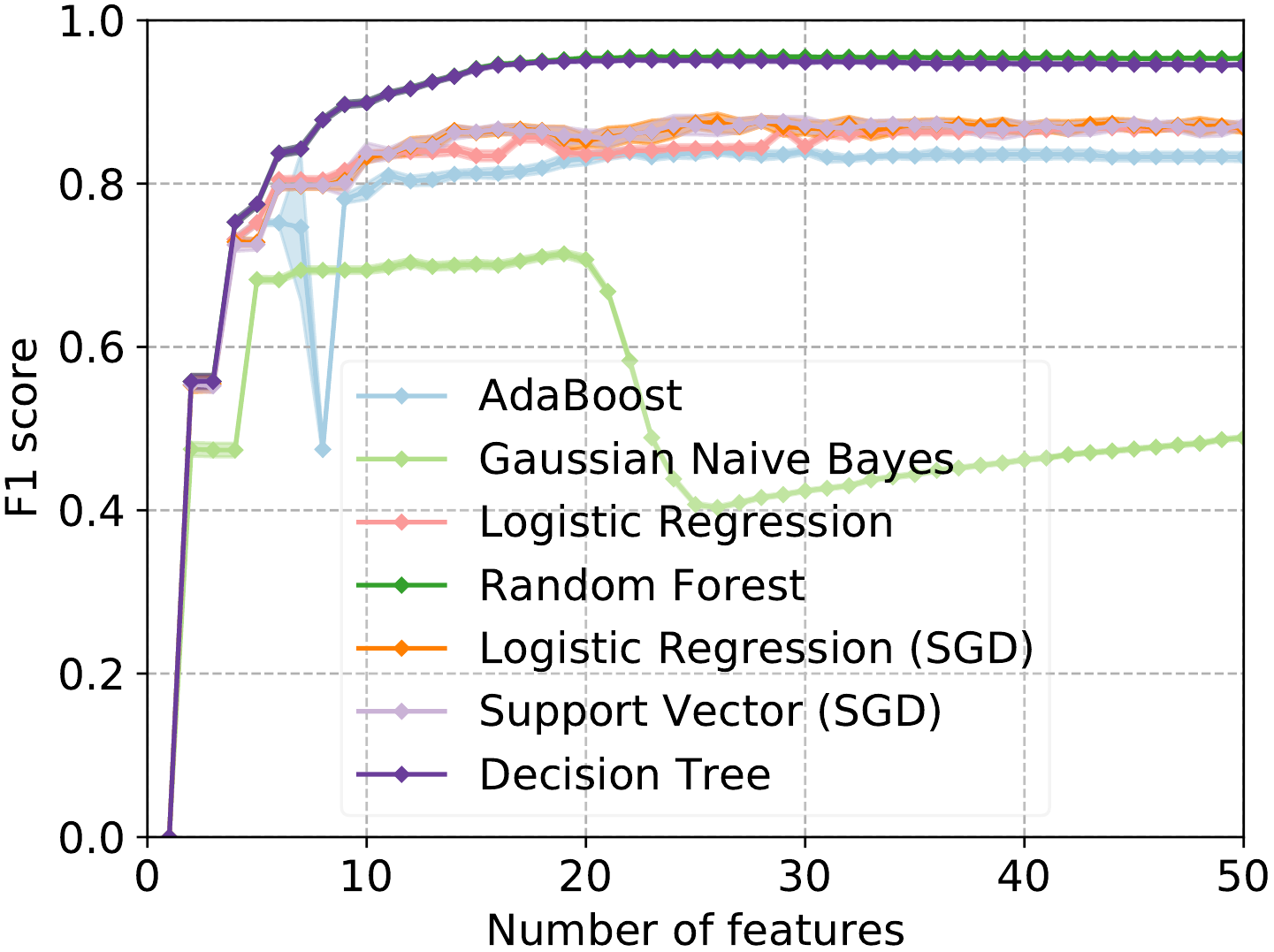} &
    \includegraphics[width=0.32\textwidth]{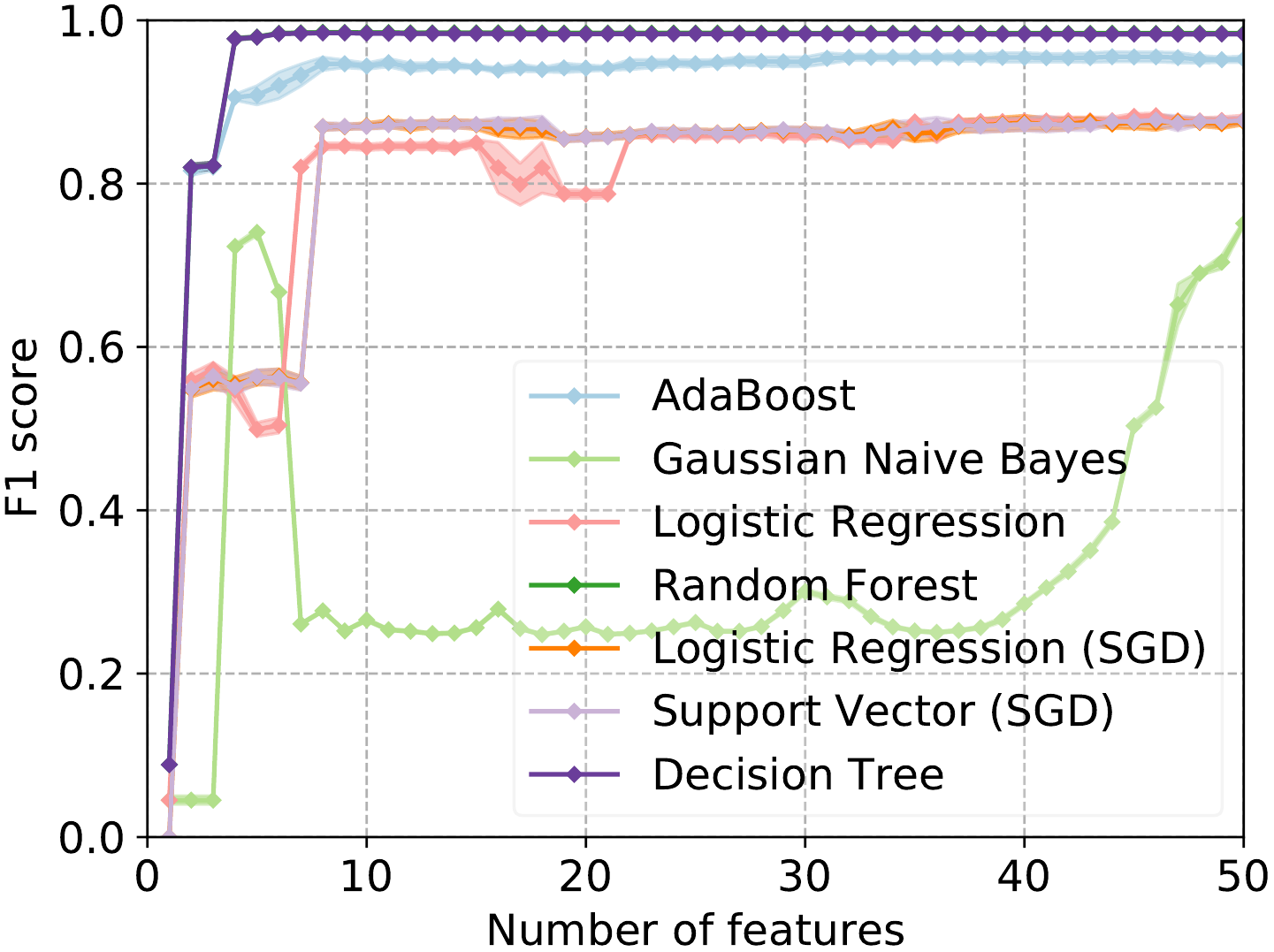} \\

    50 first packet bursts & 50 first byte bursts & 50 first IAT \\
    \includegraphics[width=0.32\textwidth]{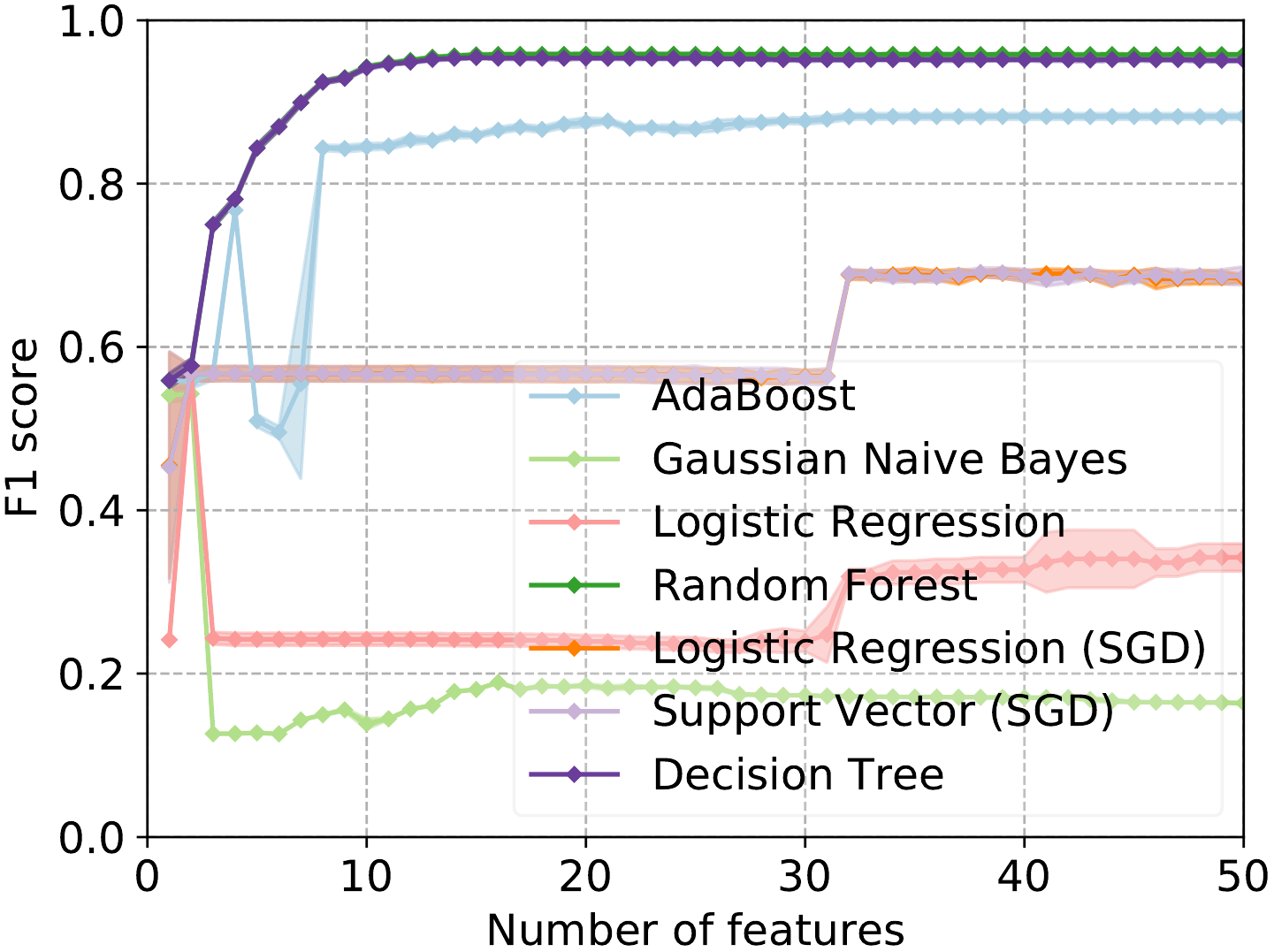} &
    \includegraphics[width=0.32\textwidth]{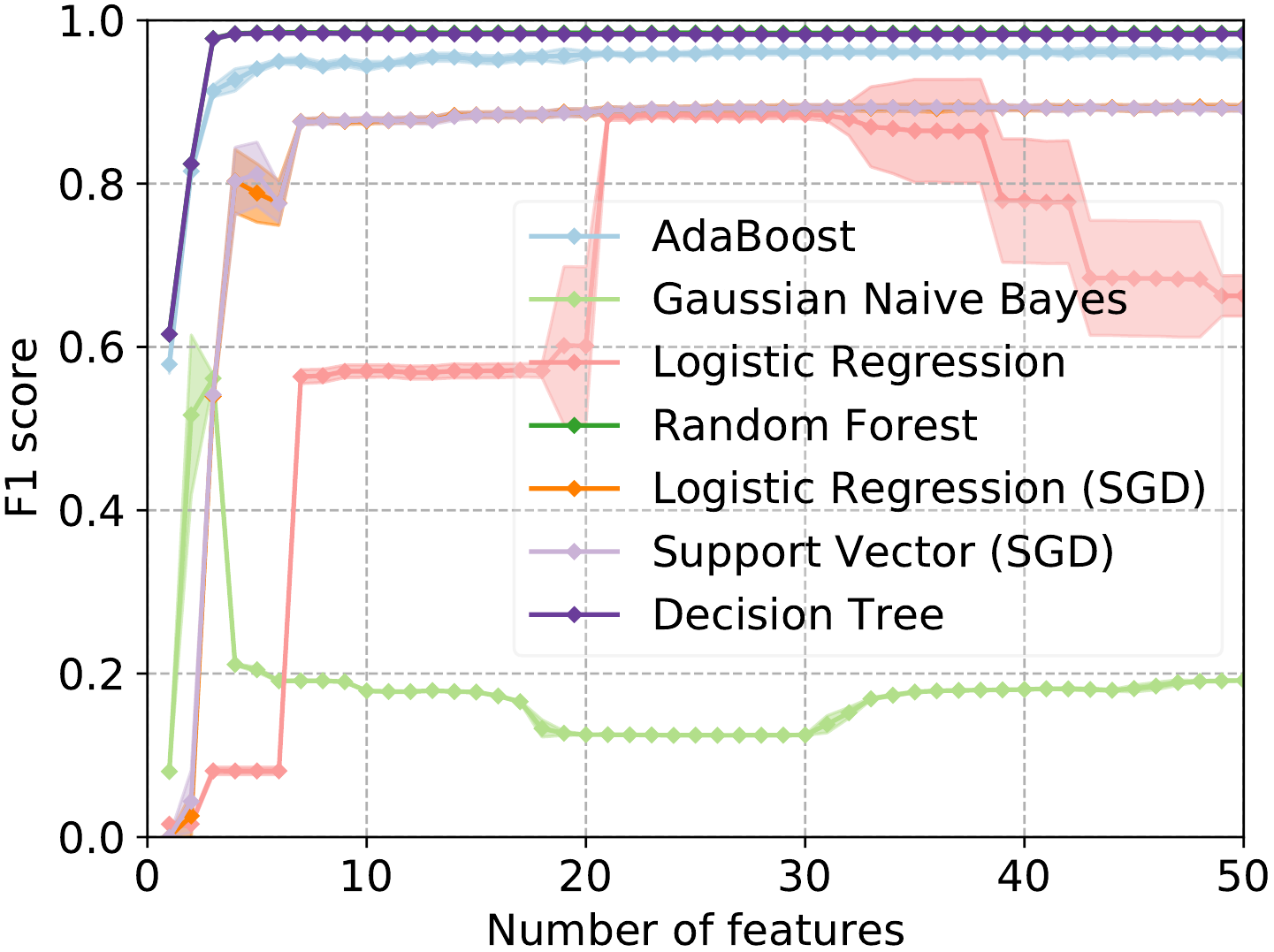} &
    \includegraphics[width=0.32\textwidth]{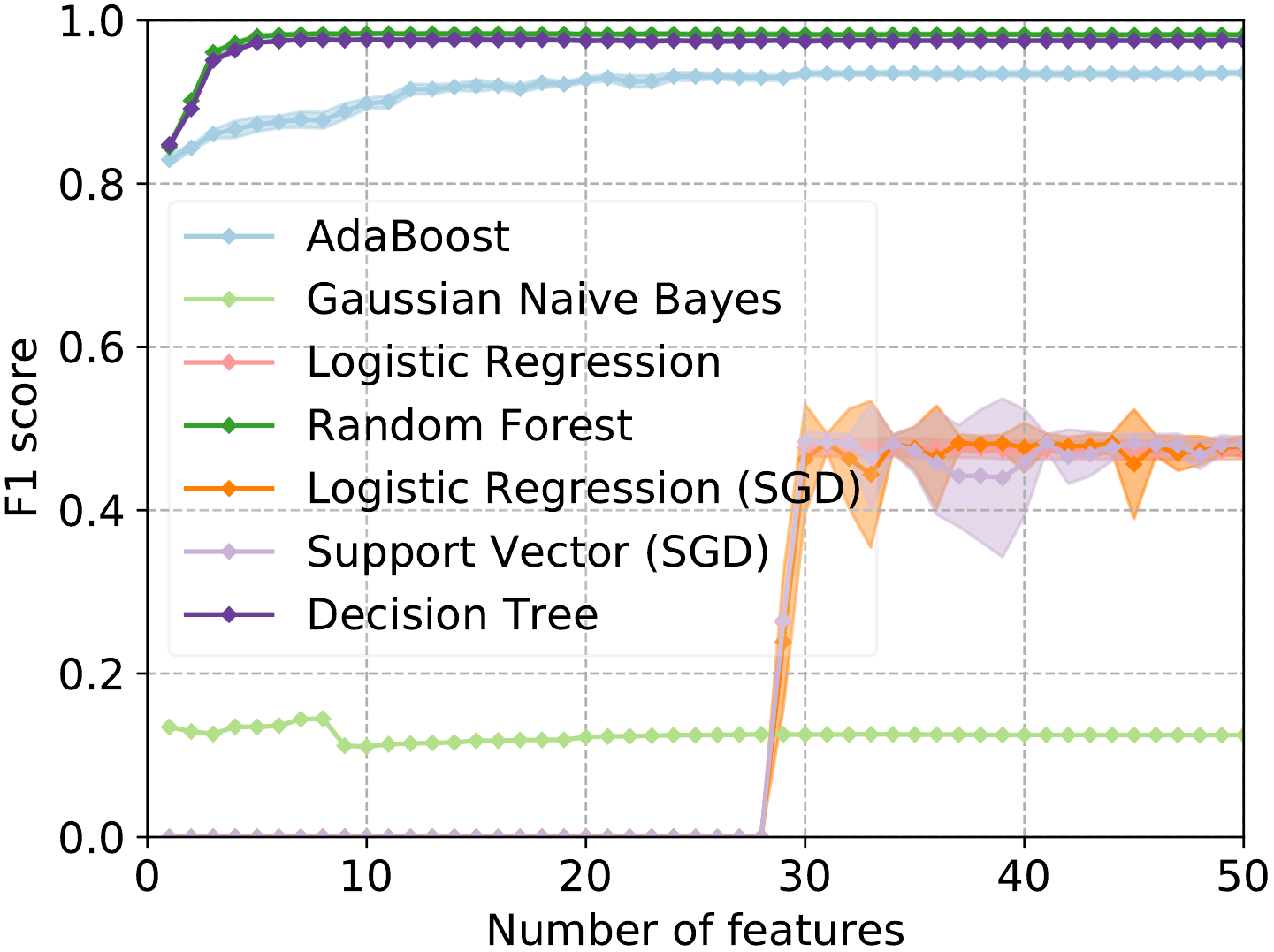} \\
  \end{tabular}

  \caption{Tunnel detection using 50 first feature.
  Shaded areas below and above the curve represent a confidence interval with 99\% confidence level.
  }
  \label{fig:feature_ow_n}
\end{figure*}

\begin{figure*}[t!]
  \centering
  \begin{tabular}{ccccc}

    50 first packet sizes & 50 first packet directions & 50 first packet sizes with direction \\
    \includegraphics[width=0.32\textwidth]{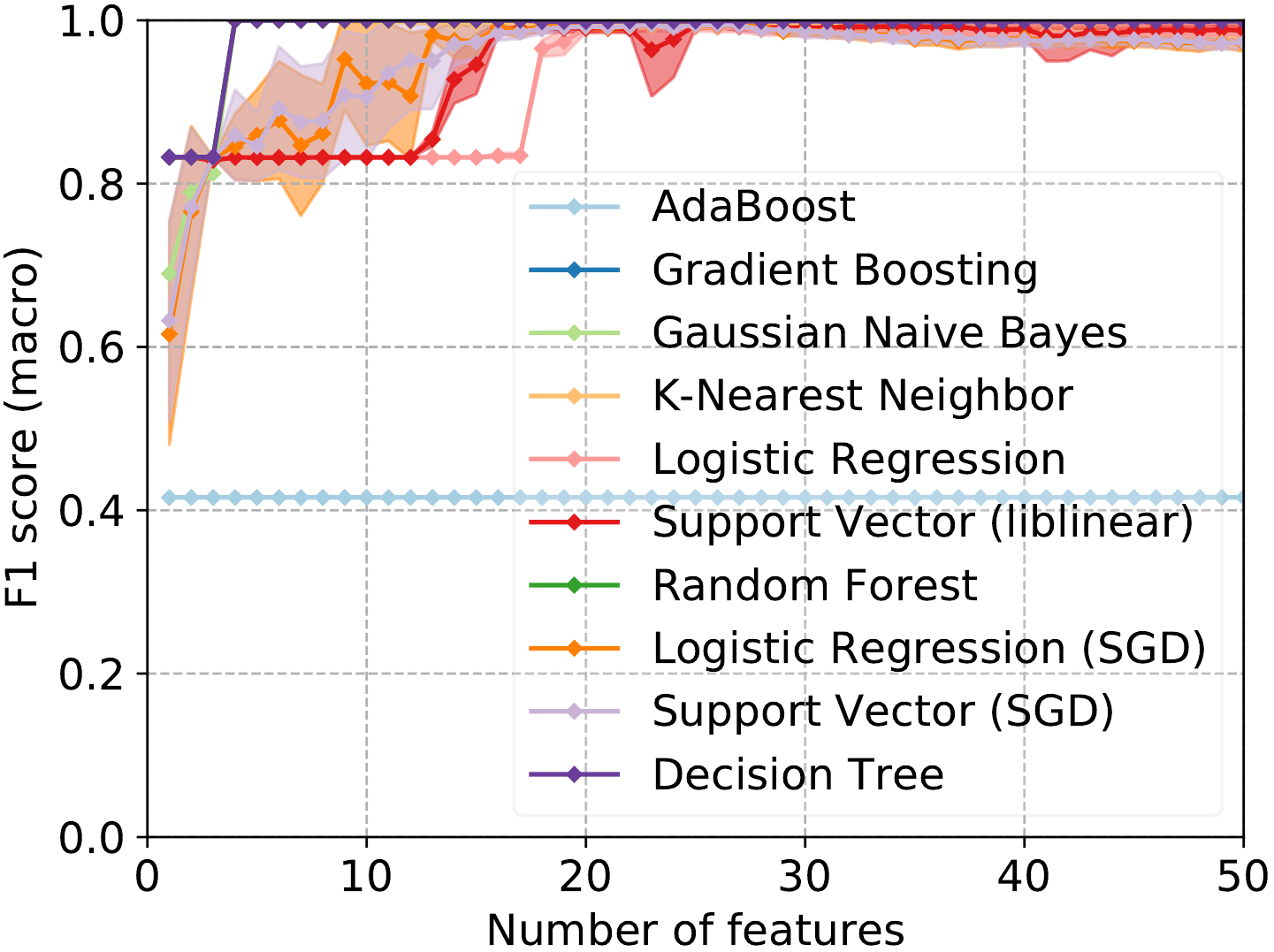} &
    \includegraphics[width=0.32\textwidth]{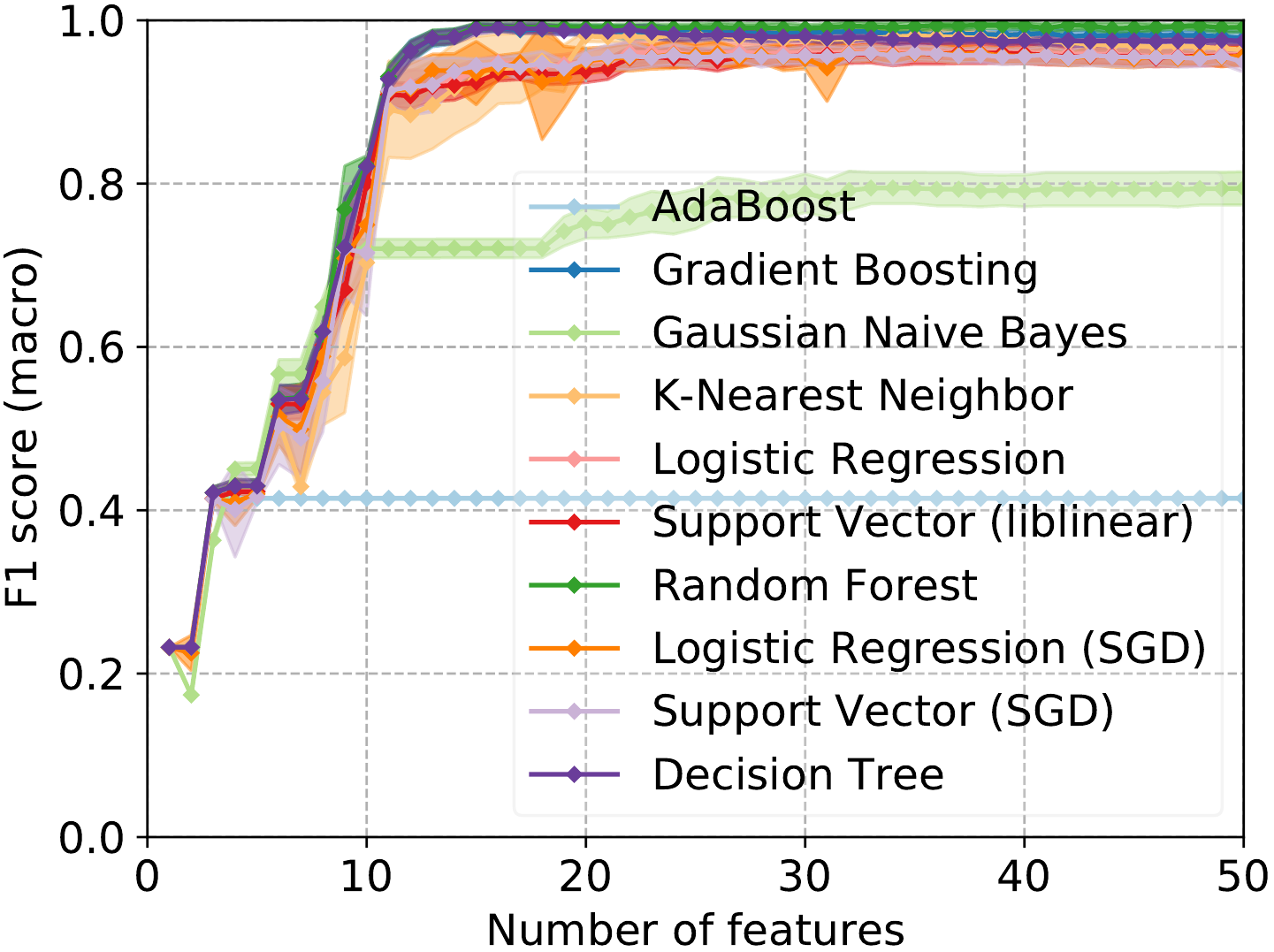} &
    \includegraphics[width=0.32\textwidth]{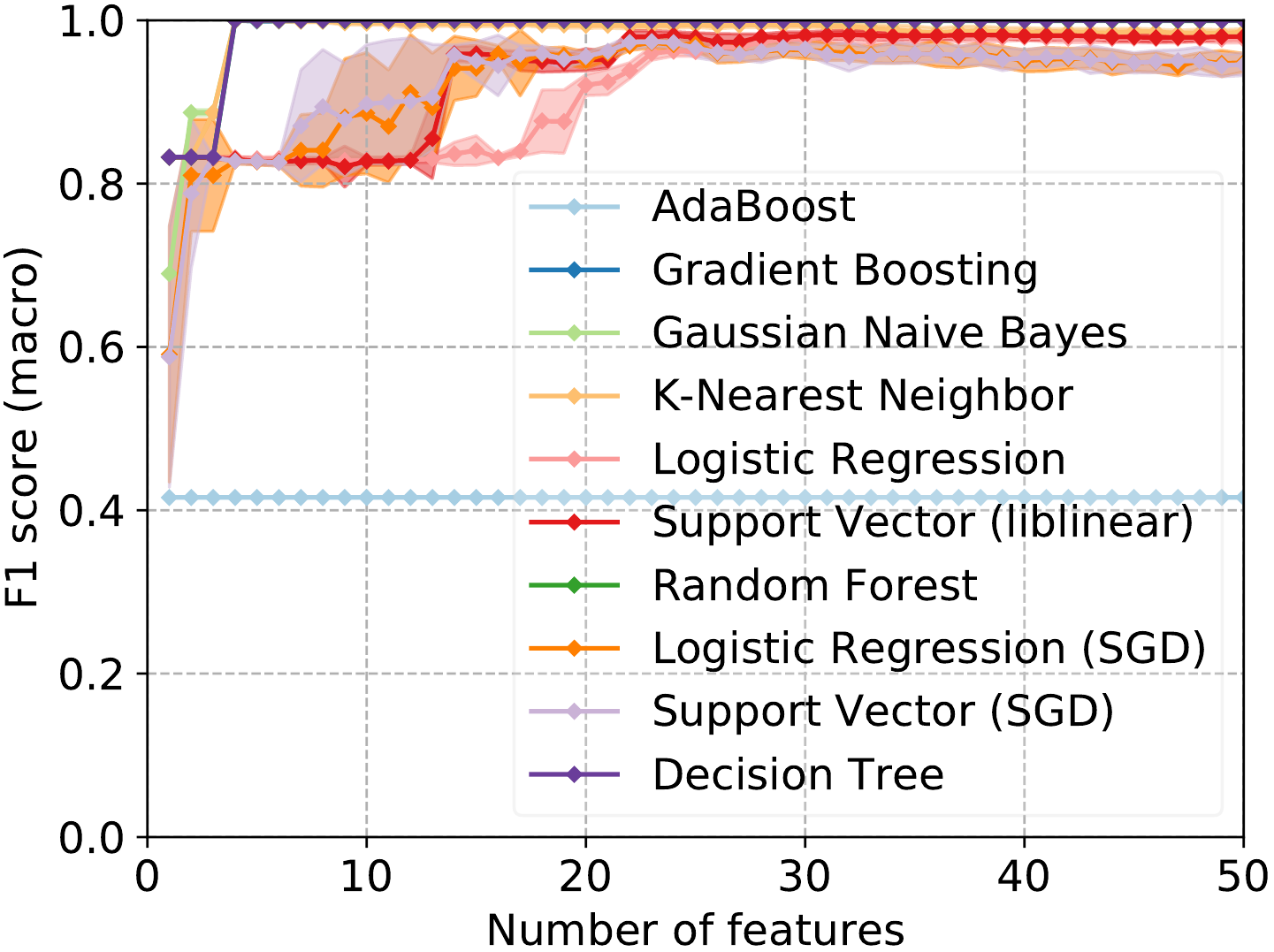} \\

    50 first packet bursts & 50 first byte bursts & 50 first IAT \\
    \includegraphics[width=0.32\textwidth]{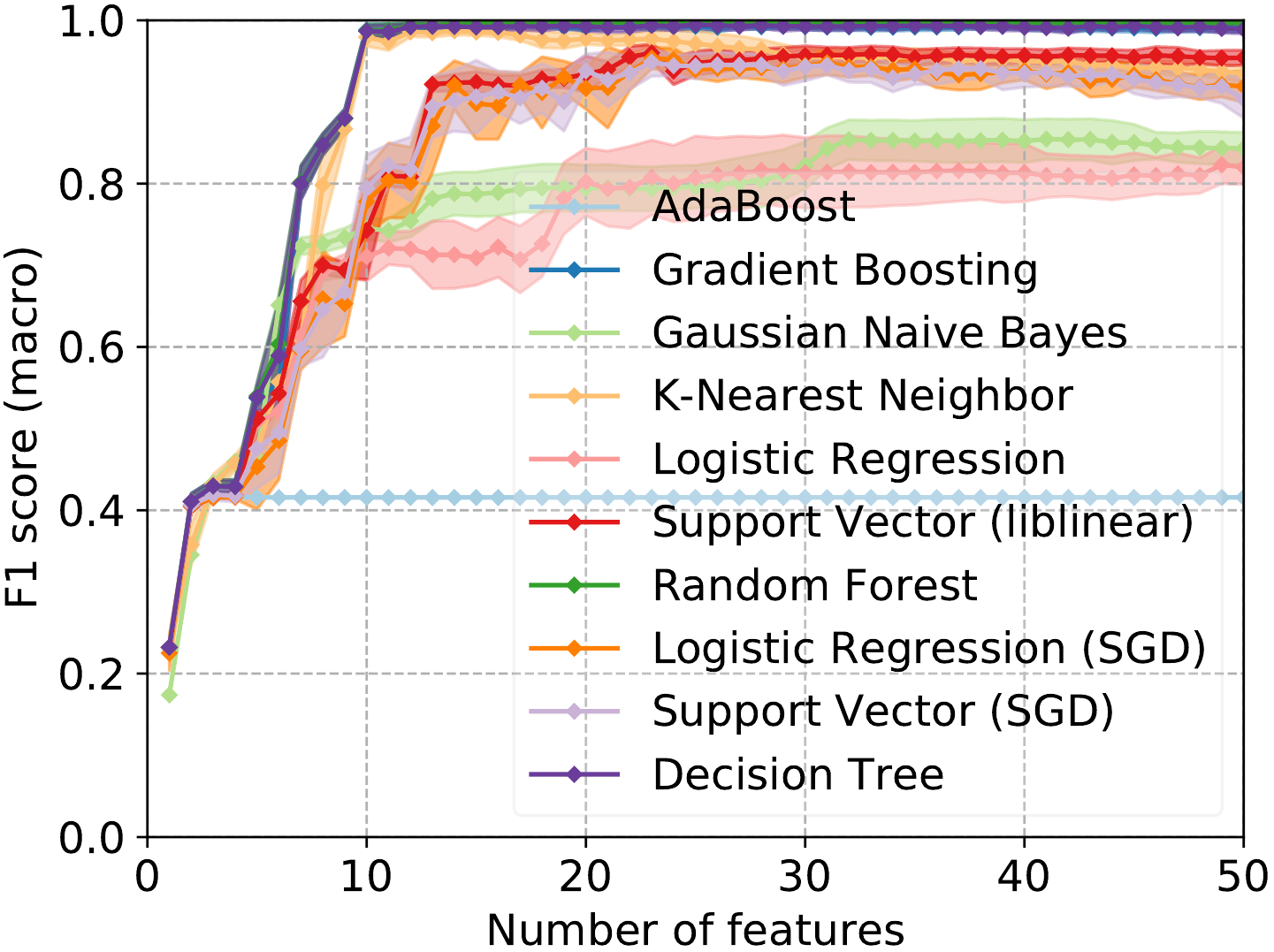} &
    \includegraphics[width=0.32\textwidth]{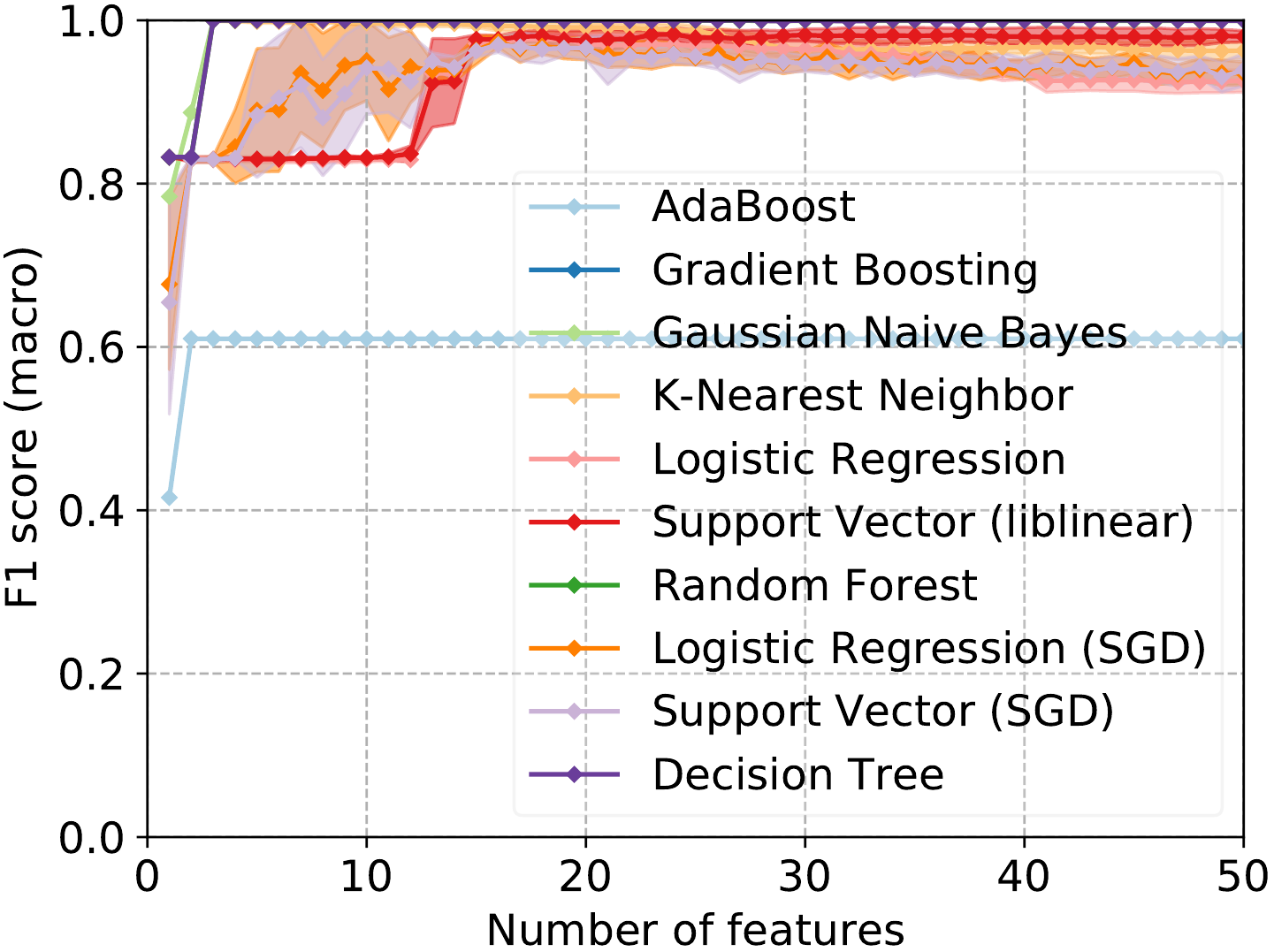} &
    \includegraphics[width=0.32\textwidth]{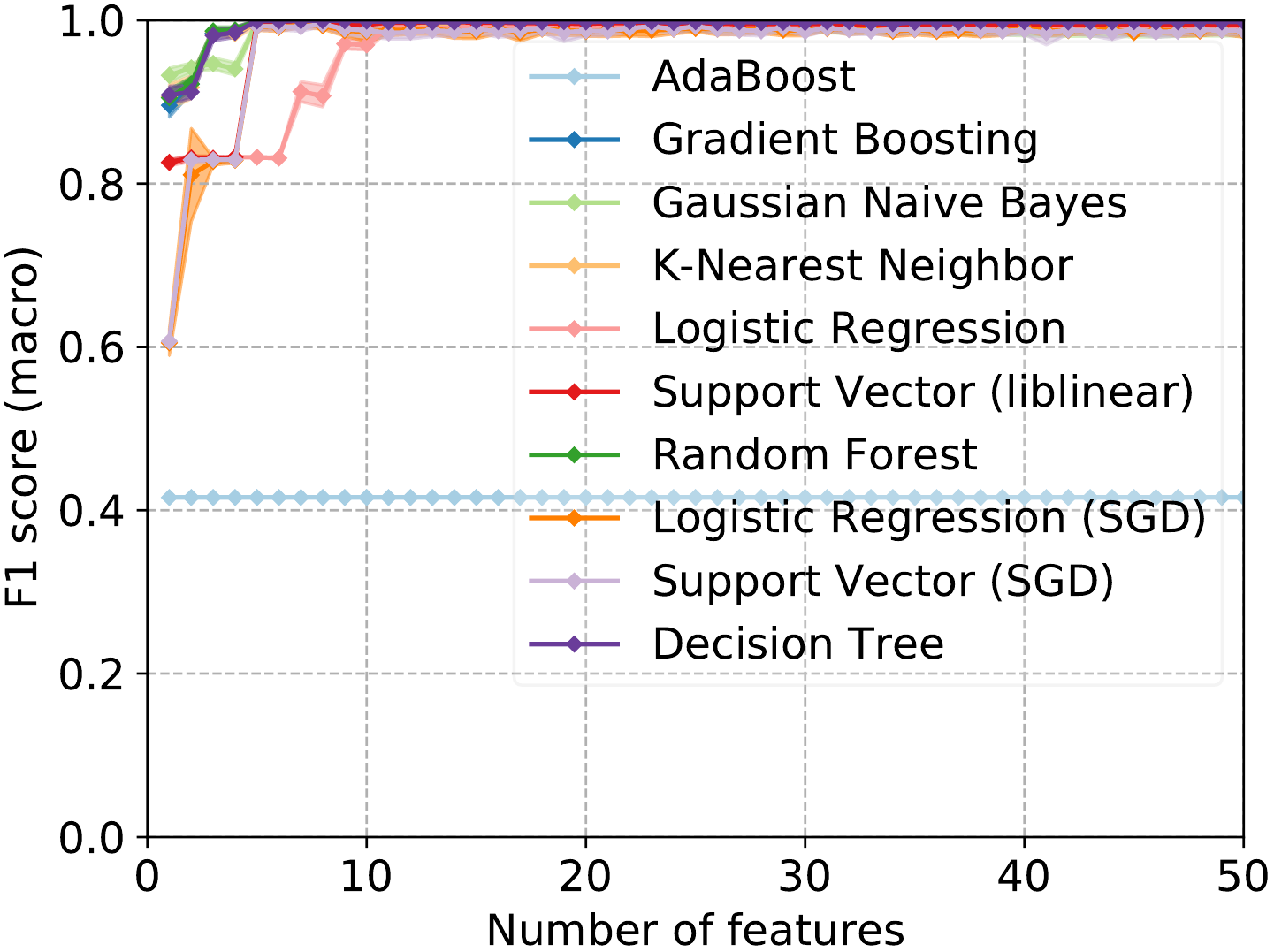} \\
  \end{tabular}

  \caption{Tunnel classification using 50 first feature.
  Shaded areas below and above the curve represent a confidence interval with 99\% confidence level.
  }
  \label{fig:feature_cwt_n}
\end{figure*}

\begin{figure*}[t!]
  \centering
  \begin{tabular}{ccccc}

    50 first packet sizes & 50 first packet directions & 50 first packet sizes with direction \\
    \includegraphics[width=0.32\textwidth]{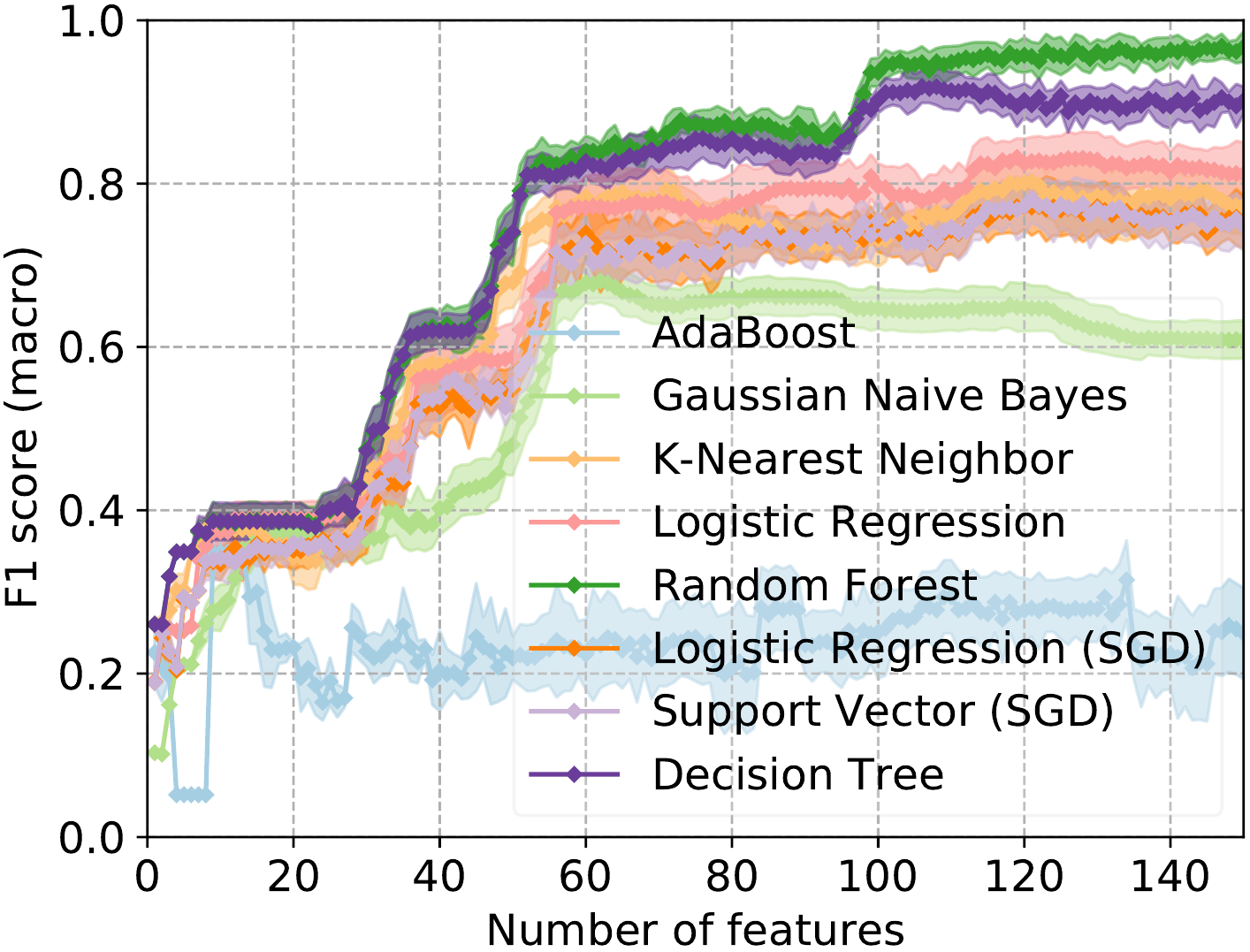} &
    \includegraphics[width=0.32\textwidth]{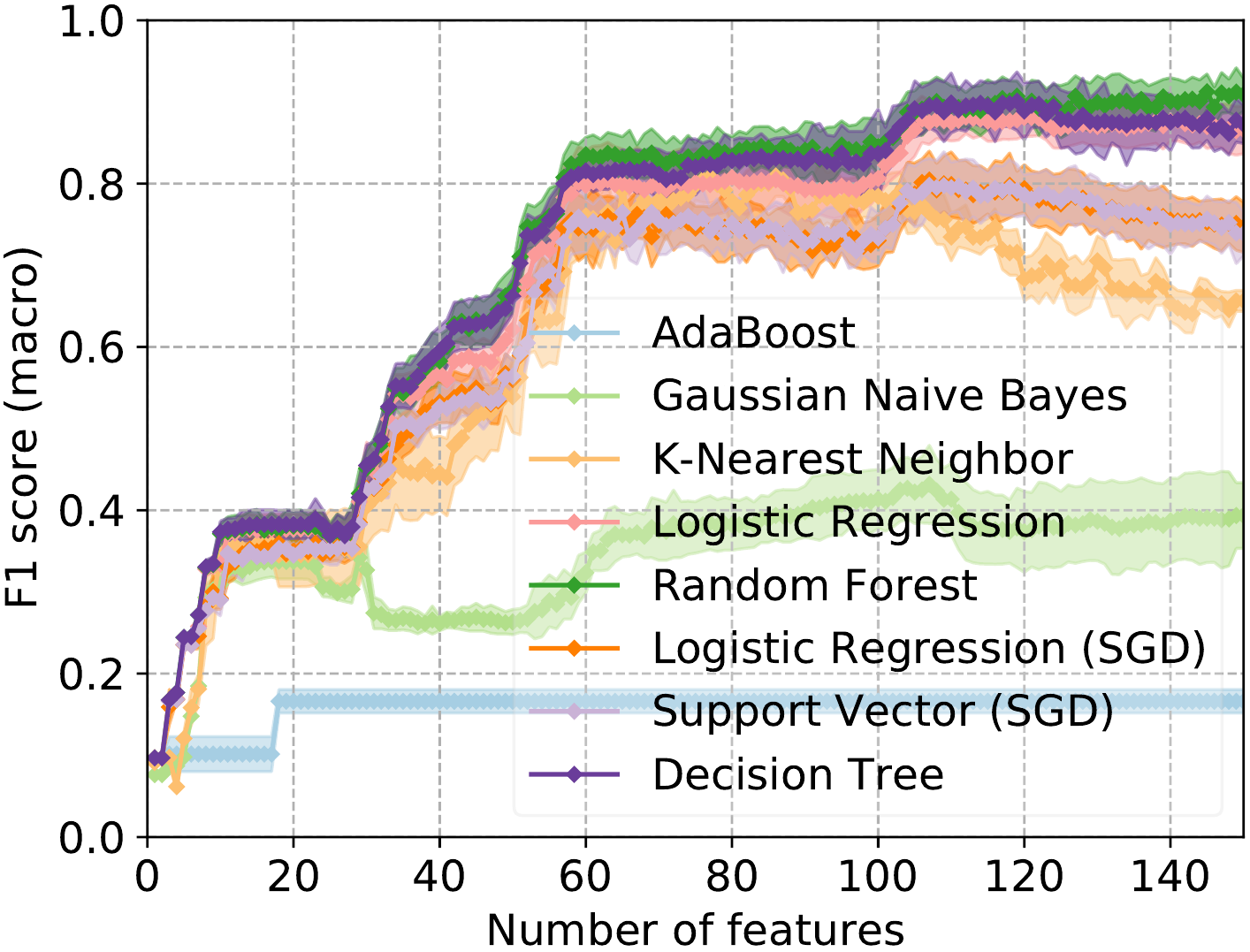} &
    \includegraphics[width=0.32\textwidth]{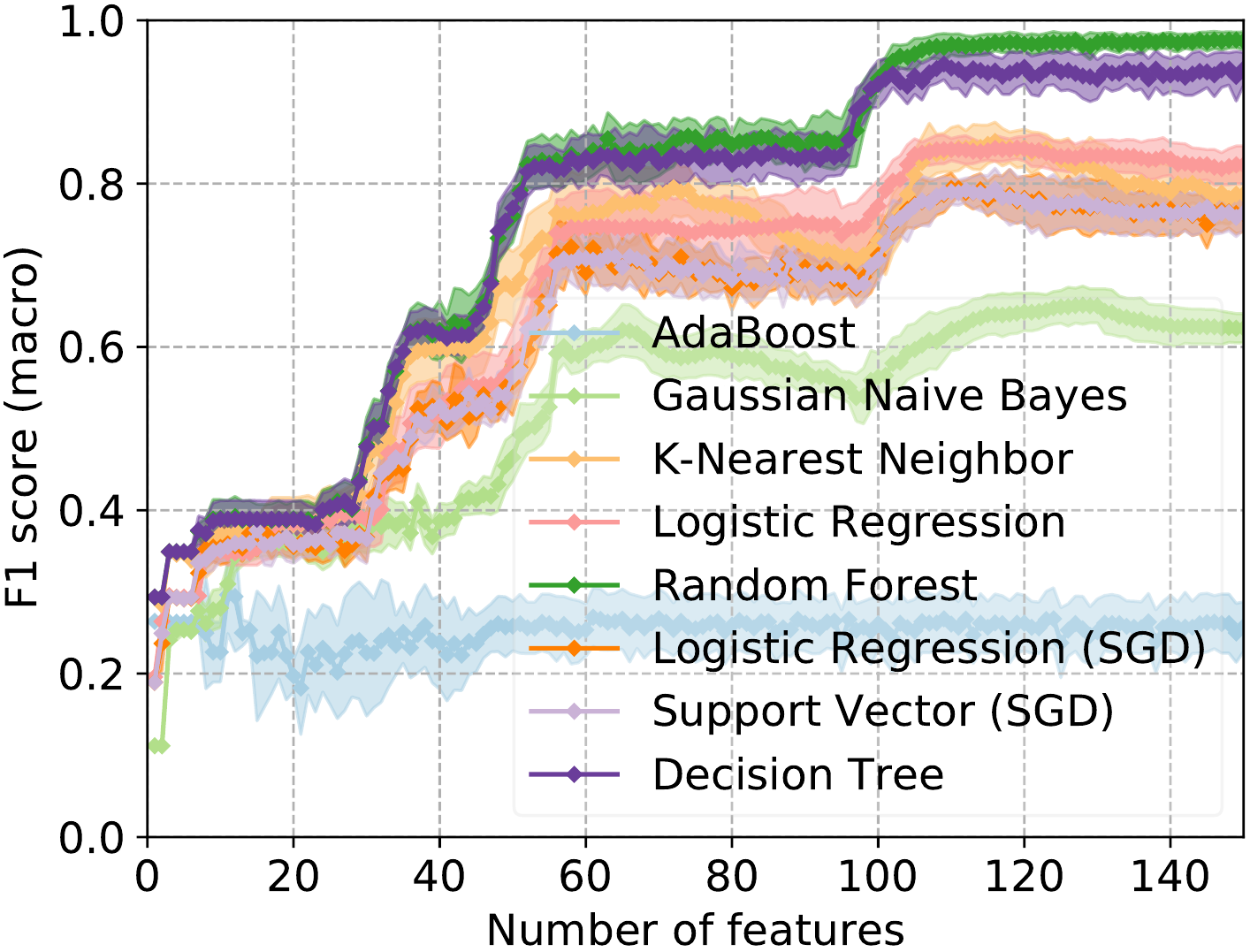} \\

    50 first packet bursts & 50 first byte bursts & 50 first IAT \\
    \includegraphics[width=0.32\textwidth]{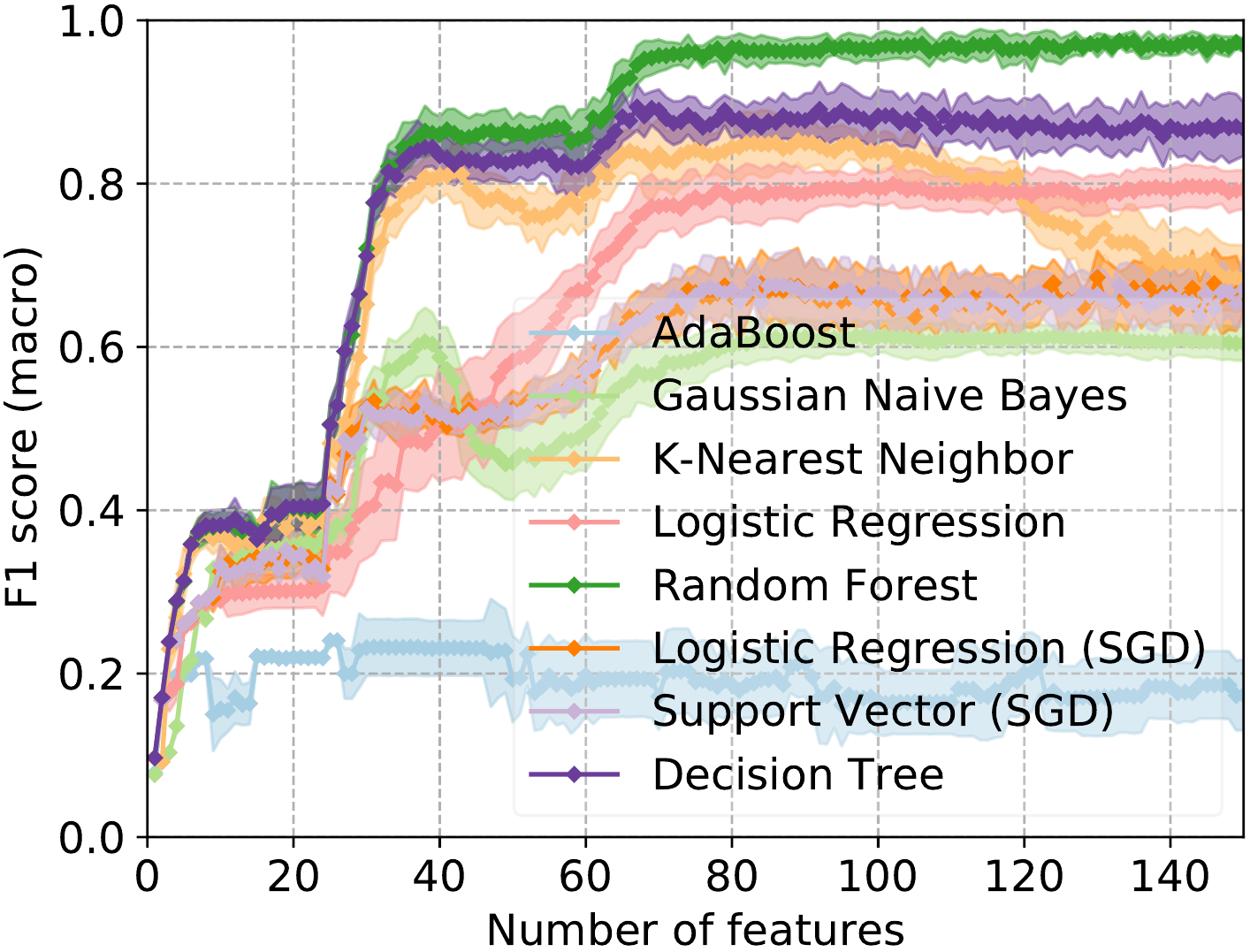} &
    \includegraphics[width=0.32\textwidth]{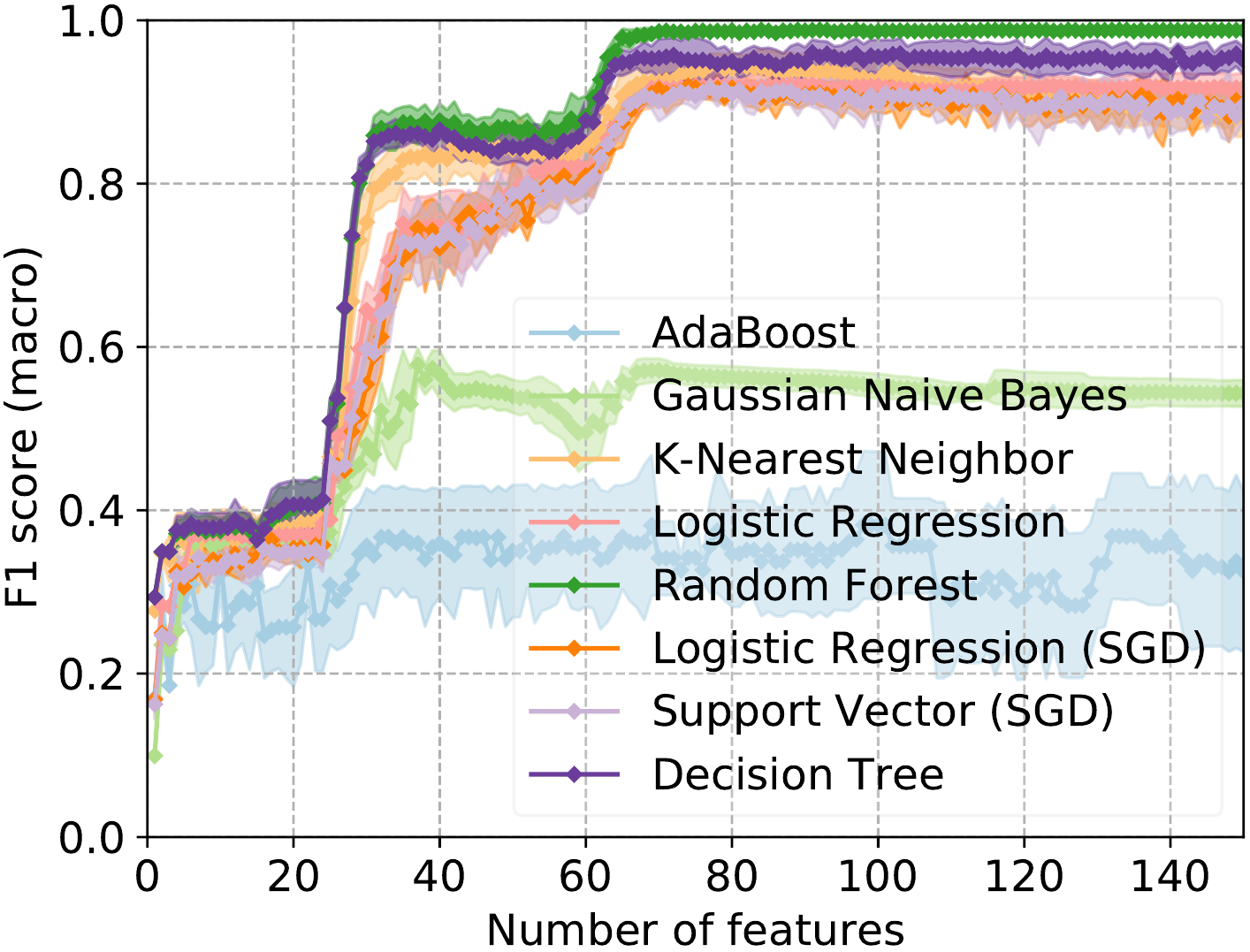} &
    \includegraphics[width=0.32\textwidth]{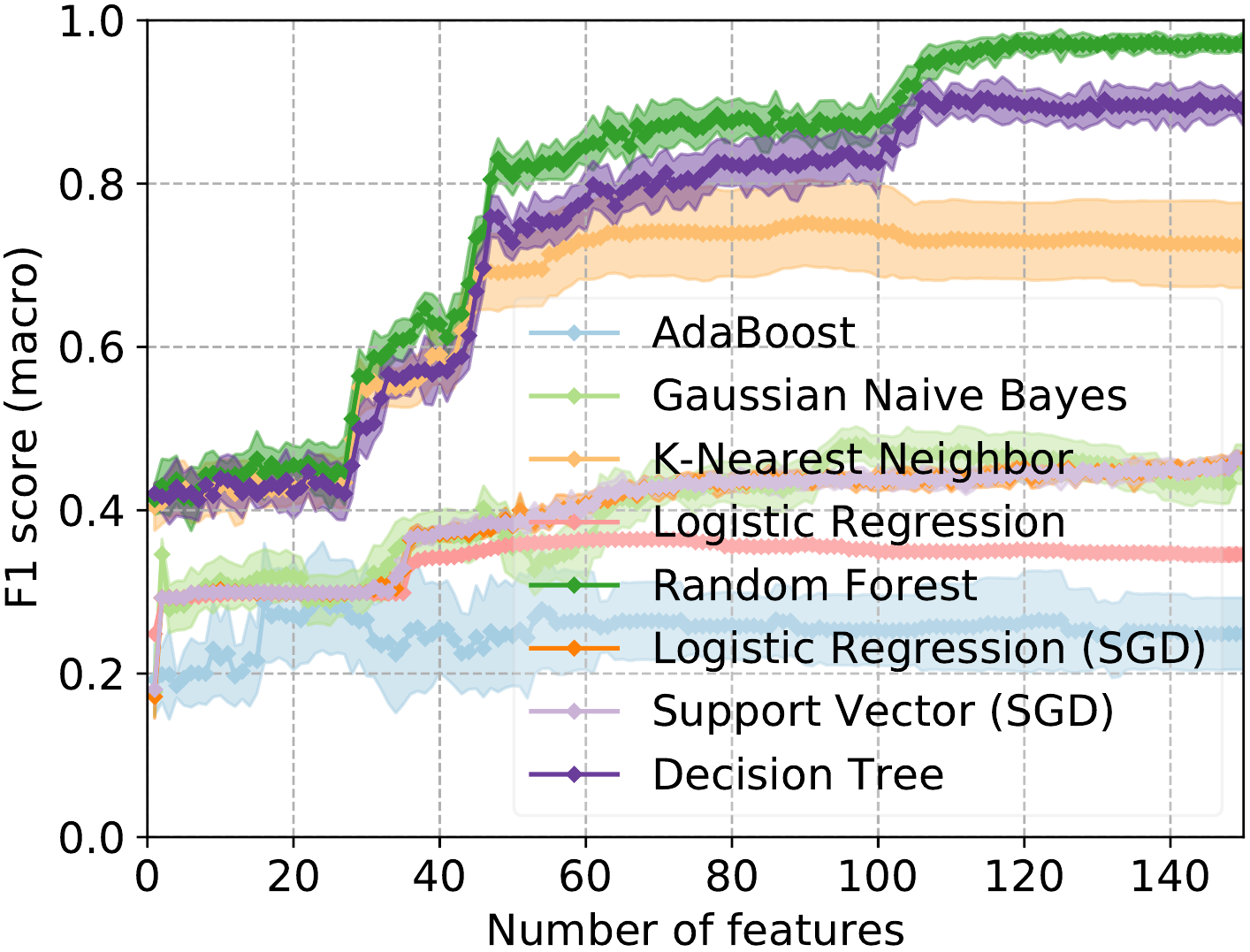} \\
  \end{tabular}
  
  \caption{Application classification inside SSH using 150 first features.
  Shaded areas below and above the curve represent a confidence interval with 99\% confidence level.
  }
  \label{fig:feature_cwa_n150_mlas_ssh}

\end{figure*}

\section{Results}
\label{sec:results}

In this section, we present the performance evaluation for all three steps 
defined in \Cref{sec:pipeline}.
We first present a study of the impact of N where N is the number N first 
feature used, e.g. packet size (see \Cref{sec:n_first_features} and 
\Cref{fig:feature}).
We then compare N first features inside feature families (here packet direction/size-related 
and burst-related) in \Cref{sec:feature_comparison_preliminary}.
Next, we reuse the best N first features inside each family and compare their 
performance in \Cref{sec:feature_comparison_global}.
We then provide a performance evaluation of these best features used 
together along with feature importance analysis (see \Cref{sec:general_results}).
Finally, we address domain generalization regarding change of background network 
traffic in \Cref{sec:dg_untunneled}, and both domain generalization and adversarial 
learning regarding MTU in \Cref{sec:dg_al_mtu}.

We only present the results of application classification inside tunnel for the 
SSH tunnel due to the lack of space.
This tunnel however exhibits the worst performance across all experiments.
Presented results thus always provide a performance lower bound for application 
classification inside tunnels.

In the next sections, we often use a subset of all extracted features (see 
\Cref{sec:feature_extraction}) to compare them.
We however always use the one-hot encoded transport protocol (TCP or UDP) as a 
additional feature to any of these feature subsets.

\begin{figure*}[t!]
  \centering
  \setlength\tabcolsep{0pt}  

  \begin{tabular}{lccccc}
    & Packet direction & Packet size & \parbox{9em}{Packet size from src} & 
    \parbox{9em}{Packet size from dst} & 
    \parbox{10em}{Packet size with direction} \\
    
    \rotatebox{90}{\hspace{1em} Tunnel detection} &
    \includegraphics[width=\widthfate\textwidth]{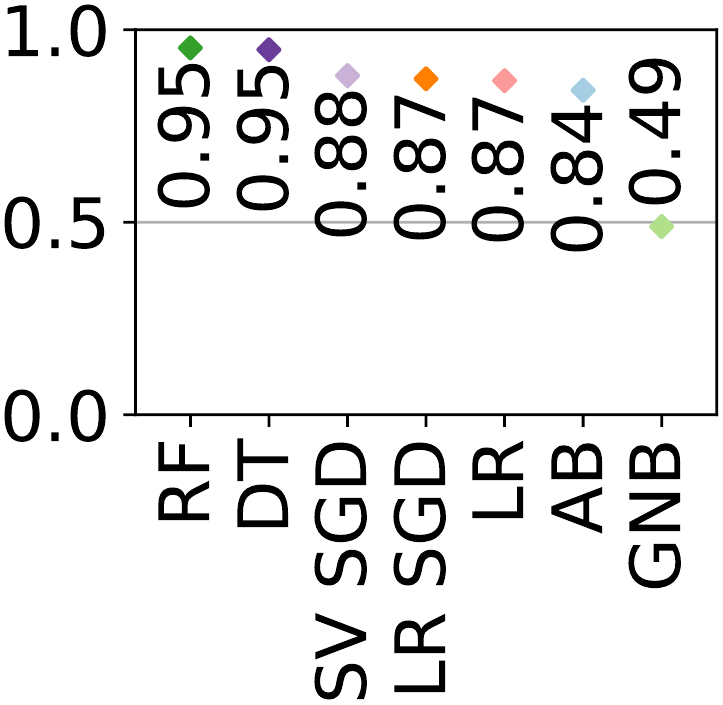} &
    \includegraphics[width=\widthfate\textwidth]{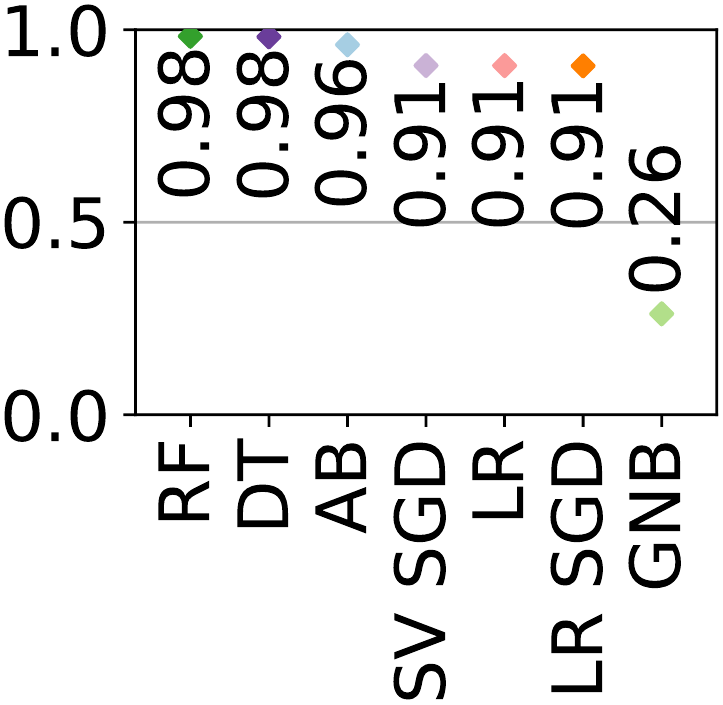} &
    \includegraphics[width=\widthfate\textwidth]{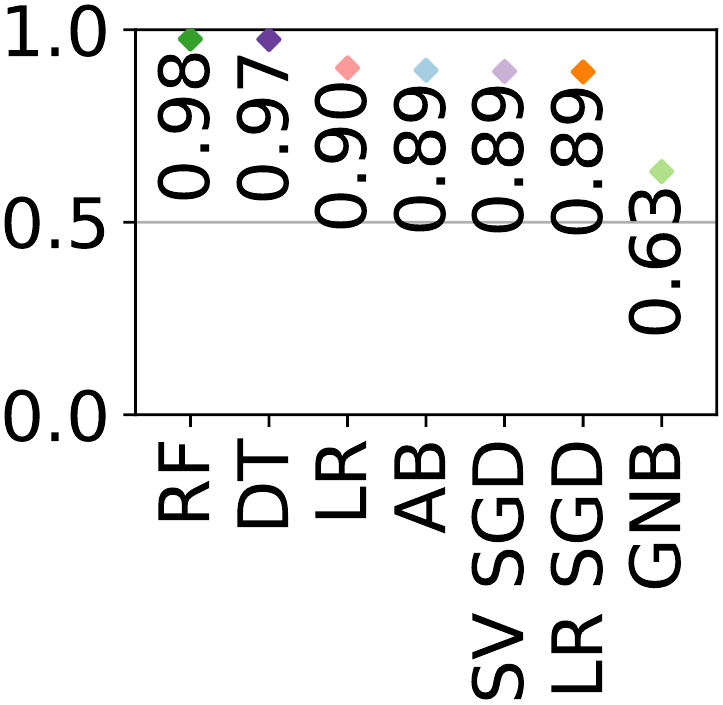} &
    \includegraphics[width=\widthfate\textwidth]{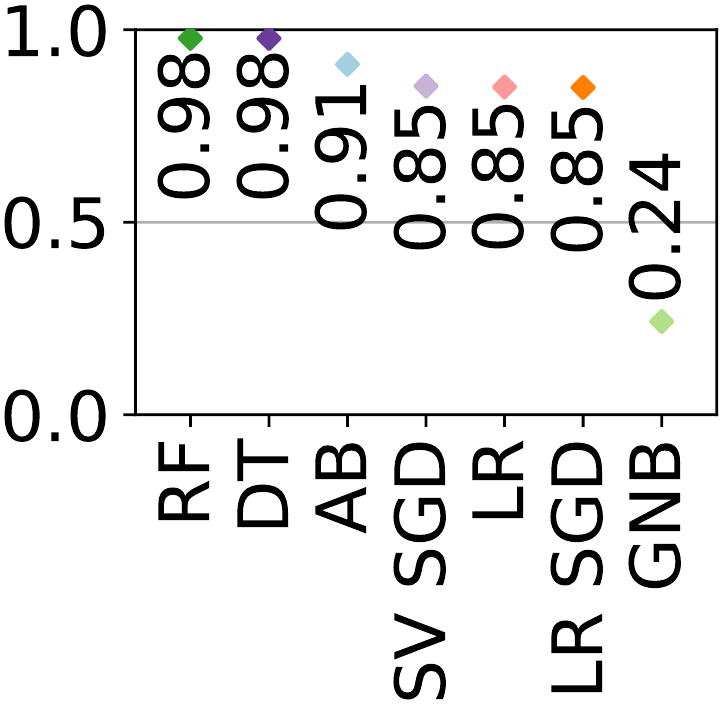} &
    \includegraphics[width=\widthfate\textwidth]{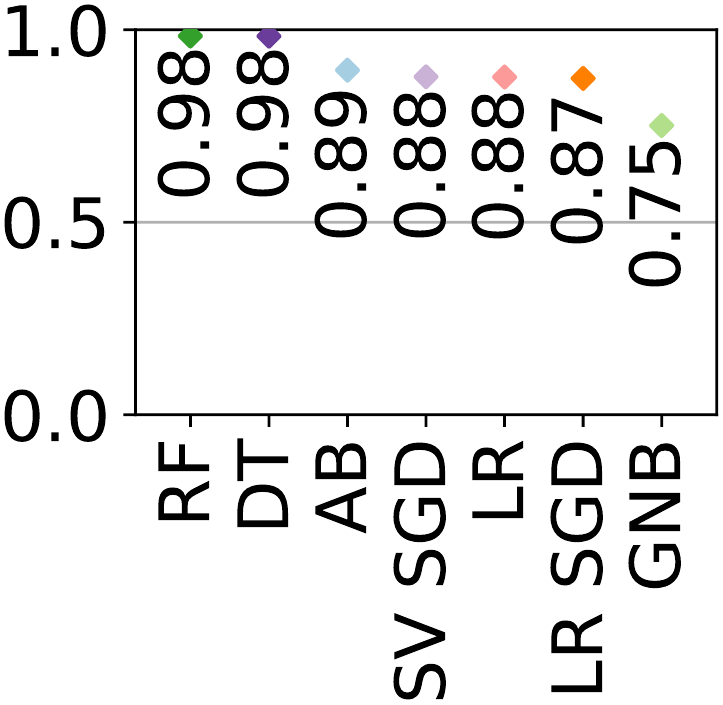} \\
    
    \rotatebox{90}{\hspace{0em} Tunnel classification} &
    \includegraphics[width=\widthfate\textwidth]{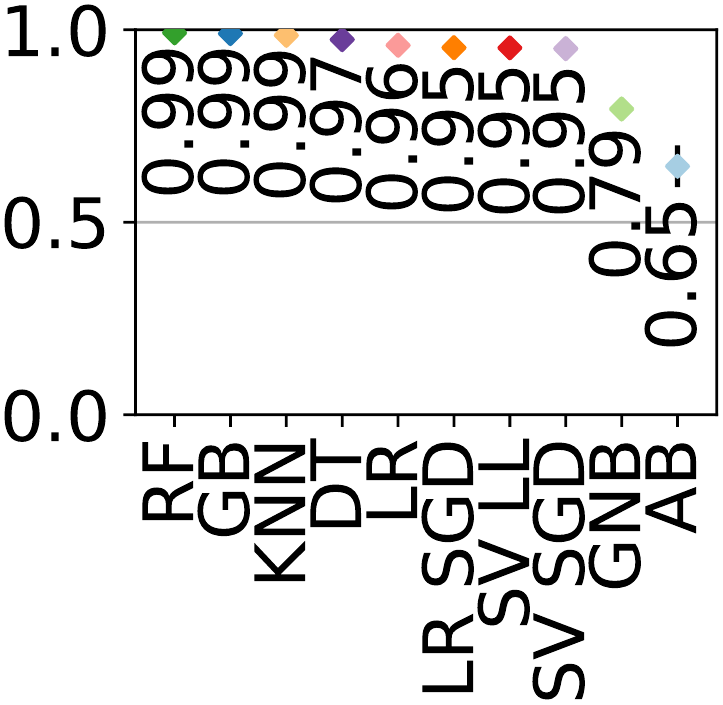} &
    \includegraphics[width=\widthfate\textwidth]{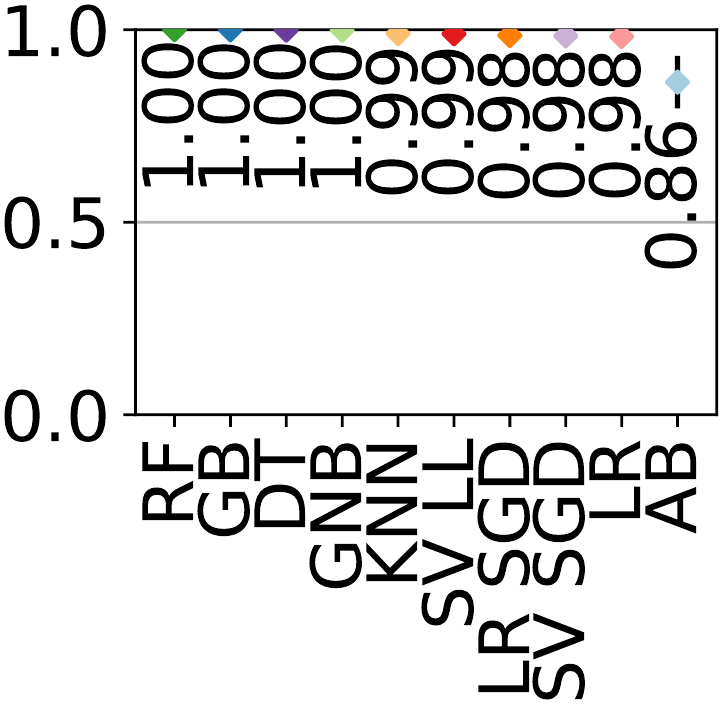} &
    \includegraphics[width=\widthfate\textwidth]{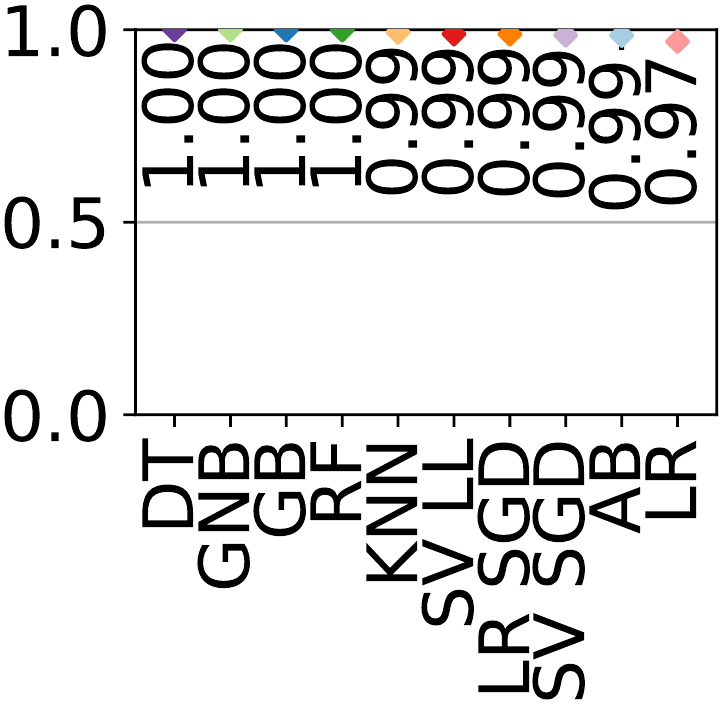} &
    \includegraphics[width=\widthfate\textwidth]{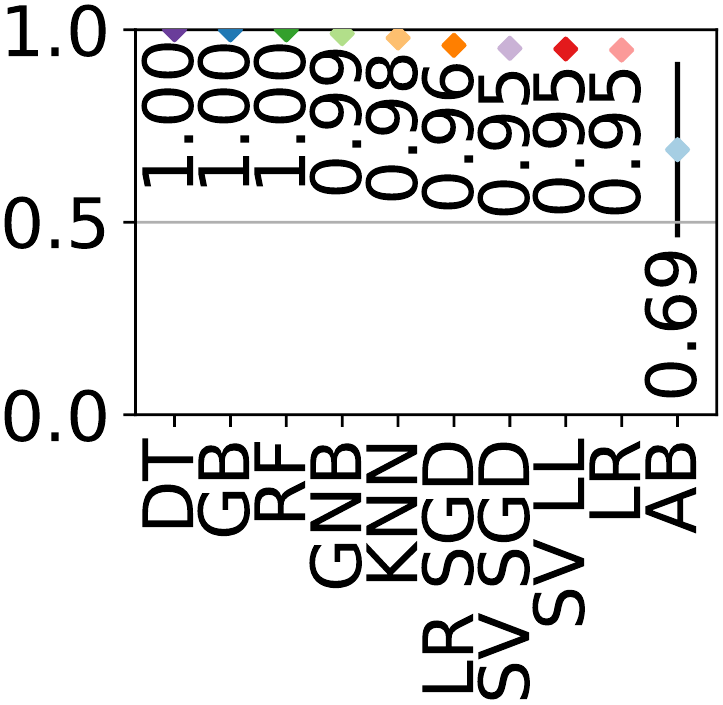} &
    \includegraphics[width=\widthfate\textwidth]{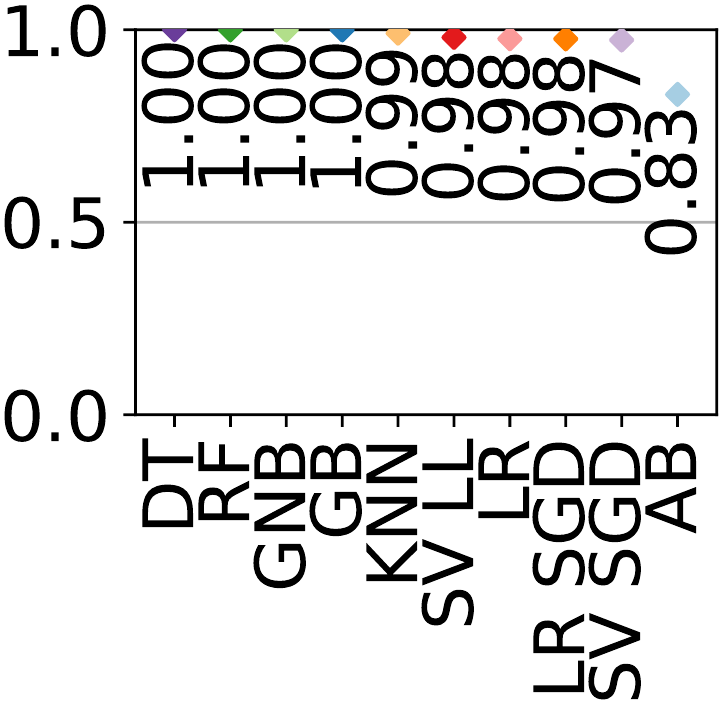} \\

    \rotatebox{90}{\parbox{9em}{\centering Application classification in SSH}} &
    \includegraphics[width=\widthfate\textwidth]{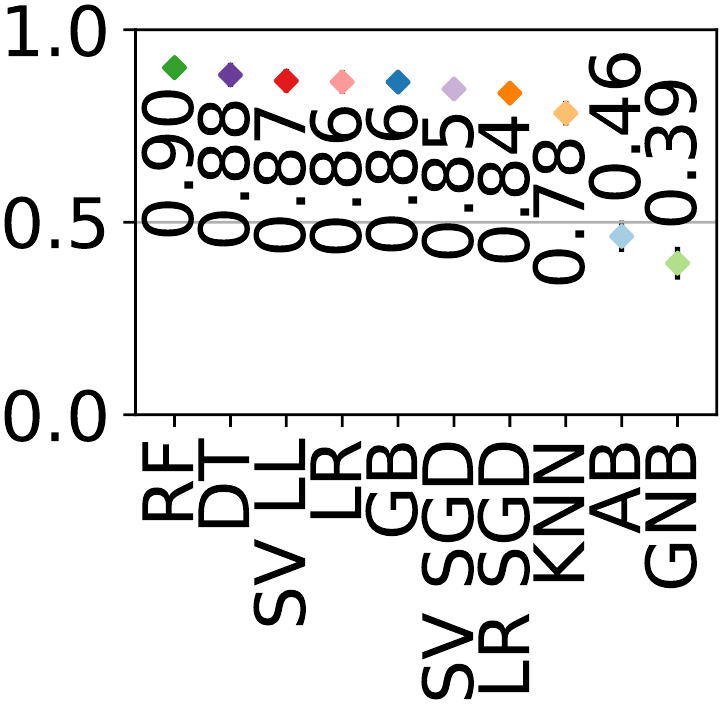} &
    \includegraphics[width=\widthfate\textwidth]{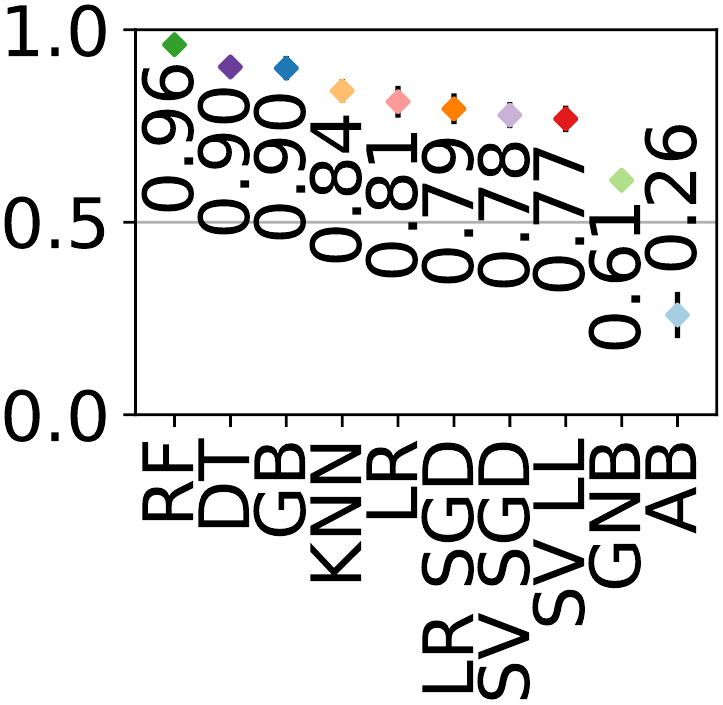} &
    \includegraphics[width=\widthfate\textwidth]{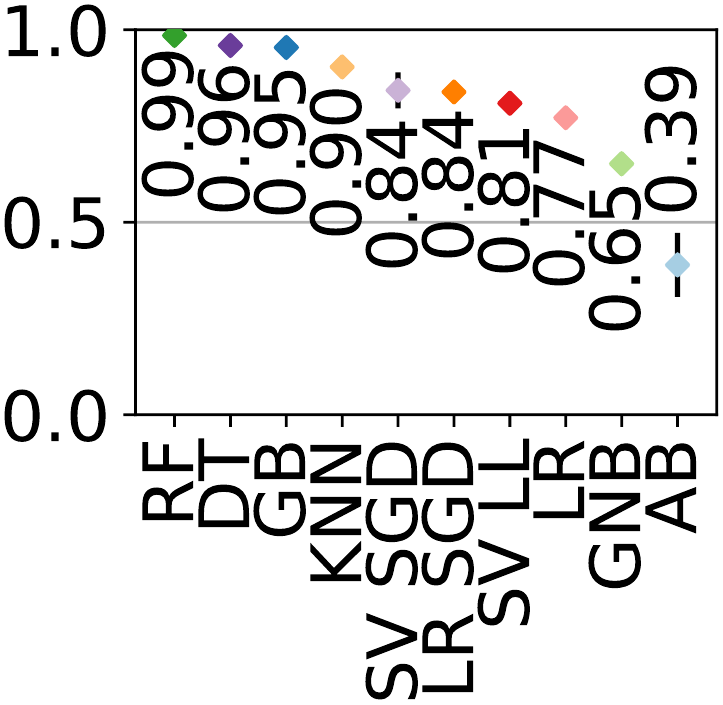} &
    \includegraphics[width=\widthfate\textwidth]{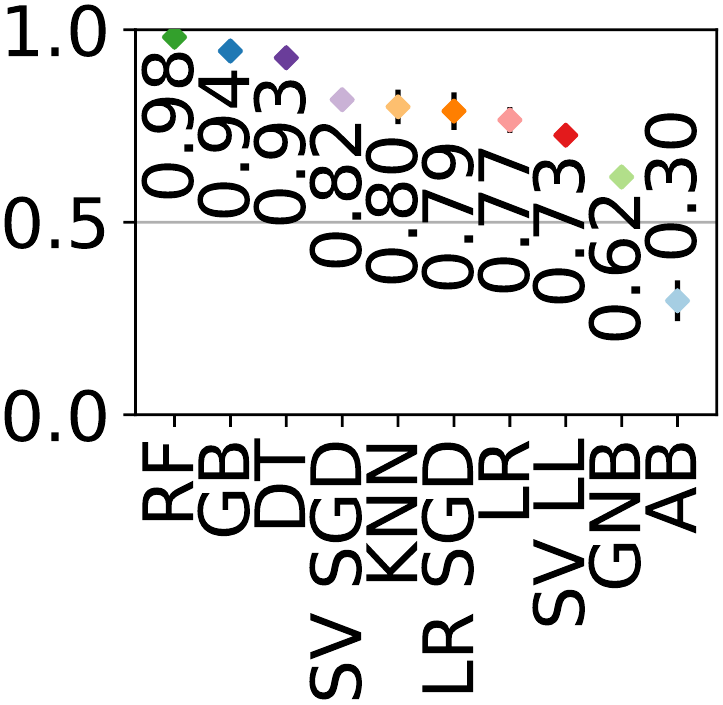} &
    \includegraphics[width=\widthfate\textwidth]{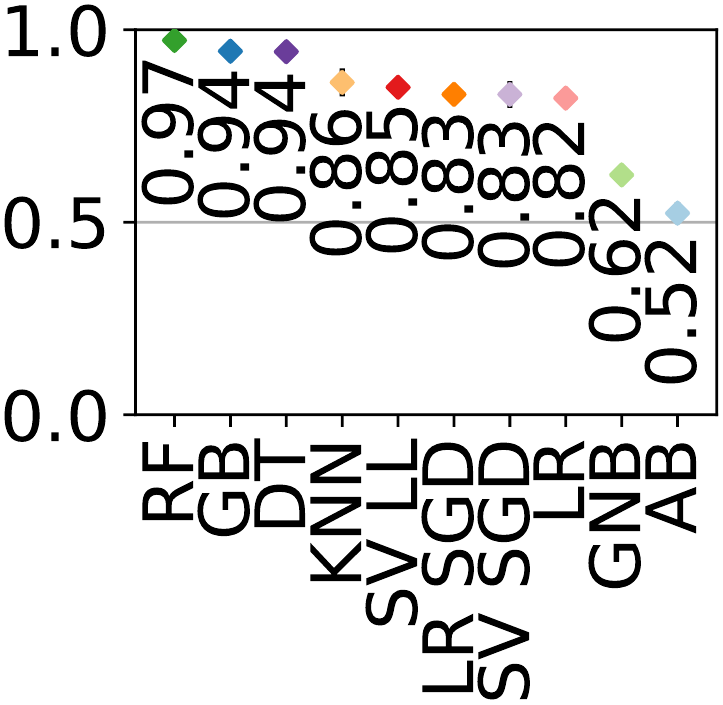} \\    
  \end{tabular}
  
  \caption{Comparison of packet size, packet size from source, packet size from destination, packet direction and packet size with direction for all pipeline steps.
  Error bars represent a confidence interval with 99\% confidence level.}
  \label{fig:feature_owcwtcwa_d_ps_pso_psr_dps}
\end{figure*}

\begin{figure*}[t!]
  \centering
  \setlength\tabcolsep{0pt}
    
  \begin{tabular}{lccccc}
    & Packet burst & Byte burst & \parbox{9em}{Byte burst from src} & \parbox{9em}{Byte burst from dst} \\
  
    \rotatebox{90}{\hspace{1em} Tunnel detection} &
    \includegraphics[width=\widthfate\textwidth]{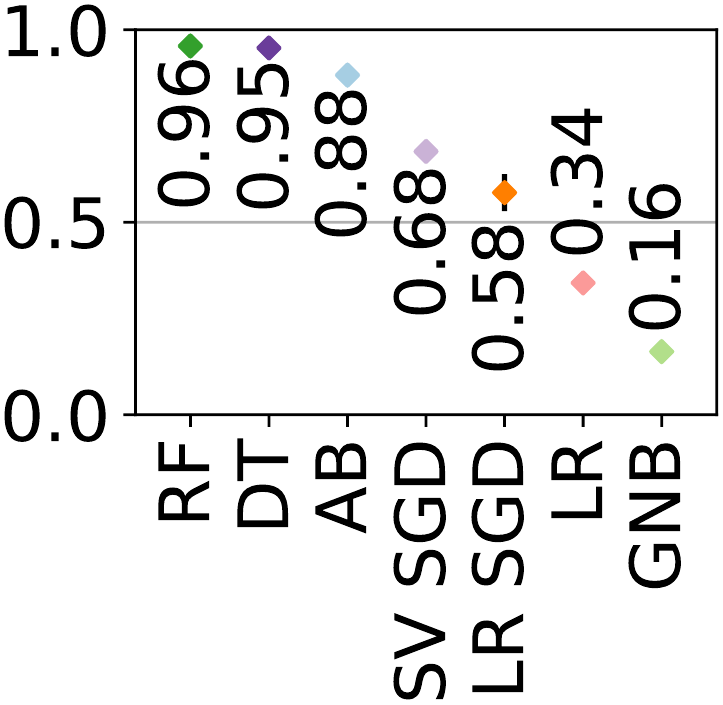} &
    \includegraphics[width=\widthfate\textwidth]{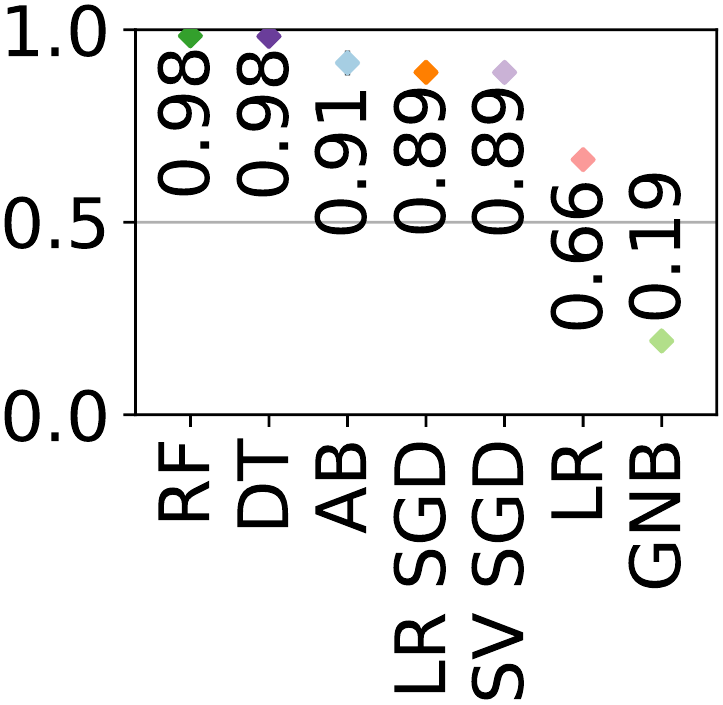} &
    \includegraphics[width=\widthfate\textwidth]{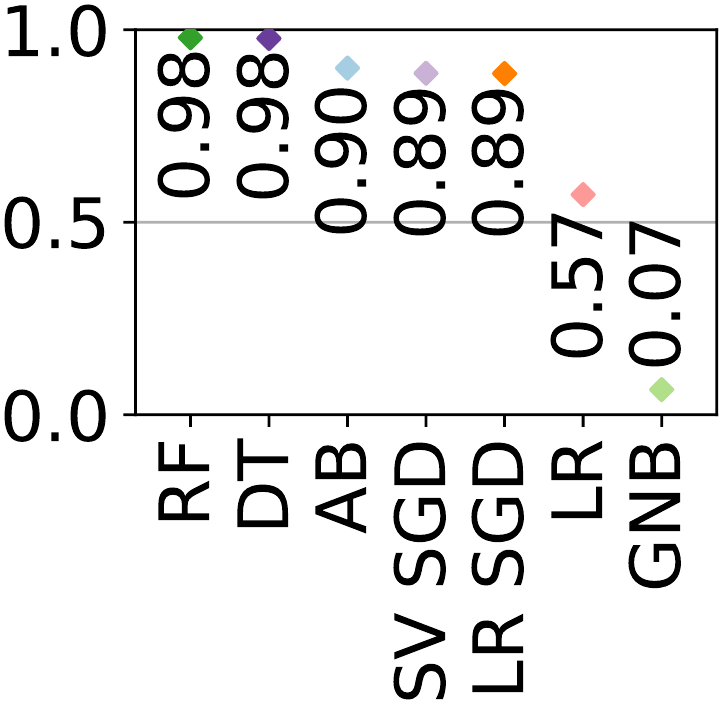} &
    \includegraphics[width=\widthfate\textwidth]{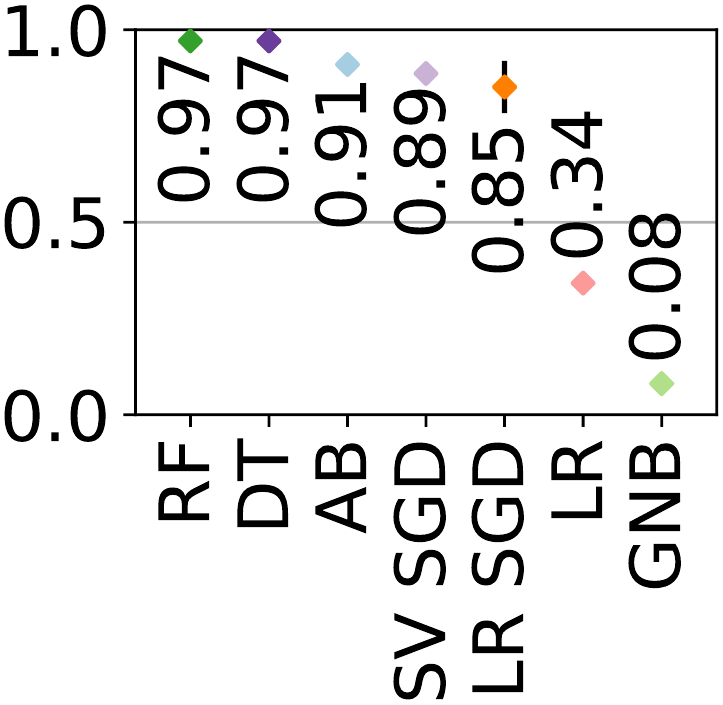} \\
    
    \rotatebox{90}{\hspace{0em} Tunnel classification} &
    \includegraphics[width=\widthfate\textwidth]{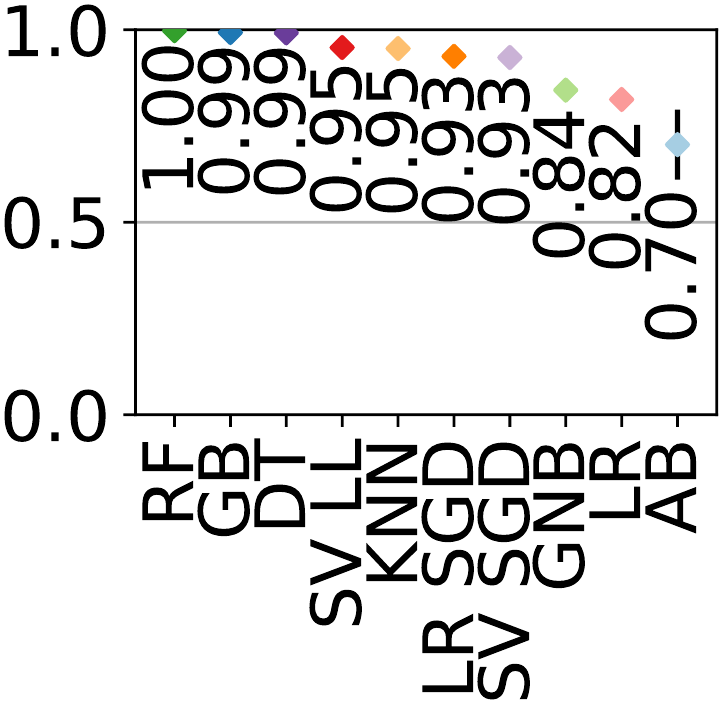} &
    \includegraphics[width=\widthfate\textwidth]{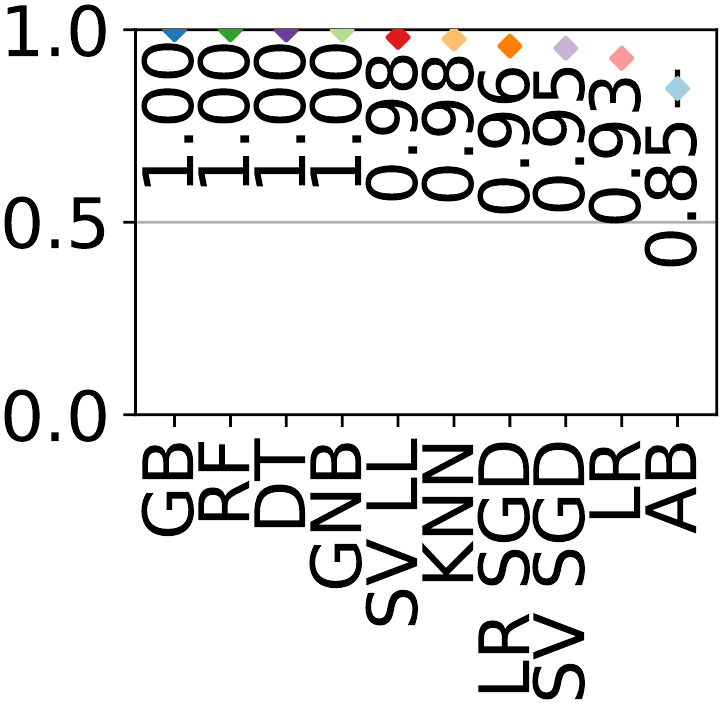} &
    \includegraphics[width=\widthfate\textwidth]{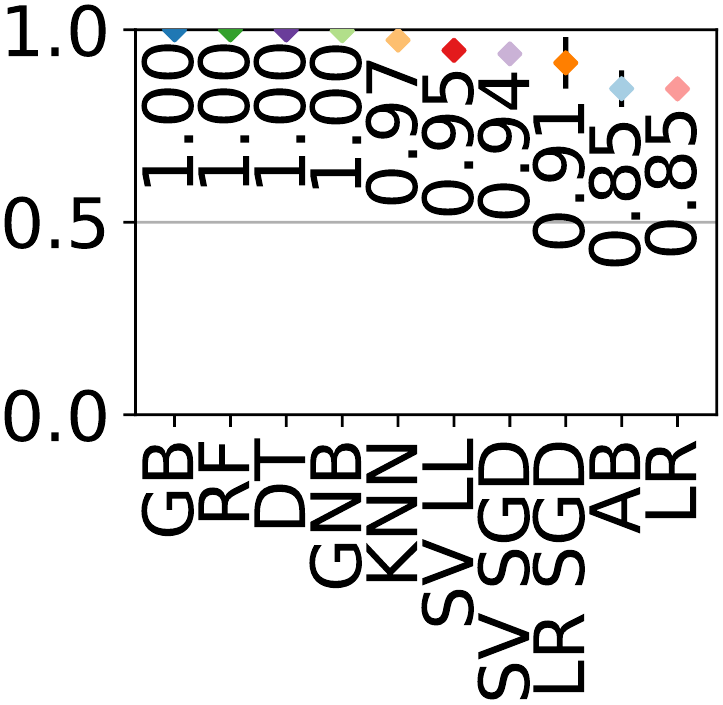} &
    \includegraphics[width=\widthfate\textwidth]{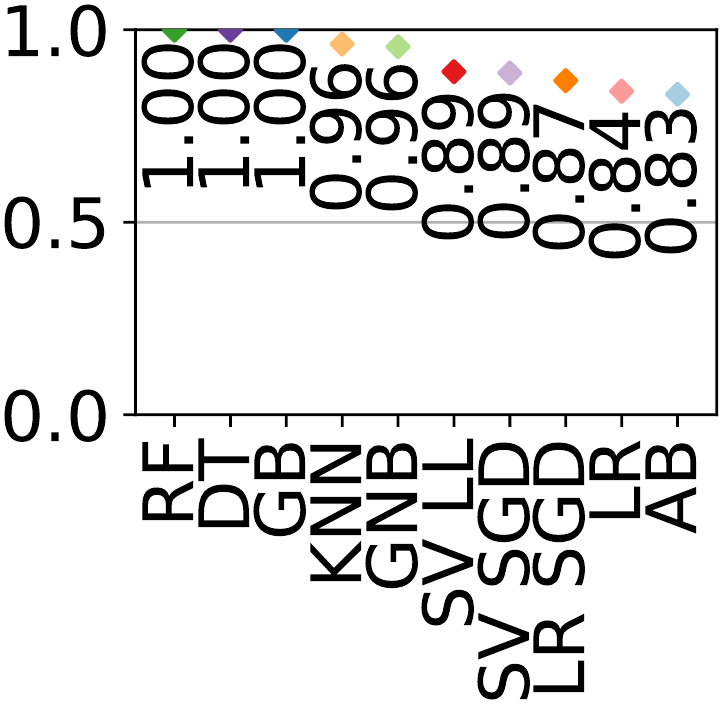} \\

    \rotatebox{90}{\parbox{9em}{\centering Application classification in SSH}} &
    \includegraphics[width=\widthfate\textwidth]{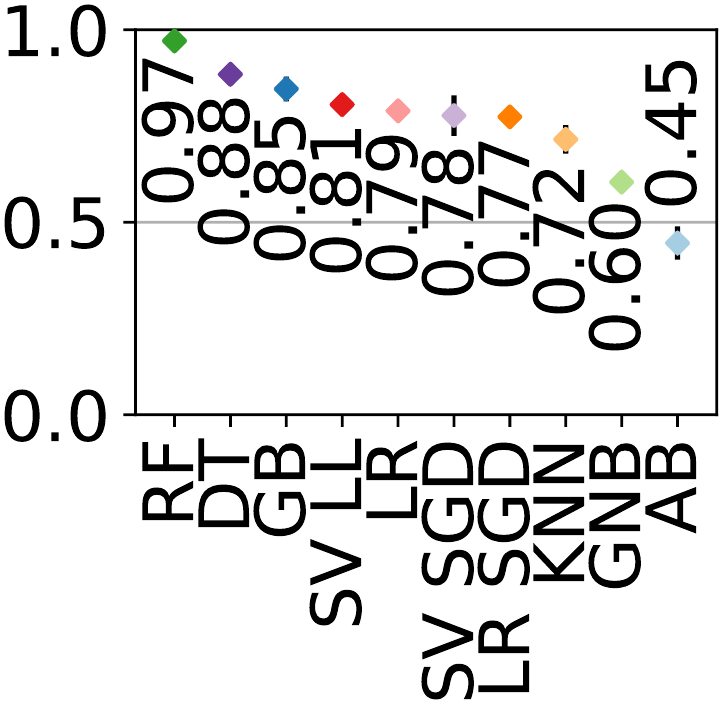} &
    \includegraphics[width=\widthfate\textwidth]{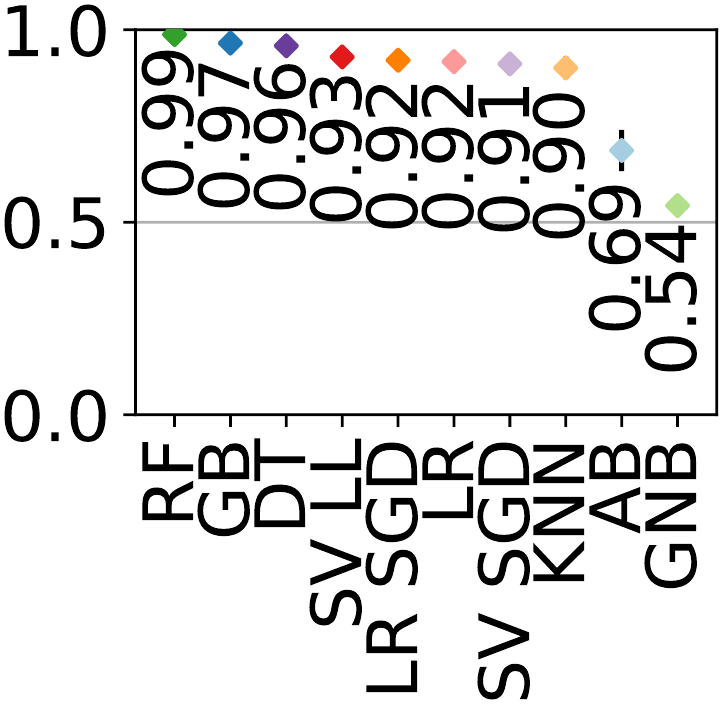} &
    \includegraphics[width=\widthfate\textwidth]{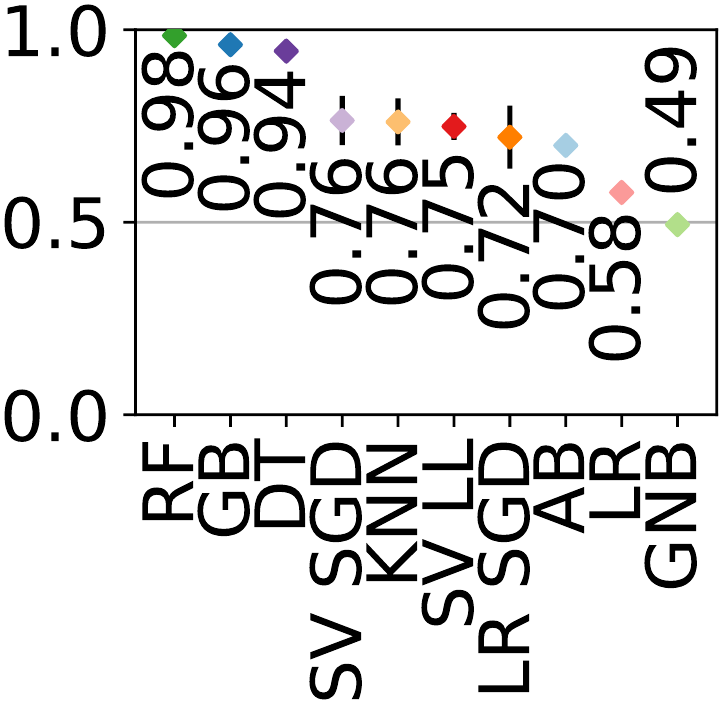} &
    \includegraphics[width=\widthfate\textwidth]{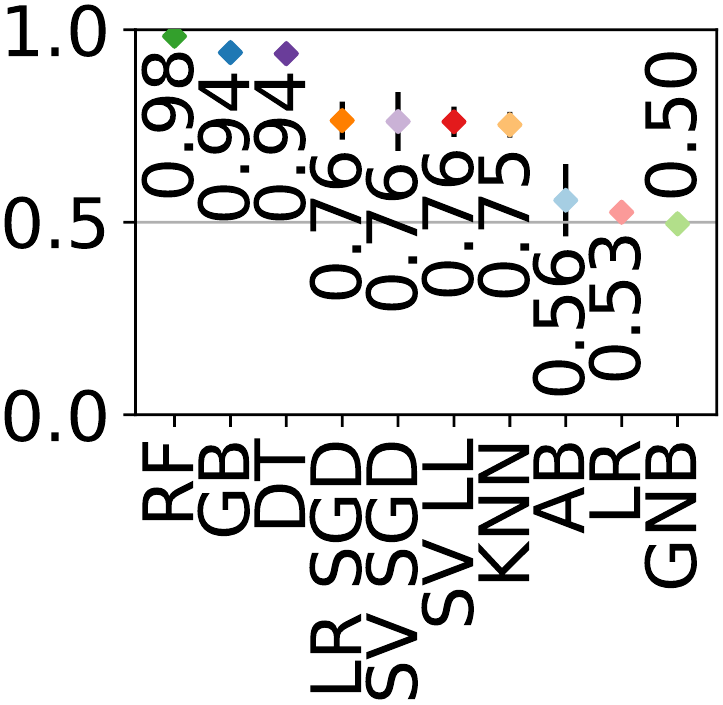} \\
    
  \end{tabular}
  
  \caption{Comparison of packet burst, byte burst, byte burst from source, and byte burst from destination for all pipeline steps.
  Error bars represent a confidence interval with 99\% confidence level.}
  \label{fig:feature_owcwtcwa_pb_bb_bbr_bbo}
\end{figure*}

\subsection{N first feature}
\label{sec:n_first_features}

\jm{TODO: compare feature importance with F1 score jump on nfirst}

In this section we analyze the impact of N on performance, where is N the number of N first 
feature used.
N first features are: packet direction, packet size, packet size with direction 
encoded in sign, packet burst, byte burst, and inter-arrival time.
We do not present elapsed time since flow start due to the lack of space and 
because its performances are always lower than IAT.
\jm{check this again}
Due to the computational cost of testing mutliple values of N, we do not 
perform a grid search to find optimal parameters inside the inner loop, and 
simply use default parameters from \Cref{table:ml_parameters}.
We address all pipeline steps: tunnel detection, tunnel classification, and 
application classification.

\jm{TODO: add text of pso/psr bbo/bbr when data is ready}

\jm{TODO: split netflow v5/v9 into base and mod (with bonus)}

\subsubsection{Tunnel detection and classification}

\Cref{fig:feature_ow_n} (resp. \Cref{fig:feature_cwt_n}) presents the F1 score obtained with N values 
between 1 and 50 for tunnel detection (resp. classification).
For both tasks, we use N between 1 and 50 because N values above 20 do 
not provide significant performance improvement for the best algorithm (random forest).

Algorithms exhibits widely varying performance.
Random forest is the best one.
When N increases, decision tree and random forest exhibit increasing performance 
for all features.
Other algorithms' performances sometimes decrease when N increases (e.g. N 
between 20 and 25 with packet direction and Gaussian naive Bayes for tunnel 
detection on \Cref{fig:feature_ow_n}).

We then compare feature performance using the F1 score for N equal to 50 with 
the best performing algorithms (decision tree and random forest).
Packet size, packet size with direction, byte burst and IAT provide very good 
overall results.
Packet direction and burst exhibit slightly lower performance.

In terms of smallest N to reach maximum performance, ie smallest possible 
memory usage to reach optimal performance, byte burst is the best 
feature, followed by packet size with direction and IAT, then packet size,
and finally packet burst and packet direction.
Byte burst is the most efficient feature in terms of memory usage.
It however does not mean that byte burst is quick time-wise to reach its best 
performance level because burst position in a flow is not directly linked 
to packet position.
Indeed, the fourth byte burst may actually be located at the 30th packet which 
would make byte burst slower than packet size with direction.

When one compares tunnel detection and tunnel classification, the smallest N 
values to reach maximum performance are similar.
Algorithm performances are however better and vary less across algorithms for 
tunnel classification than for tunnel detection.

\subsubsection{Application classification}

\Cref{fig:feature_ow_n} presents the F1 score obtained with N values 
between 1 and 150 for application classification inside SSH tunnels.
We use N between 1 and 150 because N values above 120 do not provides 
significant performance improvement for the best algorithm (random forest).
We do not present result for other tunnels due to the lack of space.
\jm{TODO: add test about TR ;TR????}

Results are similar to the tunnel detection and classification use cases.
Algorithm performance are diverse.
Random forest exhibits the best F1 scores, and its performances increase 
when N increases for all features.
This is not the case for other algorithms (see N between 100 and 150 with 
packet direction and Gaussian naive Bayes).
Overall, byte burst is the best feature.
Packet size, packet size with direction, packet burst are close second.
Packet direction exhibits the worst performance.
In terms of smallest N to reach maximum performance, byte burst is the best 
feature, closely followed by packet burst, then packet direction, packet size 
with direction, packet direction and IAT, in that order.

\subsubsection{Summary}

In this preliminary comparison, we determine that using 50 (resp. 150) features 
is enough for tunnel detection and tunnel classification (resp. application 
classification).
Overall, random forest is the best algorithm.
Byte burst is the best feature in terms of smallest N to reach the best 
performance.

\begin{figure*}[t!]
  \centering
  \setlength\tabcolsep{0pt}
    
  \begin{tabular}{lccccc}
    & \parbox{10em}{Packet size with direction} & Byte burst & IAT & Netflow v5 & Netflow v9 \\
    
    \rotatebox{90}{\hspace{1em} Tunnel detection} &
    \includegraphics[width=\widthfate\textwidth]{fa/td/te_td_am1500unaupa_n50_dps_mlad_all_cropped.pdf} &
    \includegraphics[width=\widthfate\textwidth]{fa/td/te_td_am1500unaupa_n50_bb_mlad_all_cropped.pdf} &
    \includegraphics[width=\widthfate\textwidth]{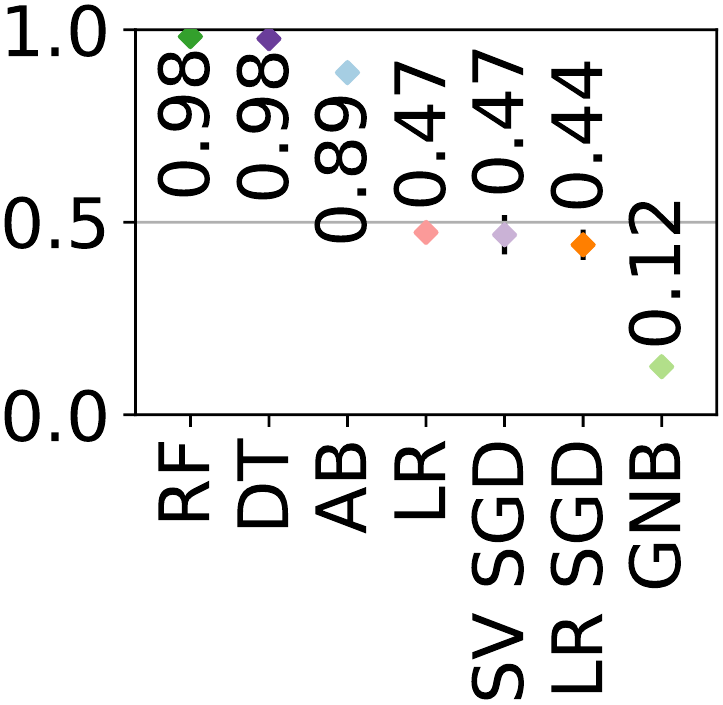} &
    \includegraphics[width=\widthfate\textwidth]{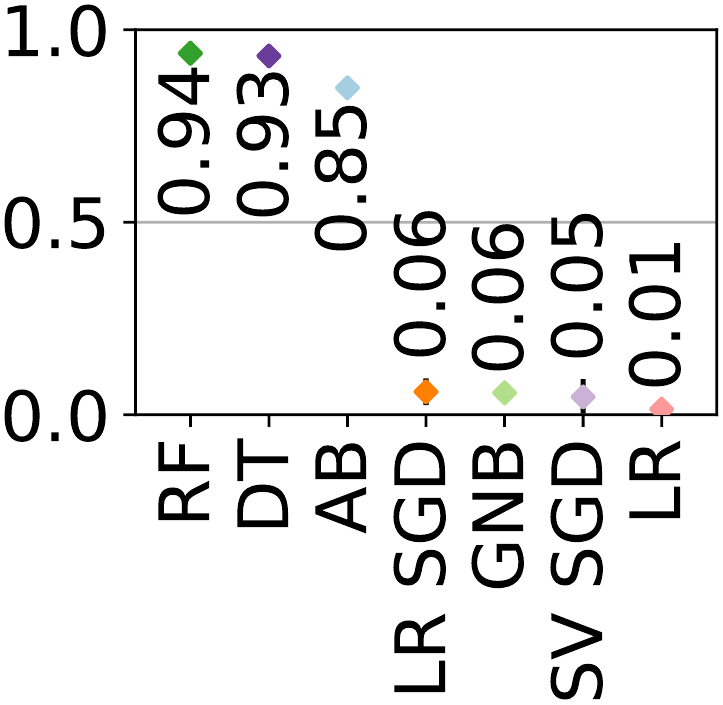} &
    \includegraphics[width=\widthfate\textwidth]{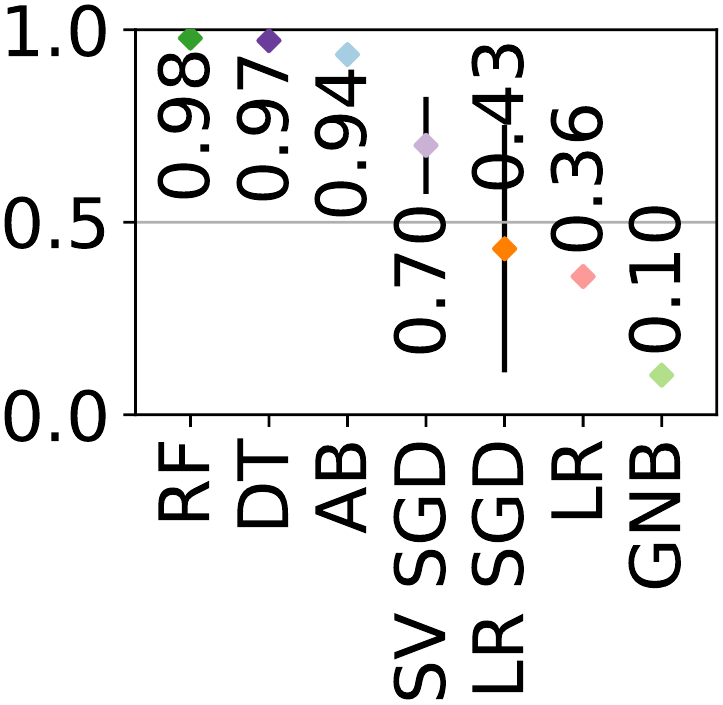} \\
    
    \rotatebox{90}{\hspace{0em} Tunnel classification} &
    \includegraphics[width=\widthfate\textwidth]{fa/tc/te_tc_am1500unaupa_n50_dps_mlad_at_cropped.pdf} &
    \includegraphics[width=\widthfate\textwidth]{fa/tc/te_tc_am1500unaupa_n50_bb_mlad_at_cropped.pdf} &
    \includegraphics[width=\widthfate\textwidth]{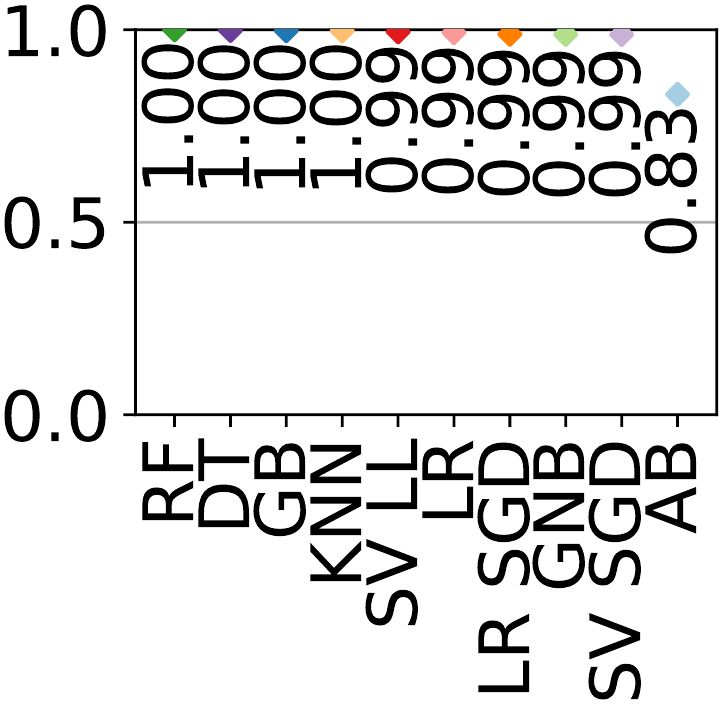} &
    \includegraphics[width=\widthfate\textwidth]{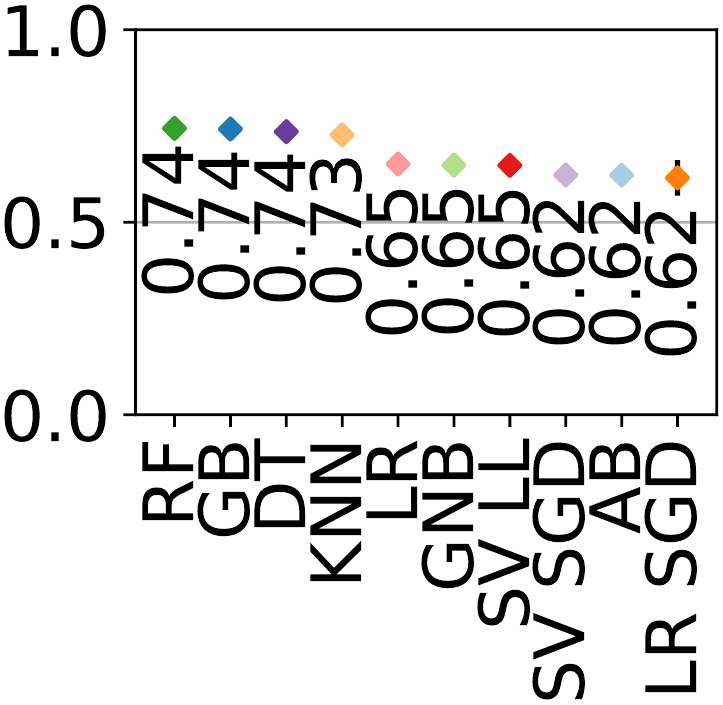} &
    \includegraphics[width=\widthfate\textwidth]{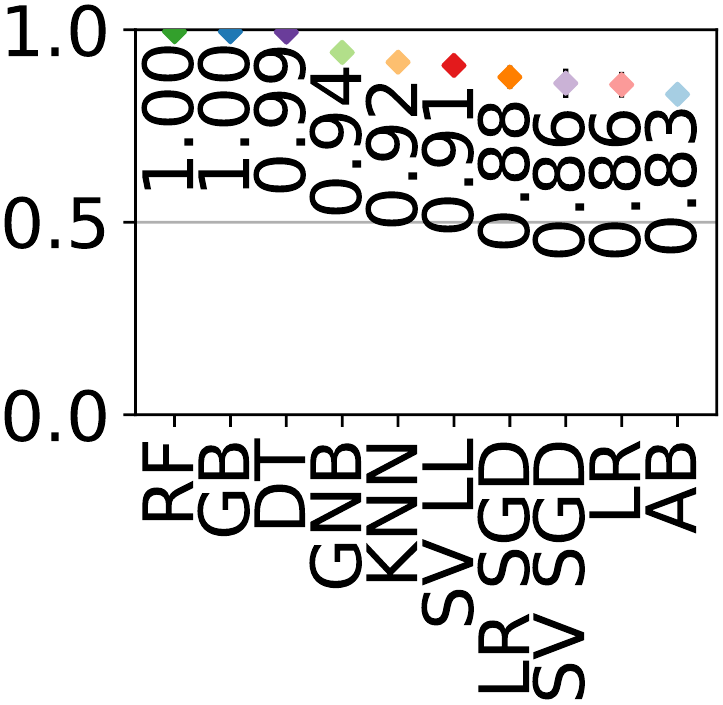} \\

    \rotatebox{90}{\parbox{9em}{\centering Application classification in SSH}} &
    \includegraphics[width=\widthfate\textwidth]{fa/ac/te_ac_am1500unaupa_n150_dps_mlad_ssh_cropped.pdf} &
    \includegraphics[width=\widthfate\textwidth]{fa/ac/te_ac_am1500unaupa_n150_bb_mlad_ssh_cropped.pdf} &
    \includegraphics[width=\widthfate\textwidth]{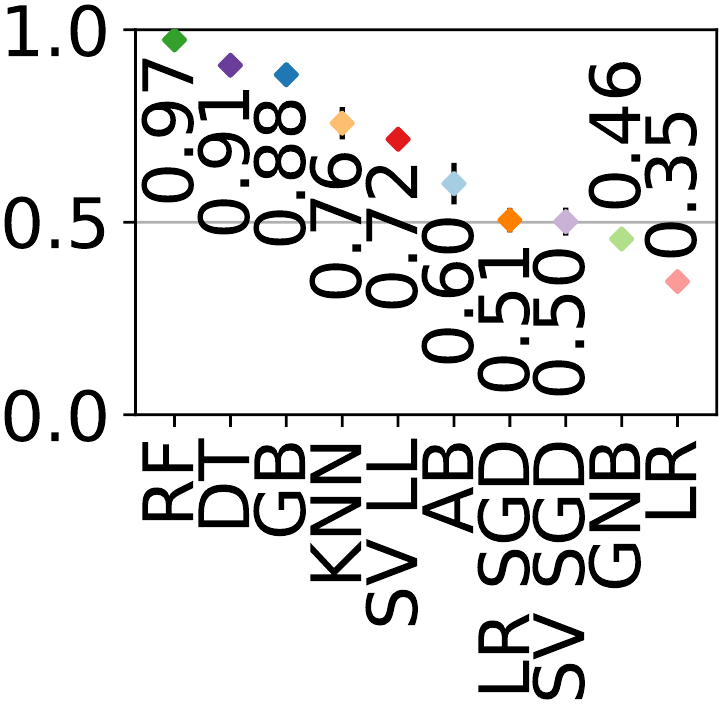} &
    \includegraphics[width=\widthfate\textwidth]{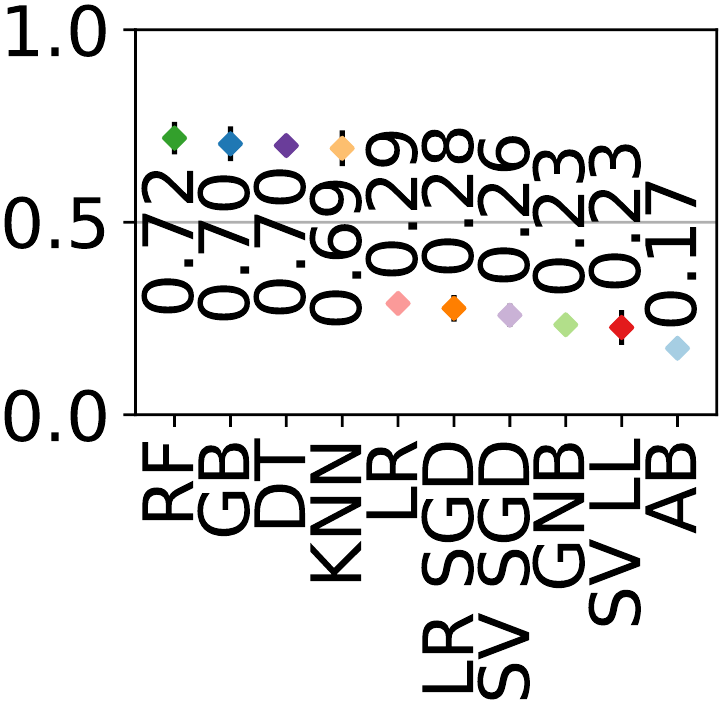} &
    \includegraphics[width=\widthfate\textwidth]{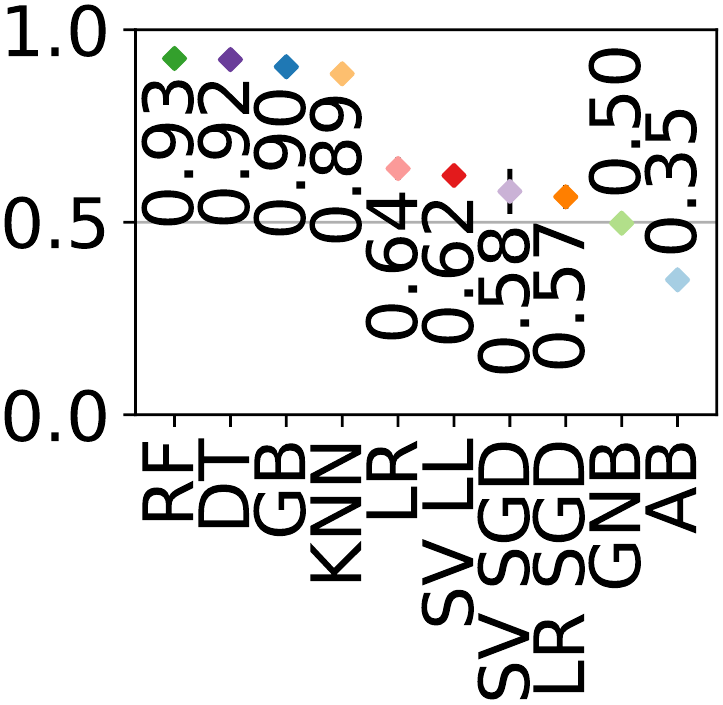} \\
    
  \end{tabular}
  
  \caption{Comparison of packet size with direction, byte burst and IAT for all pipeline steps.
  Error bars represent a confidence interval with 99\% confidence level.
  \jm{TODO: fix figure for tunnel detection and Netflow v9}
}
  \label{fig:feature_owcwt_dps_netflow_bb_i}
\end{figure*}

\subsection{Features}
\label{sec:feature_comparison}

\jm{TODO: add comparison with different payload size?}

In this section, we compare features: first, among feature families (e.g. packet 
size vs packet size with direction or packet burst vs byte burst), then across 
families using the best features from each families (e.g. packet size with 
direction vs byte burst).

\subsubsection{Preliminary feature comparison}
\label{sec:feature_comparison_preliminary}

We here analyze performance inside each feature families: packet-related 
(with packet size, packet direction, packet size from source, packet size 
from destination, packet size with direction in sign), and burst-related
(packet burst, byte burst, byte burst from source, byte burst from destination).

\paragraph{Packet direction, packet size from source, from destination, from both source and destination, and packet size with direction}

\Cref{fig:feature_owcwtcwa_d_ps_pso_psr_dps} pictures the F1 scores obtained
with ML algorithms for packet direction and feature related to packet size.
F1 scores obtained with packet size from source are always better than packet 
size from destination.
Packet size from source is also always better than packet size, except for 
tunnel detection with algorithms such as logistic regression, support vector 
and AdaBoost.
Packet direction is almost always worse than packet size from source, except for 
application classification in SSH with AdaBoost.
Packet size with direction and packet size from source exhibits very similar 
performances.
In the remainder of the paper, we arbitrarily choose to only consider packet size 
with direction.

\jm{fix missing value for app classif/adaboost/packet direction}

\paragraph{Packet burst and byte burst from source/destination/both}

\Cref{fig:feature_owcwtcwa_pb_bb_bbr_bbo} pictures the F1 scores obtained
with ML algorithms for features related to bursts.
F1 scores obtained with packet burst are always worse than byte burst 
(except for SSH application classification with Gaussian naive Bayes).
Byte burst from source and destination does not yield better results than byte burst.

\paragraph{Byte burst bigram and trigram}

We now explore the use of bigram and trigram on byte burst values.
We do not use a figure due to the lack space but one is provided in appendix 
(see \Cref{fig:feature_owcwtcwa_bb2go}).

Byte burst bigram used alone and in combination with byte burst does not 
improve results, except for IPsec-ESP with Gaussian Naive Bayes and OpenVPN TCP 
with AdaBoost and Decision Tree.
Byte burst trigram used alone and in combination) does not improve results 
(except for IPsec-ESP with all methods and OpenVPN TCP with AdaBoost and Decision 
Tree).
Considering the limited performance improvements and costly feature generation, 
we do not use byte burst bigrams or trigrams in the remainder of the paper.

\jm{remove figure and keep text?}

\subsubsection{Feature comparison}
\label{sec:feature_comparison_global}

Next we compare features selected from the first two preliminary comparisons, 
packet size with direction and byte burst, with others features such as N first 
IAT and Netflow features (v5 and v9) on \Cref{fig:feature_owcwt_dps_netflow_bb_i}.
Netflow features always yield worse results than any other features.
IAT are worse than packet size with direction and byte burst, except for tunnel 
classification and some ML algorithms such as both logistic regressions.
We do not present results for elapsed time since the start of the flow because its 
results are worse than IAT.
\jm{check this again}
Previous work \cite{novo2020flow} show that tampering with IAT and elapsed time 
since the flow start 
using a proxy yields a performance decrease.
We thus only consider packet size with direction and byte burst as relevant 
features for the remainder of the paper.

\subsubsection{Summary}

Overall, packet size with direction and byte burst offer the best performance.
We emphasize that Netflow v5 and v9 feature sets do not provide good performances.

\subsection{General results}
\label{sec:general_results}

\begin{figure}[t!]
  \centering
  \subfloat[Tunnel detection]{
    \includegraphics[width=\widthfate\textwidth]{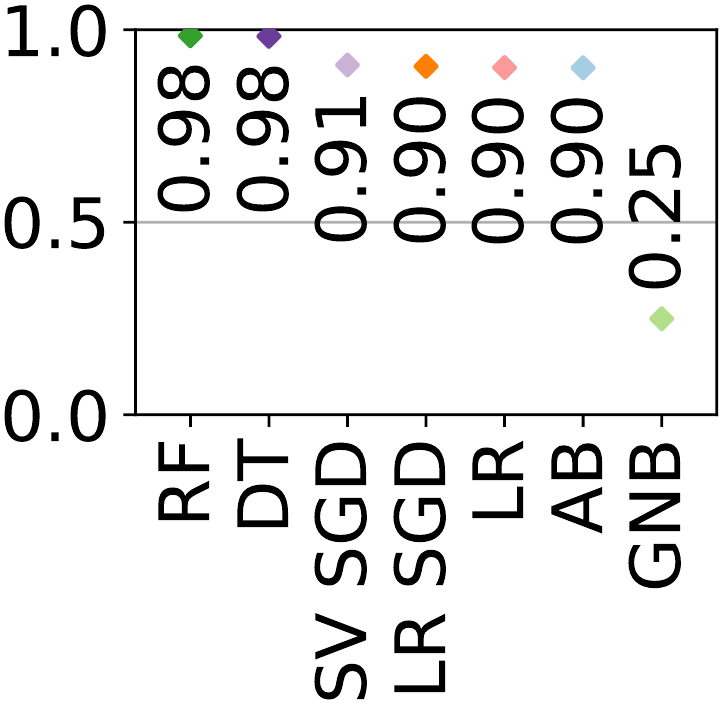}
  }
  \subfloat[Tunnel classification]{
    \includegraphics[width=\widthfate\textwidth]{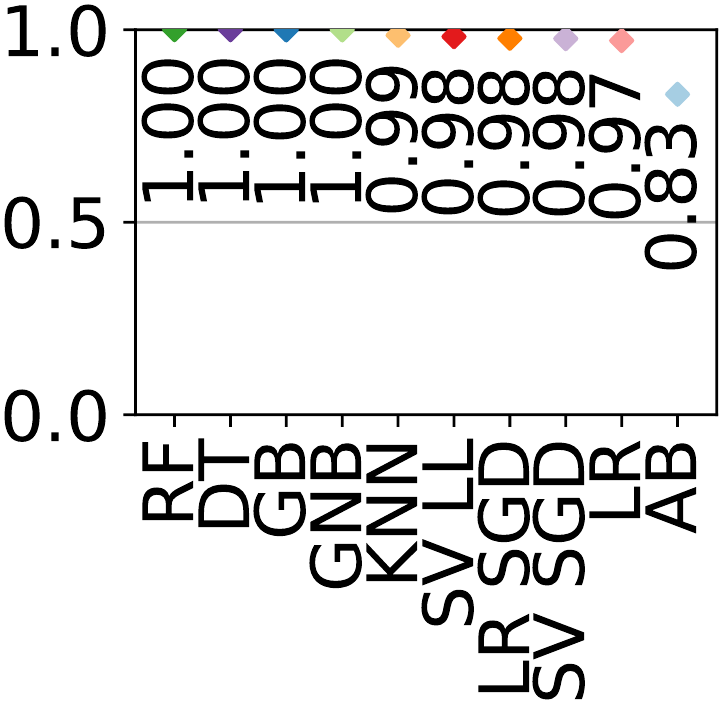}
  }
  \caption{Tunnel detection and tunnel classification using selected features (open world).
  Error bars represent a confidence interval with 99\% confidence level.}
  \label{fig:te_owcwt_fs2}
\end{figure}

\begin{figure}[t!]
  \centering
  \includegraphics[width=0.48\textwidth]{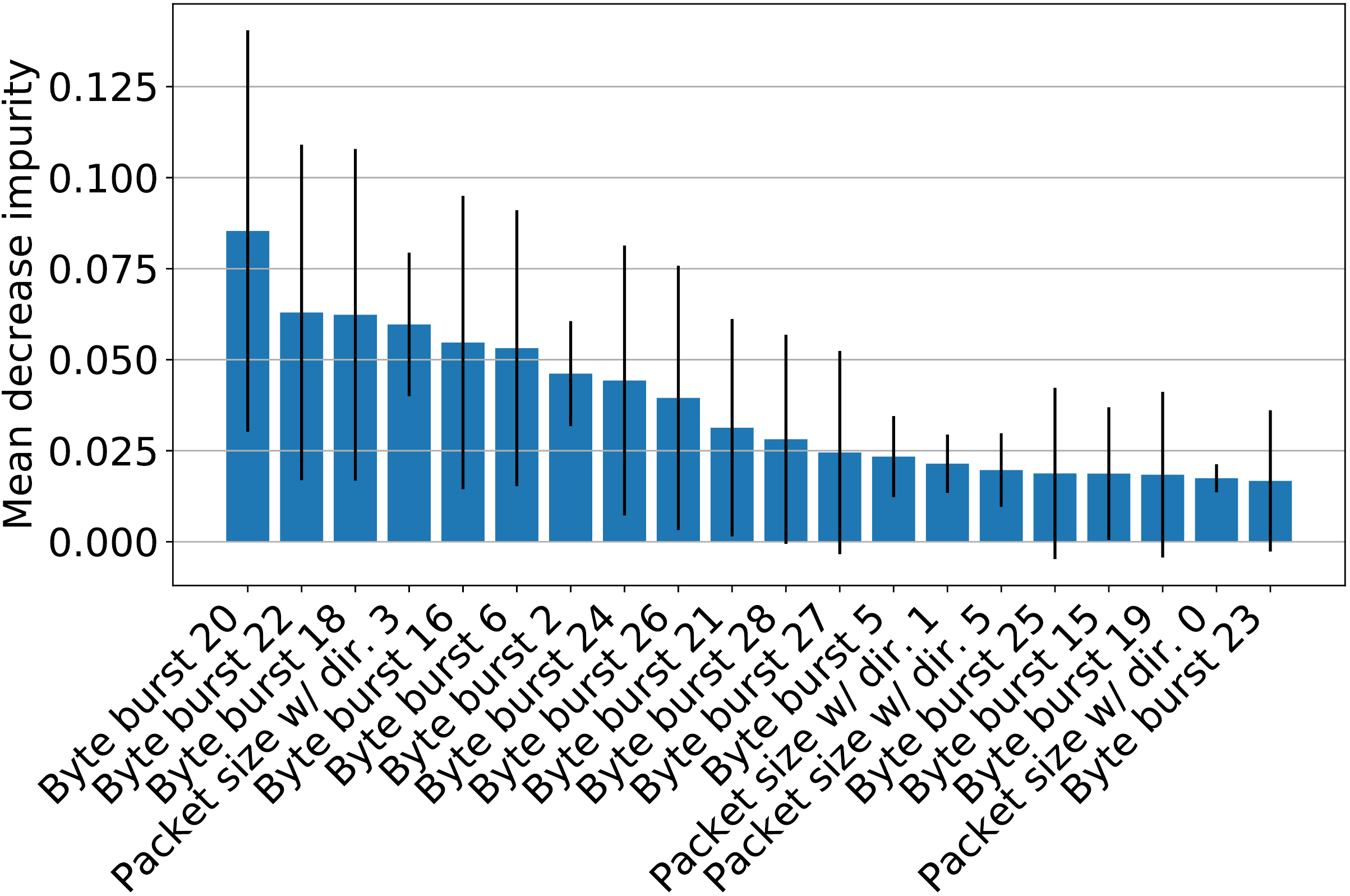}
  \caption{20 most important feature regarding Mean Decreased Impurity (MDI) 
  using Random Forest with 100 trees for tunnel detection.
  Error bars represent a confidence interval with 99\% confidence level.}
  \label{fig:fi_ow}
\end{figure}

\begin{figure}[t!]
  \centering
  \includegraphics[width=0.48\textwidth]{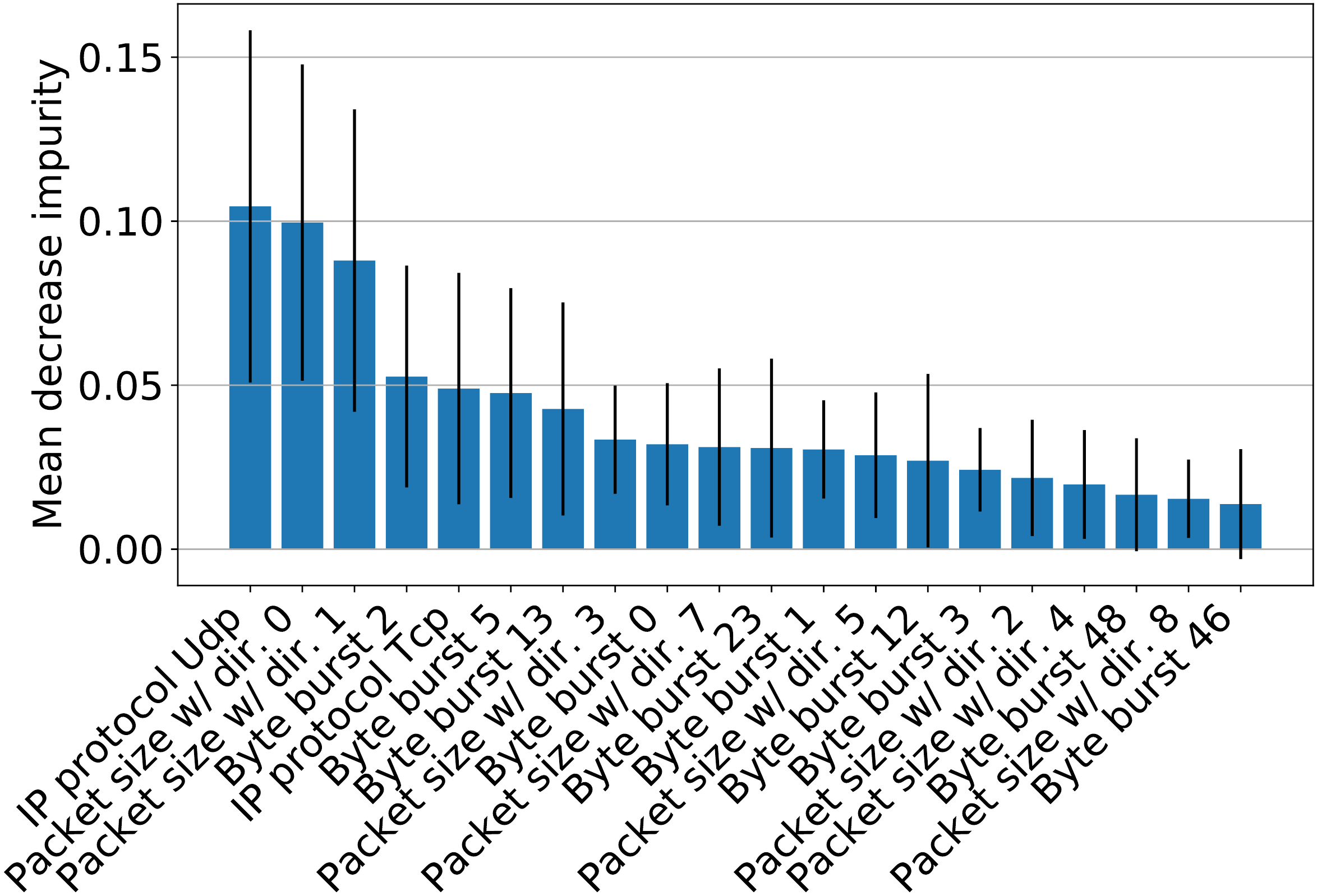}
  \caption{20 most important feature regarding Mean Decreased Impurity (MDI) 
  using Random Forest with 100 trees for tunnel classification.
  Error bars represent a confidence interval with 99\% confidence level.}
  \label{fig:fi_cwt}
\end{figure}

\begin{figure*}[t!]
  \subfloat[IPsec-ESP]{
    \includegraphics[width=\widthfate\textwidth]{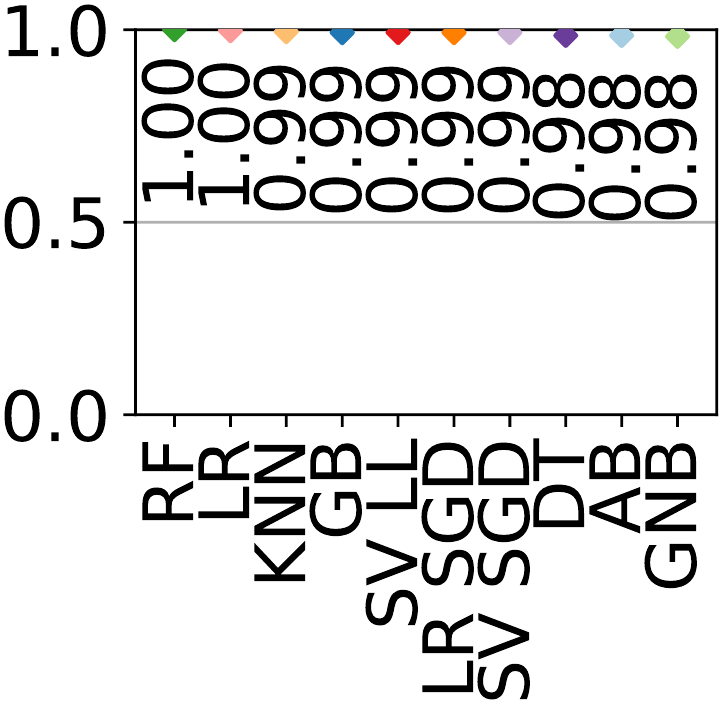}
  }
  \subfloat[OpenVPN TCP]{
    \includegraphics[width=\widthfate\textwidth]{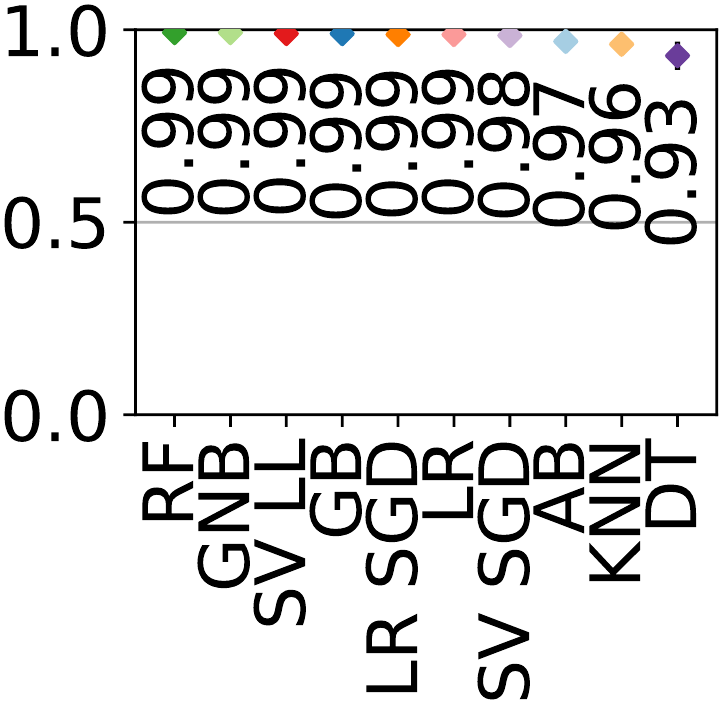}
  }
  \subfloat[OpenVPN UDP]{
    \includegraphics[width=\widthfate\textwidth]{te/ac/te_ac_am1500unaupa_n150_fs2_mlad_openvpntcp_cropped.pdf}
  }
  \subfloat[in SSH]{
    \includegraphics[width=\widthfate\textwidth]{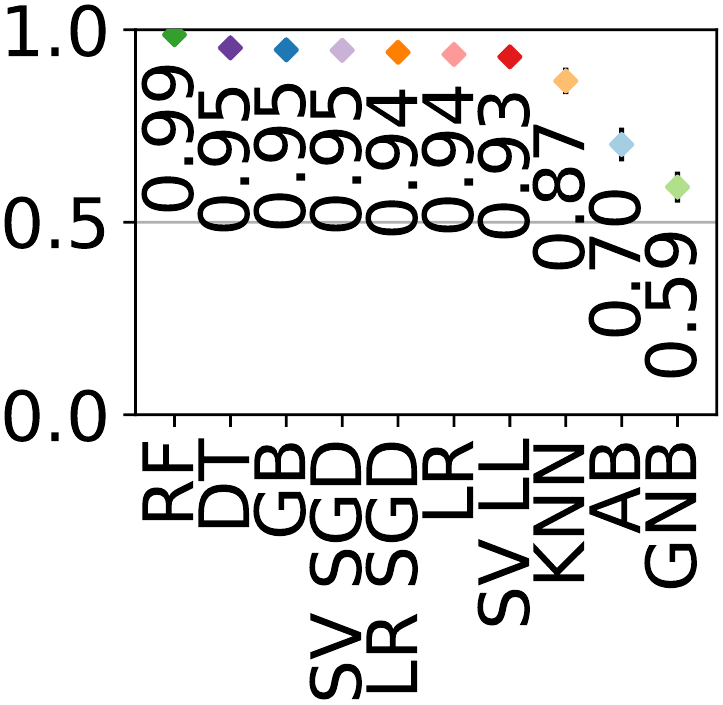}
  }
  \subfloat[in Wireguard]{
    \includegraphics[width=\widthfate\textwidth]{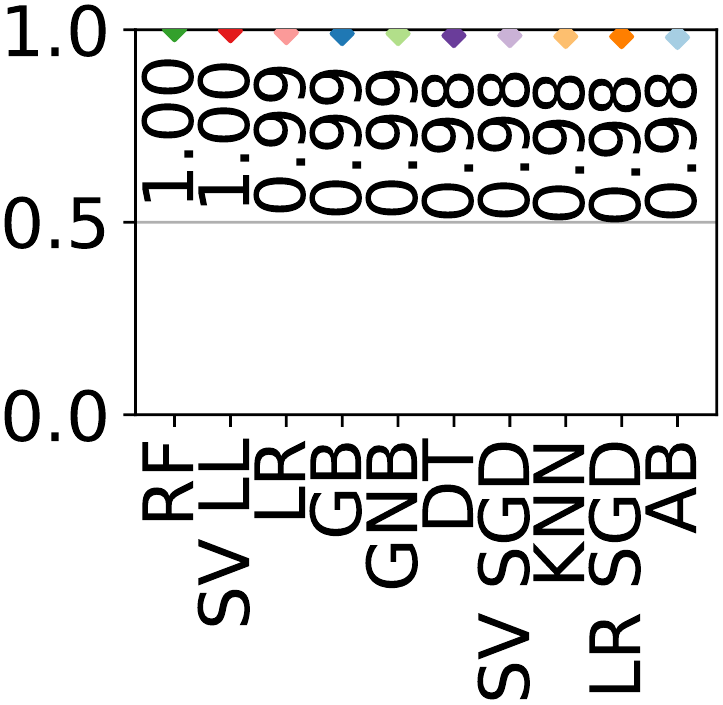}
  }
  \caption{Application classification detection using selected features.
  Error bars represent a confidence interval with 99\% confidence level.}
  \label{fig:te_cwa_fs2}
\end{figure*}

\jm{TODO: add feature importance for fs2}

We now provide a performance evaluation of packet size with direction and byte 
burst used together.
\Cref{fig:te_owcwt_fs2} pictures the F1 scores obtained for tunnel detection 
and classification.
\jm{TODO: complete after update for ow}
We observe very good performance for both tunnel detection and tunnel 
classification using the best algorithms (e.g. random forest or decision tree).
\jm{TODO: check 15 and 20 features}
\Cref{fig:fi_ow} displays the 10 most feature importance based on Mean Decreased 
Impurity (MDI) for tunnel detection.
Overall, IP protocol field-related one-hot-encoded features are not relevant.
This is consistent with the fact that tunnels here use both TCP and UDP.
16 out of the 20 most important feature are related to byte burst.
Byte bursts are thus here more important than packet sizes with direction.
\Cref{fig:fi_cwt} displays the 10 most feature importance based on Mean Decreased 
Impurity (MDI) for tunnel classification.
IP protocol field-related one-hot-encoded features are here the first and fourth 
most important features.
Beyond these two attributes, packet size with direction and byte burst are similarly important.

\Cref{fig:te_cwa_fs2} pictures the F1 scores obtained for tunnel detection 
and classification.
The most difficult tunneling protocol to classify application inside is SSH, 
followed by both OpenVPN tunnels, and finally, IPsec and Wireguard.
Similarly to \Cref{fig:te_owcwt_fs2}, using the best algorithms, random forest 
or decision tree, provides very good performances.

\jm{add text for \Cref{fig:te_cwa_fs2}}

\subsection{Domain generalization regarding untunneled traffic}
\label{sec:dg_untunneled}

\begin{table}[t!]
  \setlength\tabcolsep{6pt} 
  \centering
  \captionof{table}{Domain generalization performance between the two 
  background traffic datasets regarding F1 score using random forest for 
  tunnel detection.}
  \begin{tabular}{lllrrrrrrrrr}
    \toprule

    & & & \multicolumn{2}{c}{\textbf{Test}} \\
    \cmidrule(lr){4-5}
    & & \textbf{Feature} & \textbf{UNIBS} & \textbf{UPC} \\
    \midrule
    
    \multirow{8}{*}{\rotatebox{90}{\textbf{Train}}} &
    \multirow{4}{*}{\textbf{UNIBS}}
      & \textbf{Netflow v5}          & & 0.28\textpm0.00 \\
    & & \textbf{Netflow v9}          & & 0.37\textpm0.02 \\
    & & \textbf{Byte burst}          & & 0.61\textpm0.03 \\
    & & \textbf{Packet size w/ dir.} & & 0.75\textpm0.02 \\
    \cmidrule(lr){2-5}
    
    & \multirow{4}{*}{\textbf{UPC}}
      & \textbf{Netflow v5}          & 0.54\textpm0.01 \\
    & & \textbf{Netflow v9}          & 0.74\textpm0.01 \\
    & & \textbf{Byte burst}          & 0.98\textpm0.00 \\
    & & \textbf{Packet size w/ dir.} & 0.99\textpm0.00 \\

    \bottomrule
  \end{tabular}
  \label{table:dg_background}
\end{table}

\begin{figure*}[t!]
  \subfloat[Tunnel detection]{
    \includegraphics[width=0.32\textwidth]{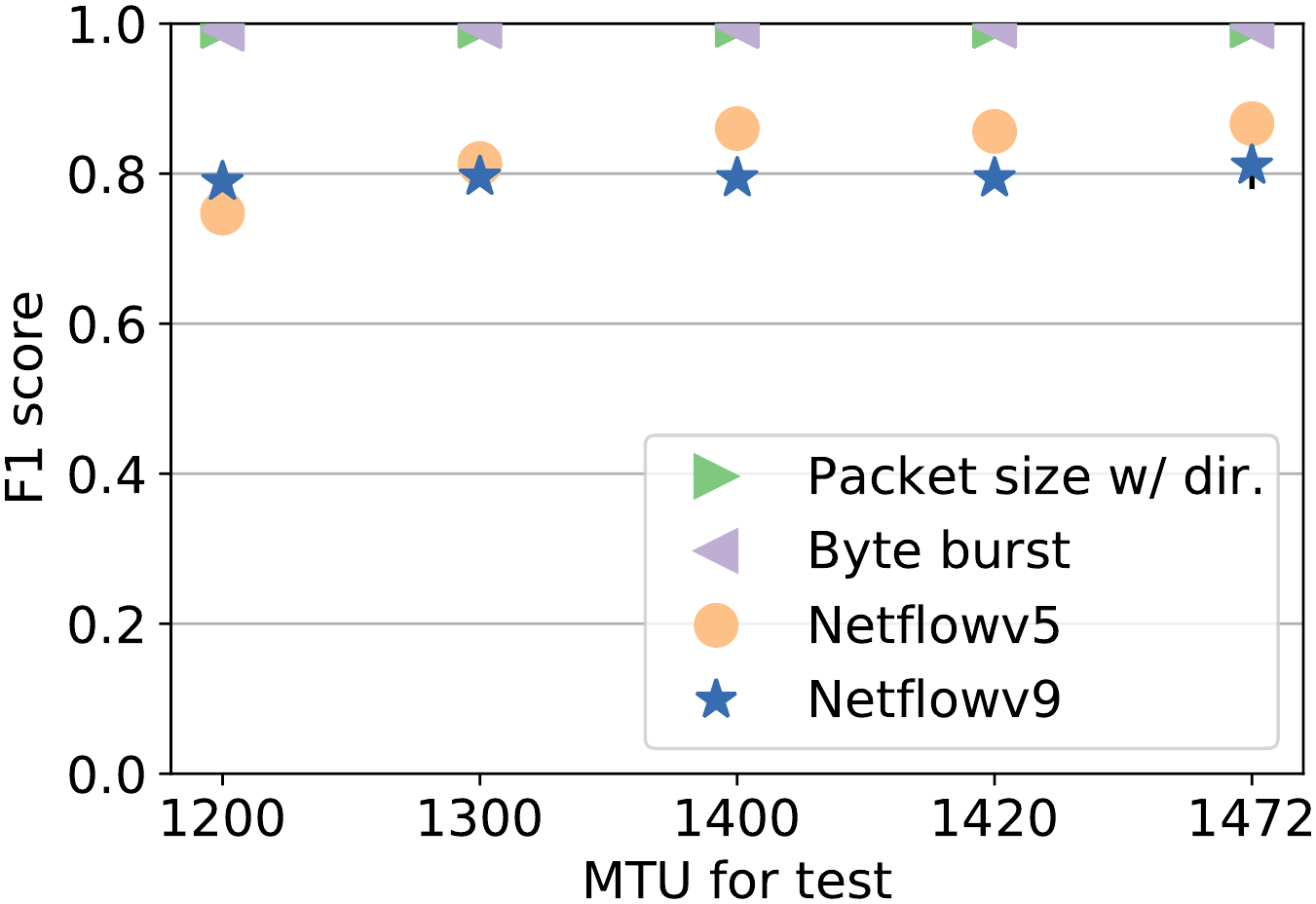}
    \label{fig:dg_ow_mtu}
  }
  \subfloat[Tunnel classification]{
    \includegraphics[width=0.32\textwidth]{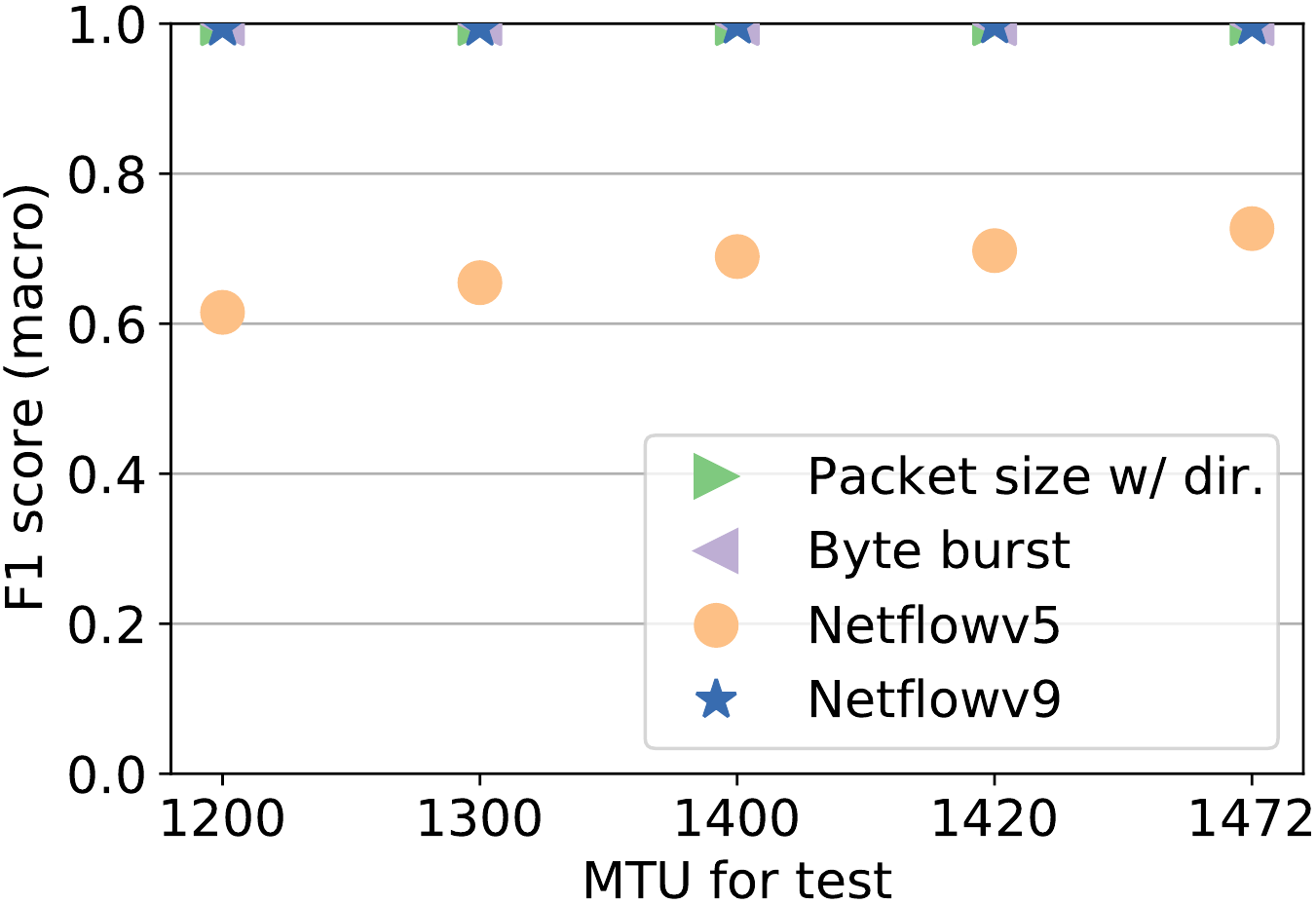}
    \label{fig:dg_cwt_mtu}
  }
  \subfloat[Application classification in SSH]{
    \includegraphics[width=0.32\textwidth]{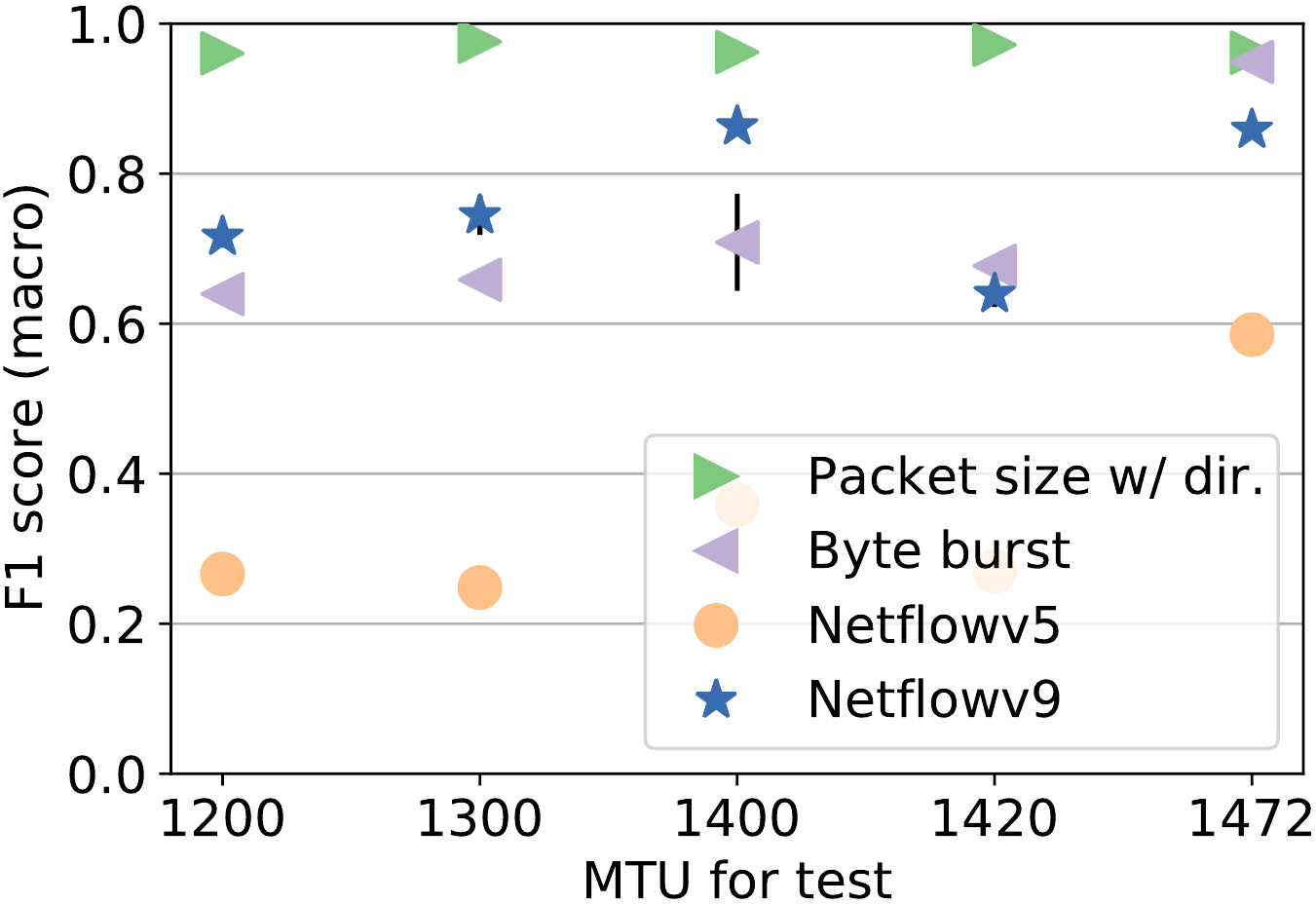}
    \label{fig:dg_cwa_mtu}
  }
  \caption{Random forest performance for training using MTU equals to 1500 and several MTU values for testing data.
  Error bars represent a confidence interval with 99\% confidence level.}
  \label{fig:dg_owcwtcwa_mtu}
\end{figure*}

\begin{figure*}[t!]
  \centering
  \subfloat[Tunnel detection]{
    \includegraphics[width=0.32\textwidth]{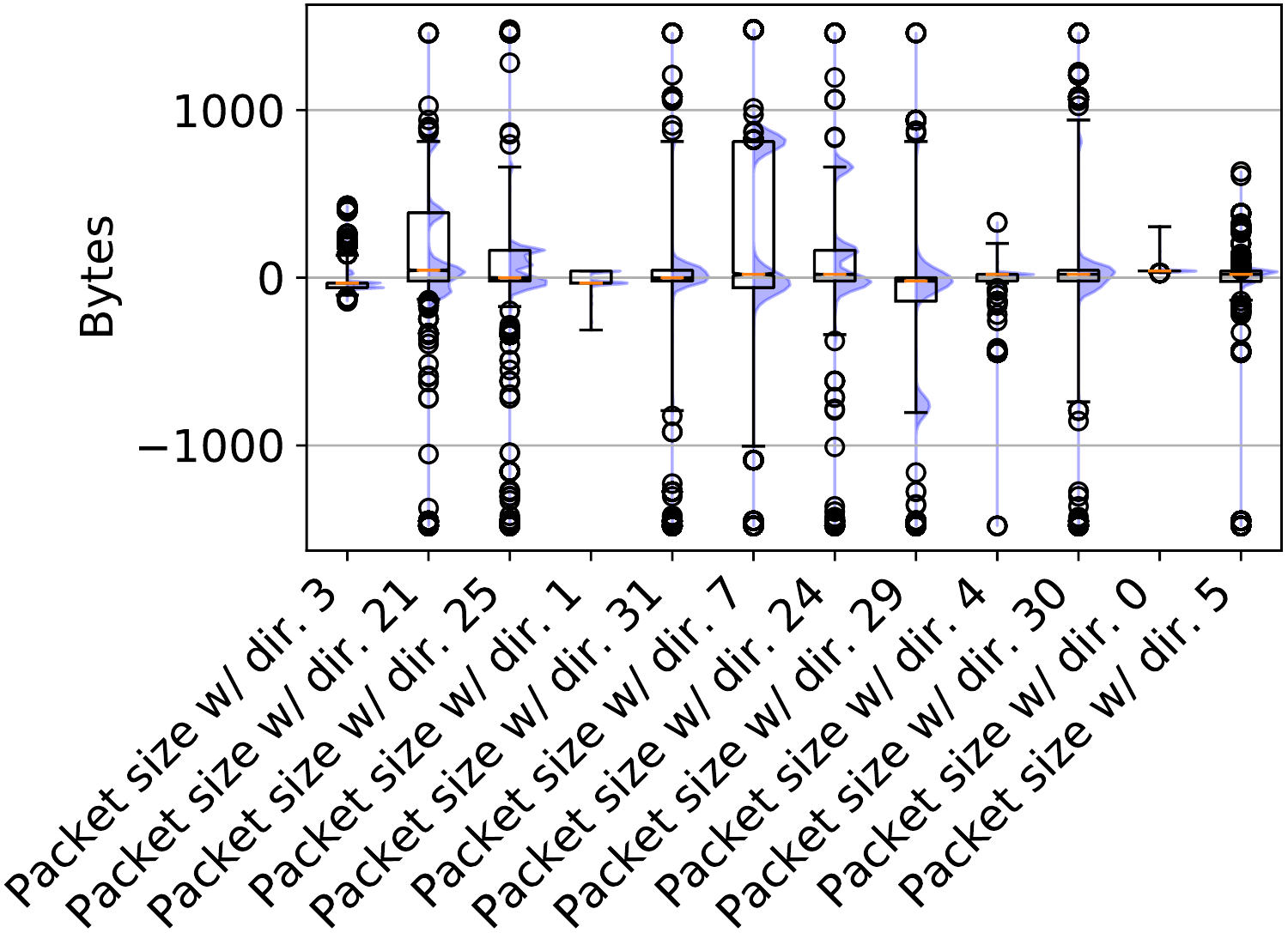}
  }
  \subfloat[Tunnel classification]{
    \includegraphics[width=0.32\textwidth]{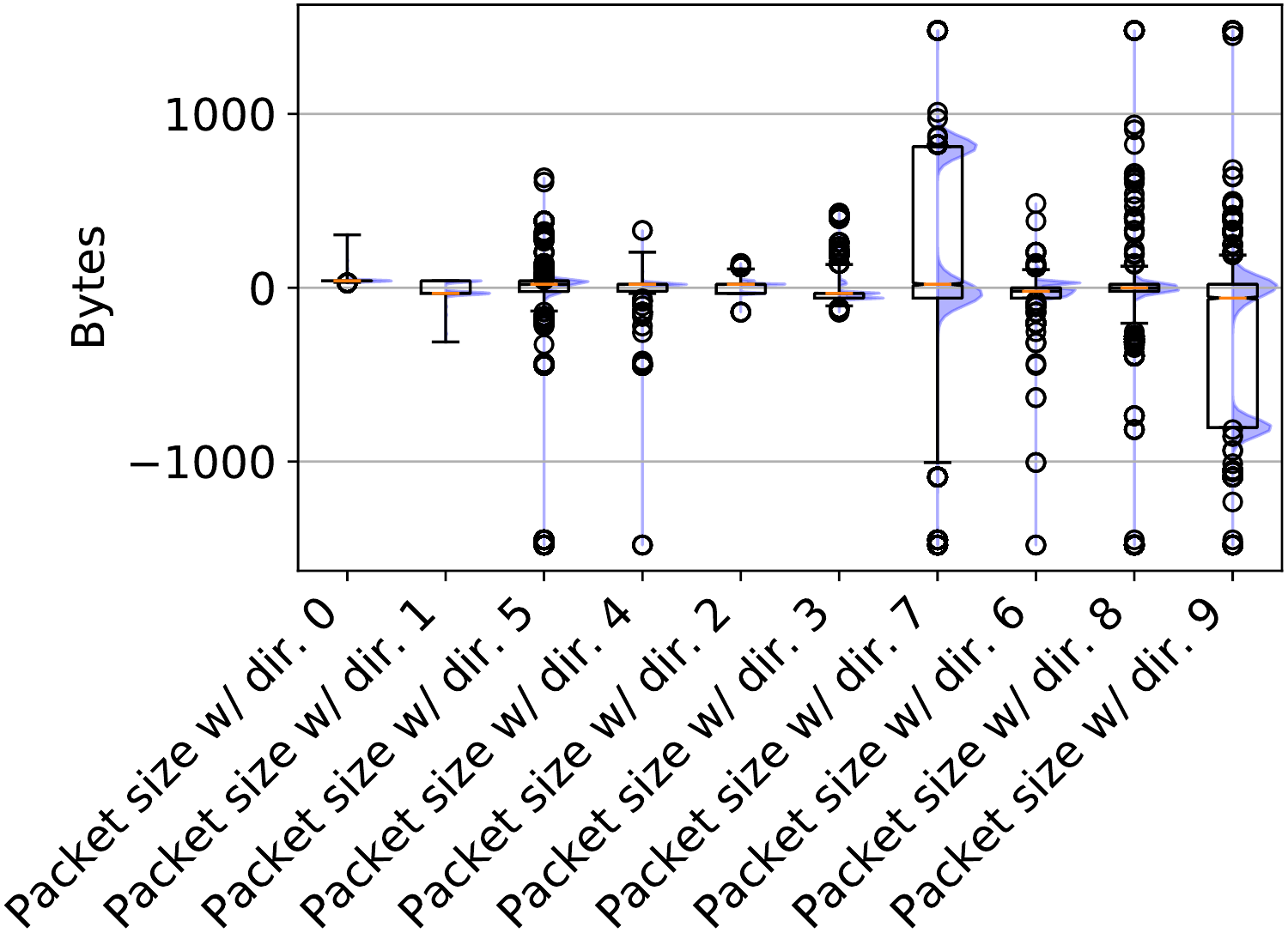}
  }
  \subfloat[Application classification in SSH]{
    \includegraphics[width=0.32\textwidth]{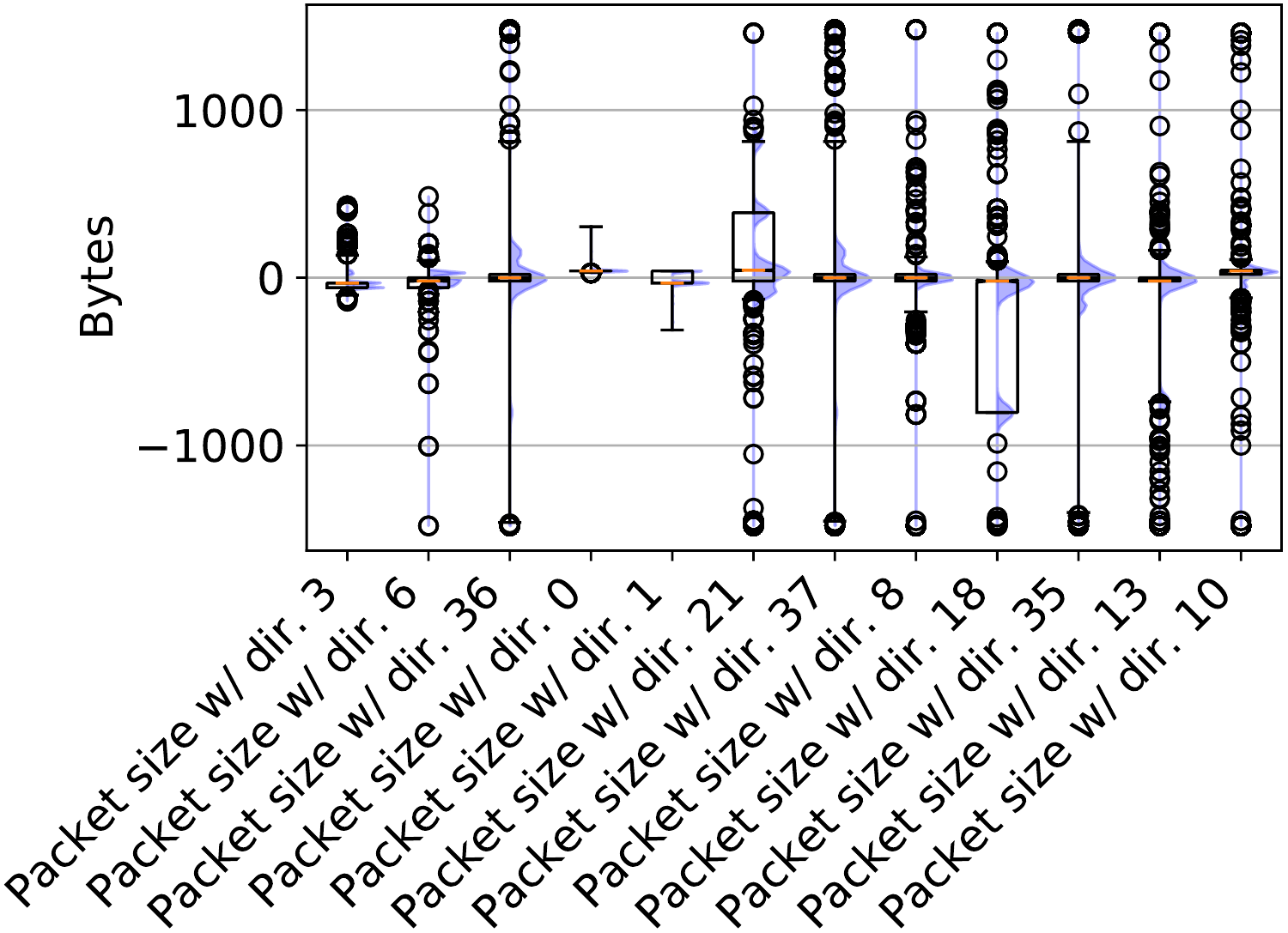}
  }
  \caption{Packet sizes with direction in byte of the 10 most important feature regarding 
  MDI using Random Forest with 100 trees for all pipeline steps.
  Boxplots encode median as the line inside the box, first and third quartile 
  as upper and lower limit of the box, whiskers as 1st and 99th quantiles, and 
  flier as points outside of these quantiles.}
  \label{fig:fi_owcwtcwa_plot}
\end{figure*}

\begin{figure*}[t!]
  \subfloat[Tunnel detection]{
    \includegraphics[width=0.32\textwidth]{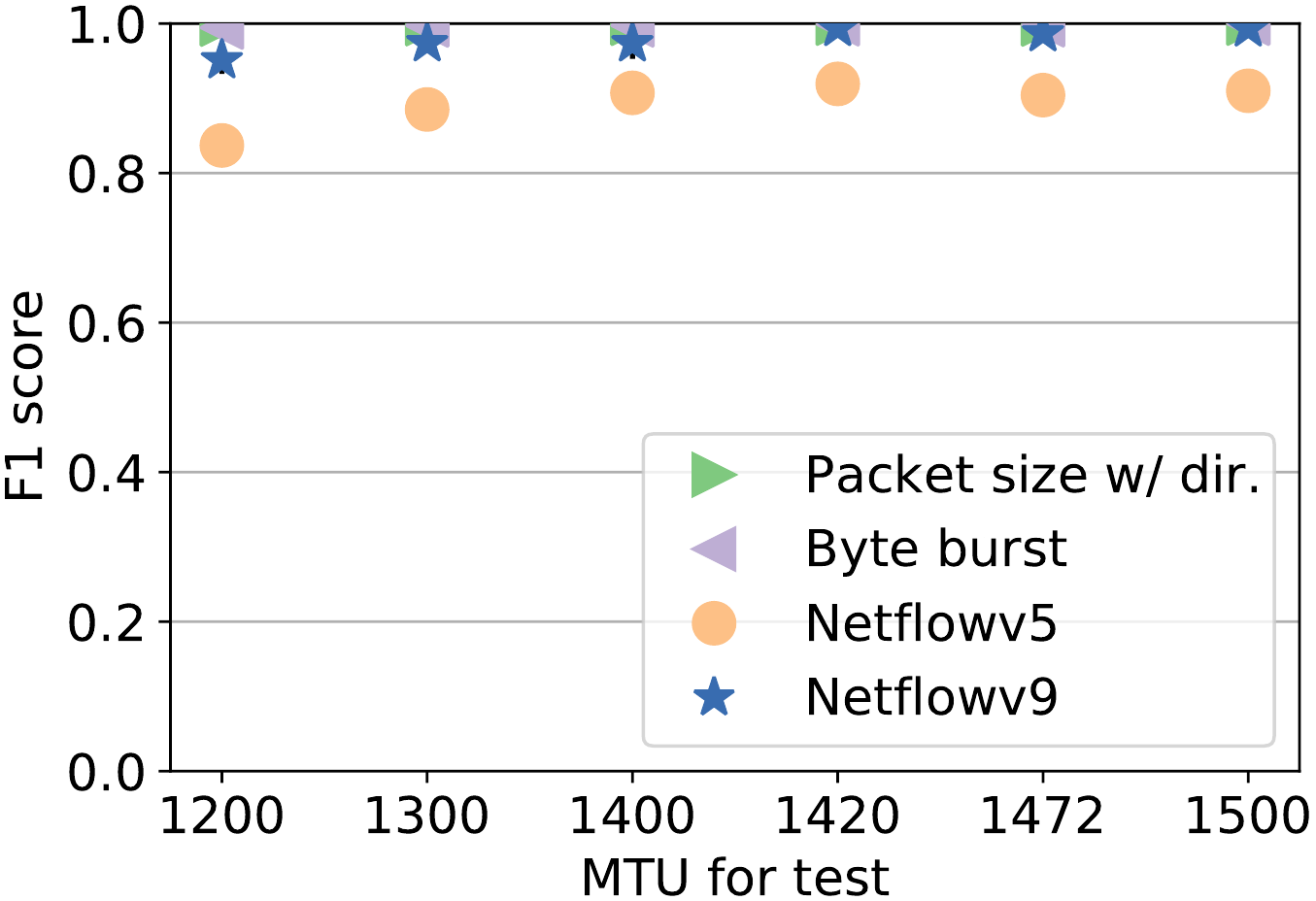}
    \label{fig:dg_ow_mtu_hrd}
  }
  \subfloat[Tunnel classification]{
    \includegraphics[width=0.32\textwidth]{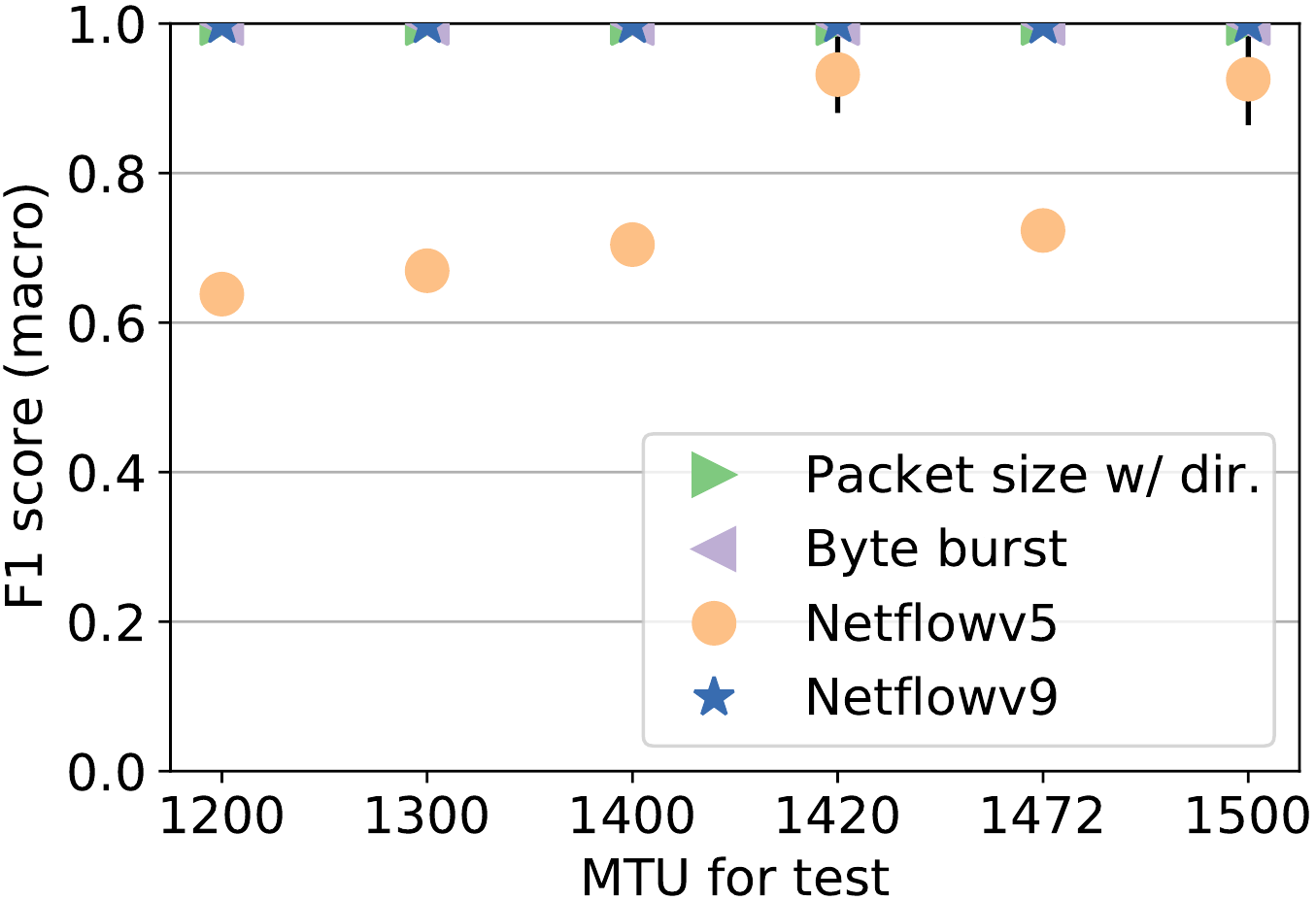}
    \label{fig:dg_cwt_mtu_hrd}
  }
  \subfloat[Application classification in SSH]{
    \includegraphics[width=0.32\textwidth]{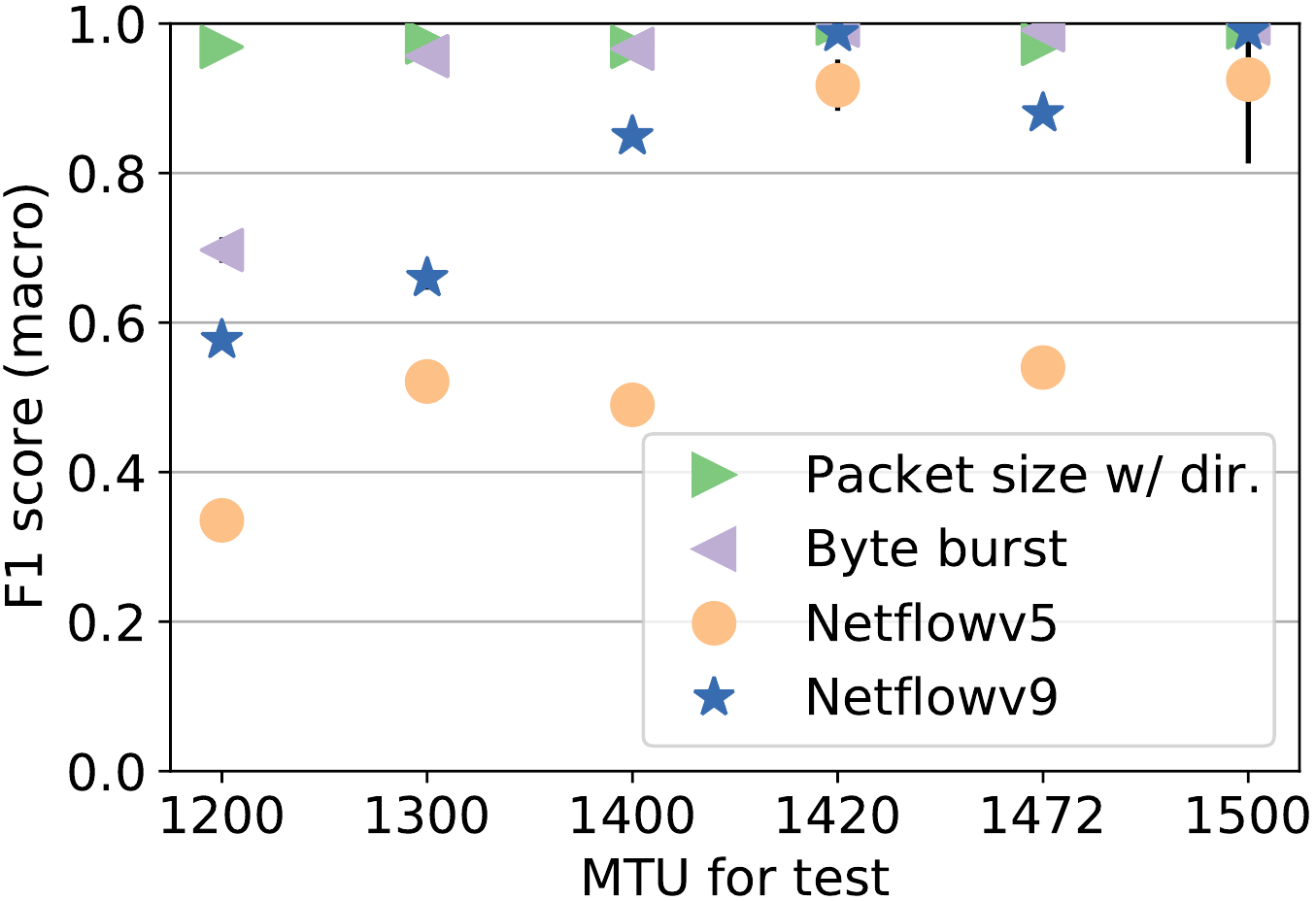}
    \label{fig:dg_cwa_mtu_hrd}
  }
  \caption{Random forest performance for training using MTU equals to 1500 and several MTU values for testing data.
  Error bars represent a confidence interval with 99\% confidence level.}
  \label{fig:dg_owcwtcwa_mtu_hrd}
\end{figure*}

We here address the impact of the untunneled traffic on tunnel detection 
performance.
We here only consider tunnel detection because untunneled traffic is
only used in this step of the ML pipeline (see \Cref{fig:pipeline}).
To this end, we use the UNIBS dataset for training and the UPC dataset for 
testing, and then, the other way around.
We here do not use tunneling protocol-related flows from the UNIBS and UPC 
datasets.
\Cref{table:dg_background} present F1 scores for packet sizes with direction, 
byte bursts, and Netflow v5 and Netflow v9 features.
Although we previously show in  \Cref{sec:feature_comparison_global} that Netflow features
exhibit poor performances, we still include them here as classic flow-based features.
When we train a random forest model on the UPC (resp. UNIBS) dataset and test 
it on the UNIBS (resp. UPC) dataset using Netflow v5, we obtain a 0.54 (0.28) F1 score.
These results are very poor, especially compared to what we previously obtained in
\Cref{sec:feature_comparison_global,sec:general_results}.
Netflow v9 exhibit the second worst performance decrease, followed by byte burst 
and then, packet size with direction.
We hypothesize that training with UPC yield better results than training with 
UNIBS because of two reasons.
First, the number of instances in the UPC dataset is much bigger than in the 
UNIBS one (see \Cref{table:datasets}) which helps generalization of models 
trained with UPC data.
Second, UPC dataset is more recent than UNIBS (2014 vs 2009).
This means that traffic categories present in UNIBS are probably also in UPC while
new protocols, applications, or usages that appeared between 2009 and 2014 may be 
present in UPC but not in UNIBS.
Although we do not actually try to detect these new applications, 
we hypothesize that their presence inside training data may be enough to modify 
learned model.

The observed performance decrease is consistent with existing work targeting 
traffic classification \cite{Pietrzyk2009Challenging}.
This section shows that training data must be as diverse as possible to ensure 
a limited performance decrease when the trained model is deployed in a new 
context.

\subsection{Domain generalization and adversarial learning regarding MTU}
\label{sec:dg_al_mtu}

The Maximum Transmission Unit (MTU) is the maximum amount of data that can be 
sent on a physical medium.
Its value is 1500 for Ethernet and much of Internet \cite{custura2018exploring}.
This value is configurable by OSes on network interfaces.
Reducing MTU decreases the size of sent and received packets.
This in turns impacts both the number sent and received packets, and the total 
amount of data sent (because of the increasing overhead).
We here address the impact of MTU modification on our three pipeline steps.
We first consider common MTU used on Internet beyond the default value of 1500 
which are, according to \cite{custura2018exploring}, 1472 and 1420.
Then, we also consider an attacker that leverages a purposely lowered MTU 
to generate adversarial samples.
We here use MTU values of 1400, 1300, and 1200.
We thus obtain six MTU values (1500, 1472, 1420, 1400, 1300 and 1200) that we use to generate 
traffic (see \Cref{sec:traffic_generation}) for the following experiments.
We here do not use the SSH flows from UNIBS because we do not know which MTU was used.

\jm{explain specific k-fold cross validation}

\Cref{fig:dg_owcwtcwa_mtu} presents our results for all three pipeline steps,
packet size with direction, byte burst and Netflow v5 and v9 features when a 
1500 bytes MTU is used for training.
Results for MTU testing equal to 1500 are different than 
\Cref{sec:feature_comparison_global} because UNIBS SSH is not used.
Similarly to \Cref{sec:dg_untunneled}, we include Netflow feature as classic flow-based features.
Performances modifications regarding testing MTU change across pipeline steps 
and features: tunnel classification performances only decrease for Netflow v5,
tunnel detection is impacted with both Netflow feature sets,
while application classification's F1 scores deteriorate for all features 
except packet sizes with direction.
MTU lowering reduces the amount of data sent in a single packet.
This, in turn, increases the packet number in a flow.
Netflow v5 and v9 feature sets are impacted because they contains packet 
number-related features, either for all packet of flow or a subset (e.g. only 
from the source in Netflow v9).
We observe that byte burst performances decrease as the testing MTU is smaller
for application classification, while this does not happen with packet sizes 
with direction.
We hypothesize that byte burst is actually able to look further in time into 
the flow, and is thus more sensitive to traffic changes caused by MTU reduction.
\jm{TODO: dig into actual position of byte burst in terms of packets using packet bursts}
Contrary to byte burst, we observe that performances for packet sizes with 
direction do not deteriorate as testing MTU decreases.
\Cref{fig:fi_owcwtcwa_plot} displays boxplots and violinplots of packet 
sizes with direction which exhibit the ten biggest MDI feature importance when
random forest is used.
It shows that most feature values are far from the testing MTU.
We thus hypothesize that packet size with direction performances do not 
decrease because feature values are not modified when testing MTU changes.

We do not display the results obtained with a training MTU of 1420 and 1470
due to the lack of space.
The results are similar to what is obtained with a training MTU of 1500: 
best results are obtained when testing MTU is equal to the training one, and 
performances deteriorate otherwise.

\section{Discussions and future work}

We identify two main limitations to our work.
The first limitation is related to tunneling protocols and application coverage.
Although our tunneling protocol set is more complete than existing work, we 
did not address tools such as PPTP.
We also did not deploy DNS-based tunelling which is commonly used for obfuscation.
Regarding applications, we use web browsing and file transfer as applications 
inside tunnels.
Common usage such as command line is not included in our work.
We want to improve both deployed tunneling and application in future work.
The other limitation is that we only use a single application inside each tunnel.
Users however do not use tunelling this way, and often use several applications 
inside a tunnel.
Our results regarding tunnel detection and classification using N first features 
such as packet size or byte burst are however not impacted by this limitation 
because they need a very small of features N to perform well (see 
\Cref{sec:n_first_features}).

Many recent on network traffic metadata analysis work use deep learning-related
techniques.
We however achieve very good performance with classical ML algorithms such as 
random forest or decision tree and thus decided to not use deep learning at all.
\jm{TODO: add motivation on runtime}

We also plan to extend this work along two other axes.
First, we use each tunelling protocol with a single cipher suite.
We want to extend this work by exploring other cipher suites (e.g. a different 
block size) which may alter network traffic metadata.
Second, we plan to investigate prediction intervals for both aleatoric and epistemic 
uncertainties using \cite{mapie}.

We think that our contribution on domain generalization and adversarial learning 
regarding MTU is very important for all use cases that leverage network traffic 
metadata.
Regarding security and privacy, we hypothesize that use cases such as malware C2 
detection, website fingerprinting, DoH fingerprinting, Tor detection are impacted by MTU 
manipulation.

\section{Conclusion}

In this work, we design a methodology to detect tunneling protocols, classify 
them once detected, and identify application used inside tunnels.
We provide a thorough analysis of both features and machine learning algorithms.
We also address domain generalization evaluation regarding the untunneled traffic 
for tunnel detection, and domain generalization and adversarial 
learning analysis regarding MTU for all three pipeline steps.

\newpage

\newpage

\newpage

\vspace{50em}

\appendix

\section{Appendix}

\subsection{Learning curve}

\begin{figure*}[t!]
  \subfloat[Tunnel detection]{
    \includegraphics[width=0.32\textwidth]{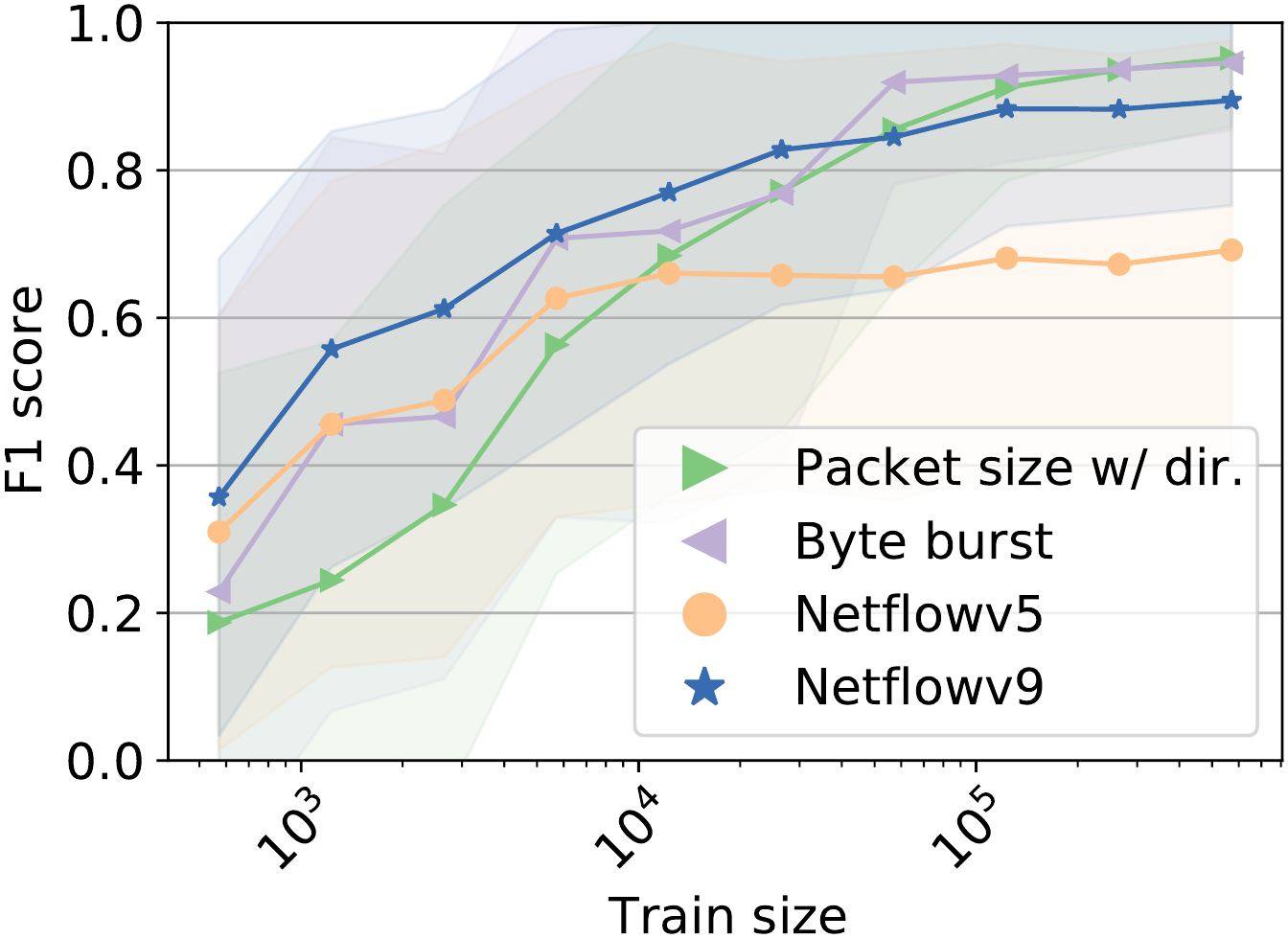}
    \label{fig:lc_ow}
  }
  \subfloat[Tunnel classification]{
    \includegraphics[width=0.32\textwidth]{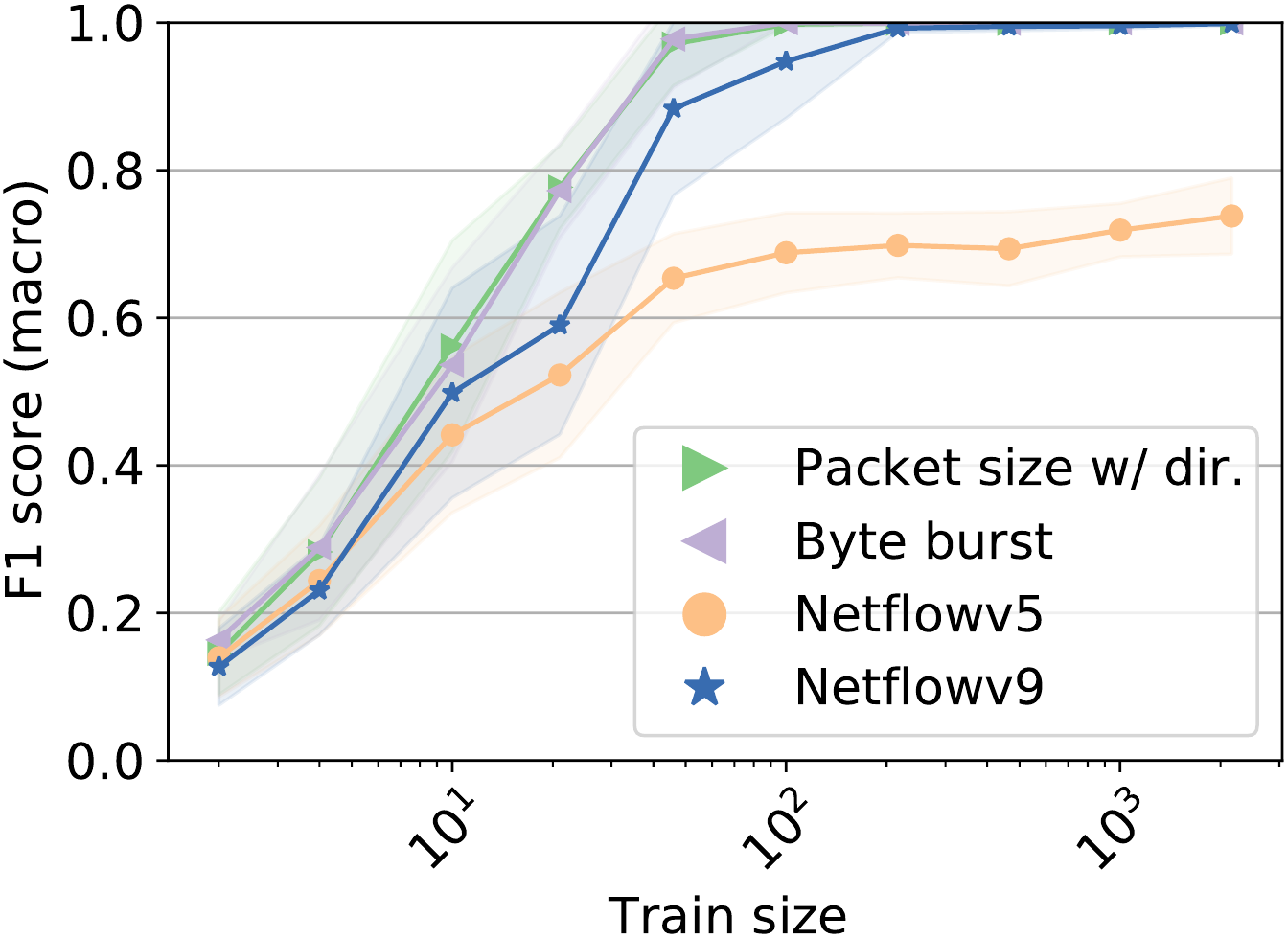}
    \label{fig:lc_cwt}
  }
  \subfloat[Application classification in SSH]{
    \includegraphics[width=0.32\textwidth]{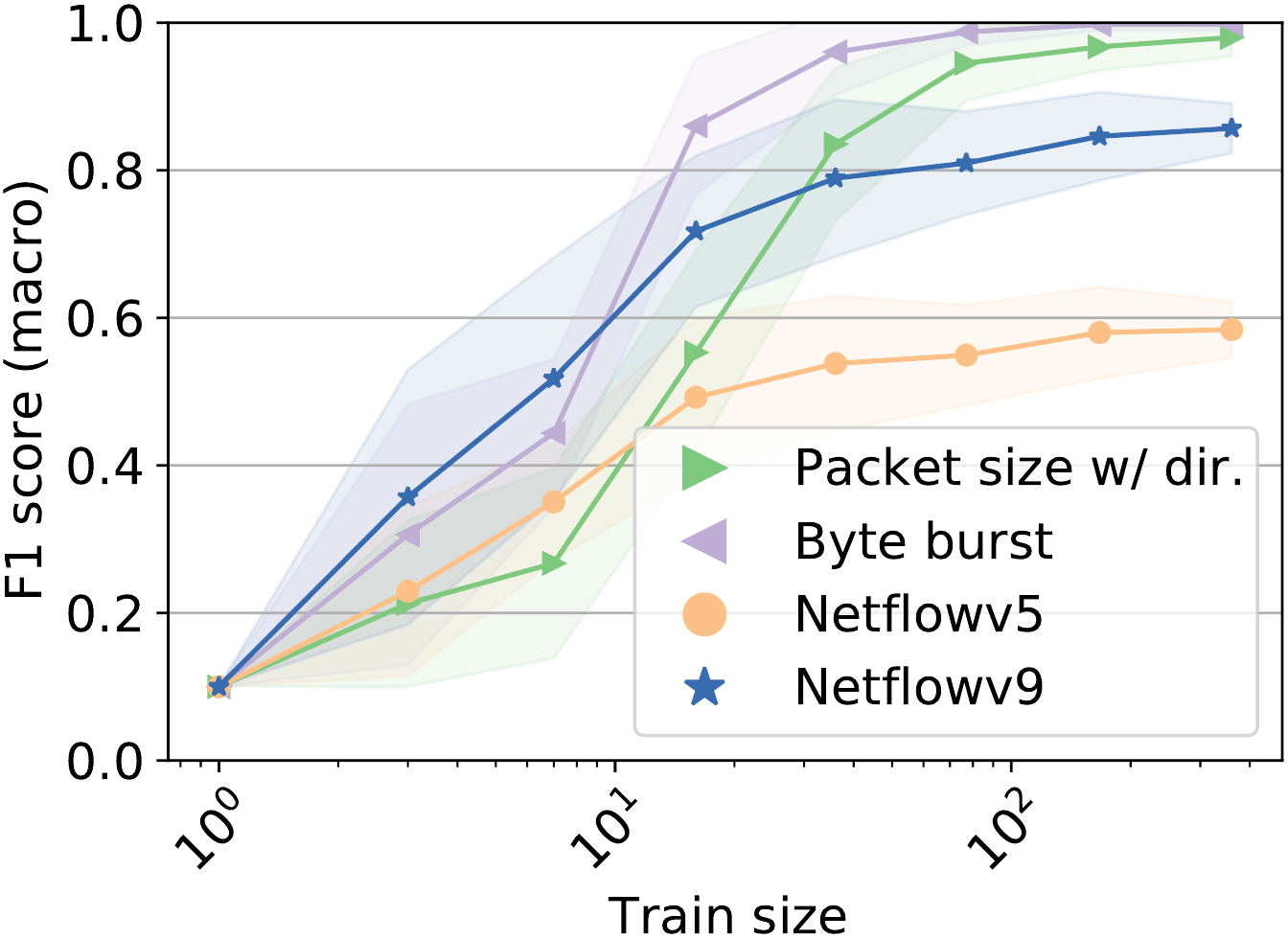}
    \label{fig:lc_cwa}
  }
  \caption{Random forest performance for several training dataset sizes.
  Error bars represent a confidence interval with 99\% confidence level.}
  \label{fig:lc_owcwtcwa}
\end{figure*}

\Cref{fig:lc_owcwtcwa} pictures the F1 scores obtained with random forest for 
several trainign dataset sizes.
For all scenario, performance increase more important for smaller increase of 
training data size than 
This performance increase is negligible for the biggest increase in training 
data size (cf point to the right of each figure and log scale).

\subsection{Netflow}

\begin{figure*}[t!]
  \centering
  \setlength\tabcolsep{0pt}
    
  \begin{tabular}{lccccc}
    & Netflow v5 base & Netflow v5 ext. & Netflow v9 base & Netflow v9 ext. \\
    
    \rotatebox{90}{\hspace{1em} Tunnel detection} &
    \includegraphics[width=\widthfate\textwidth]{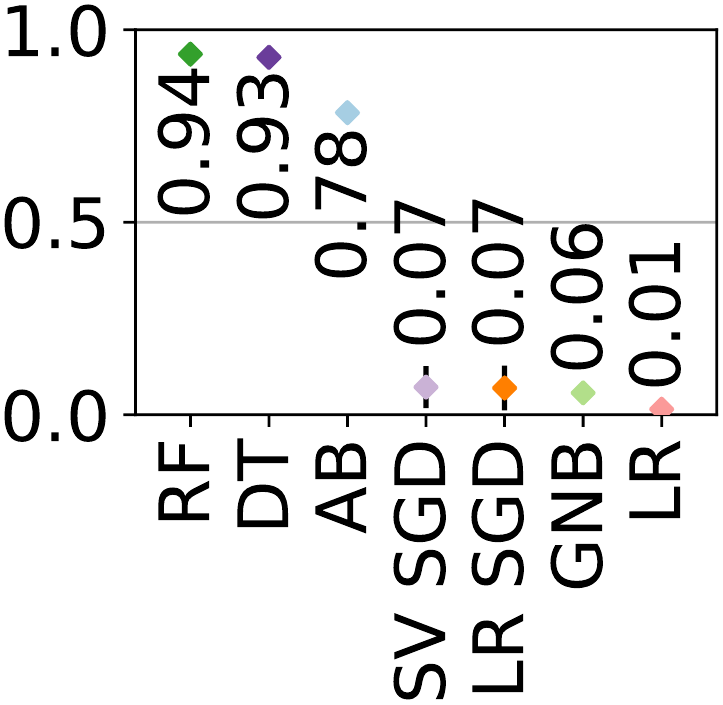} &
    \includegraphics[width=\widthfate\textwidth]{fa/td/te_td_am1500unaupa_n50_netflowv5e_mlad_all_cropped.pdf} &
    \includegraphics[width=\widthfate\textwidth]{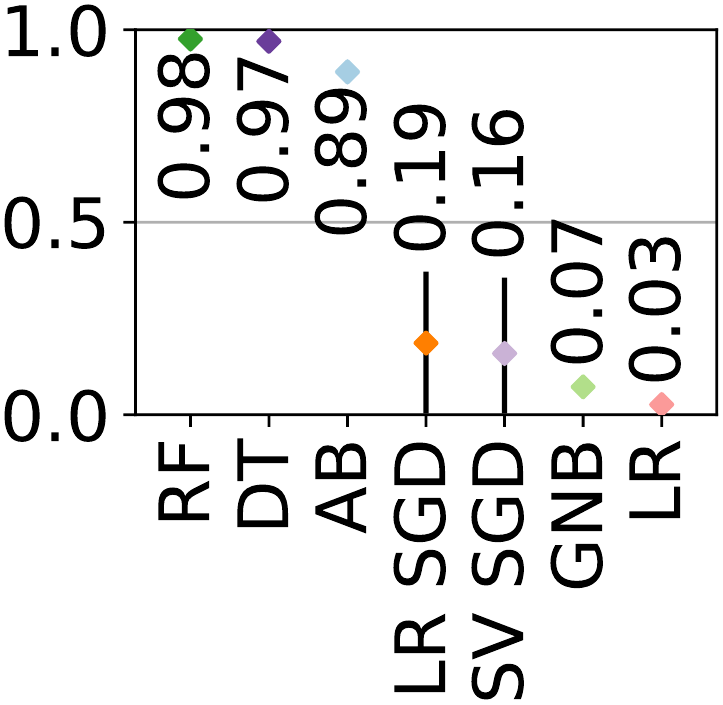} &
    \includegraphics[width=\widthfate\textwidth]{fa/td/te_td_am1500unaupa_n50_netflowv9e_mlad_all_cropped.pdf} \\
    
    \rotatebox{90}{\hspace{0em} Tunnel classification} &
    \includegraphics[width=\widthfate\textwidth]{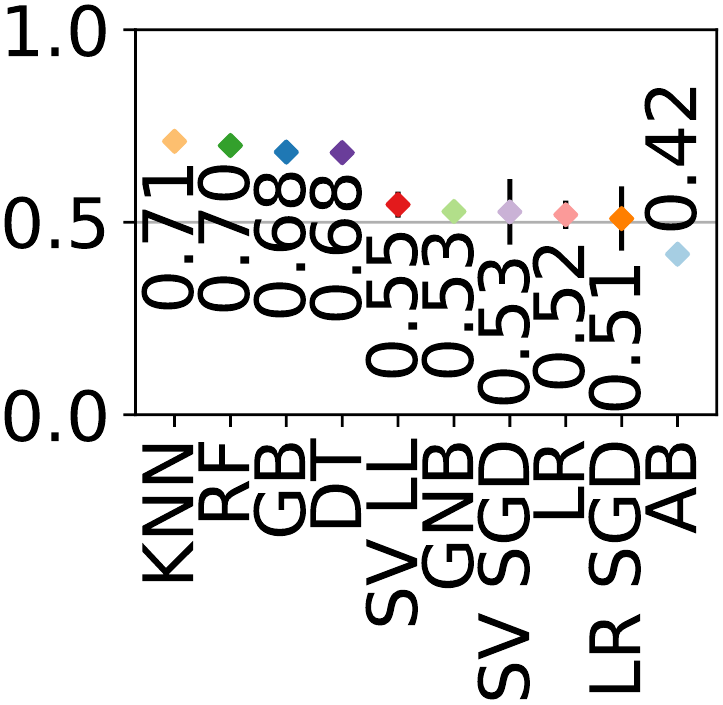} &
    \includegraphics[width=\widthfate\textwidth]{fa/tc/te_tc_am1500unaupa_n50_netflowv5e_mlad_at_cropped.pdf} &
    \includegraphics[width=\widthfate\textwidth]{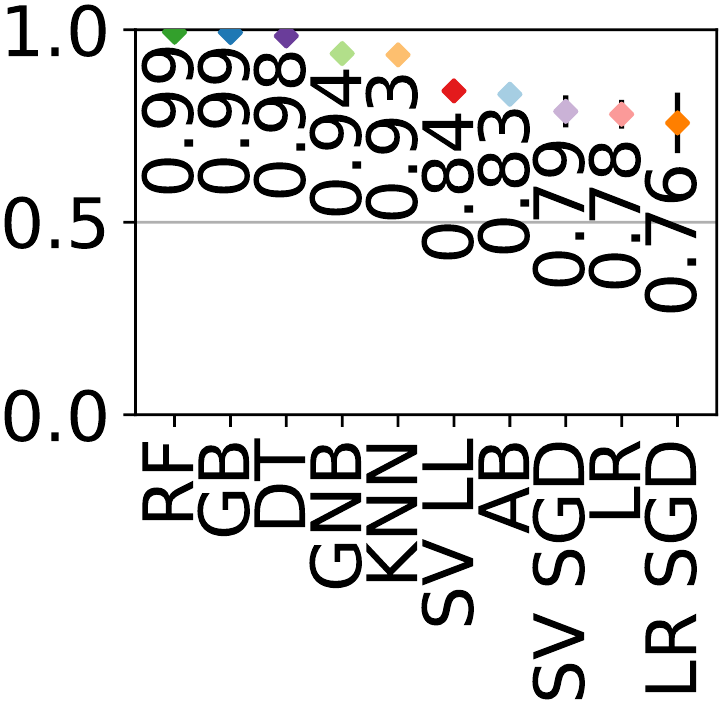} &
    \includegraphics[width=\widthfate\textwidth]{fa/tc/te_tc_am1500unaupa_n50_netflowv9e_mlad_at_cropped.pdf} \\

    \rotatebox{90}{\parbox{9em}{\centering Application classification in SSH}} &
    \includegraphics[width=\widthfate\textwidth]{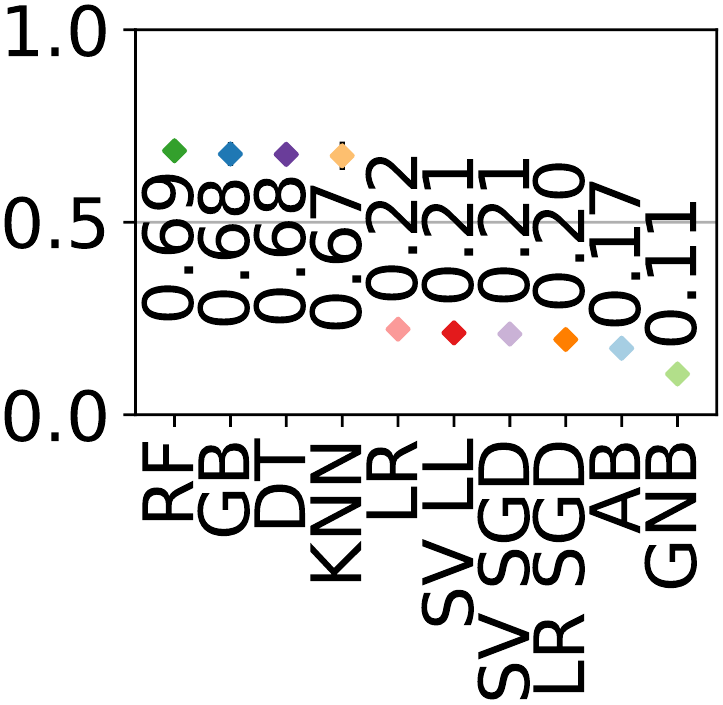} &
    \includegraphics[width=\widthfate\textwidth]{fa/ac/te_ac_am1500unaupa_n150_netflowv5e_mlad_ssh_cropped.pdf} &
    \includegraphics[width=\widthfate\textwidth]{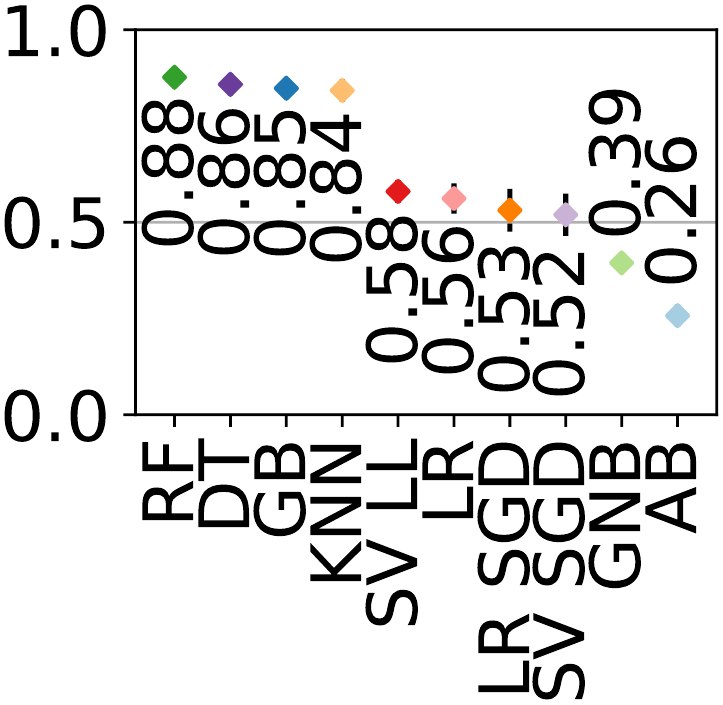} &
    \includegraphics[width=\widthfate\textwidth]{fa/ac/te_ac_am1500unaupa_n150_netflowv9e_mlad_ssh_cropped.pdf} \\
    
  \end{tabular}
  
  \caption{Comparison of packet size with direction, byte burst and IAT for all pipeline steps.
  Error bars represent a confidence interval with 99\% confidence level.
  \jm{TODO: fix figure for tunnel detection and Netflow v9}
}
  \label{fig:feature_tdtcac_netflow}
\end{figure*}

\subsection{Byte burst bigram and trigram}

\begin{figure*}[t!]
  \centering
  \setlength\tabcolsep{0pt}
    
  \begin{tabular}{lccccc}
    & Byte burst & Byte burst bigram only & Byte burst with bigram & Byte burst trigram only & Byte burst with trigram \\
  
    \rotatebox{90}{\hspace{1em} Tunnel detection} &
    \includegraphics[width=\widthfate\textwidth]{fa/td/te_td_am1500unaupa_n50_bb_mlad_all_cropped.pdf} &
    \includegraphics[width=\widthfate\textwidth]{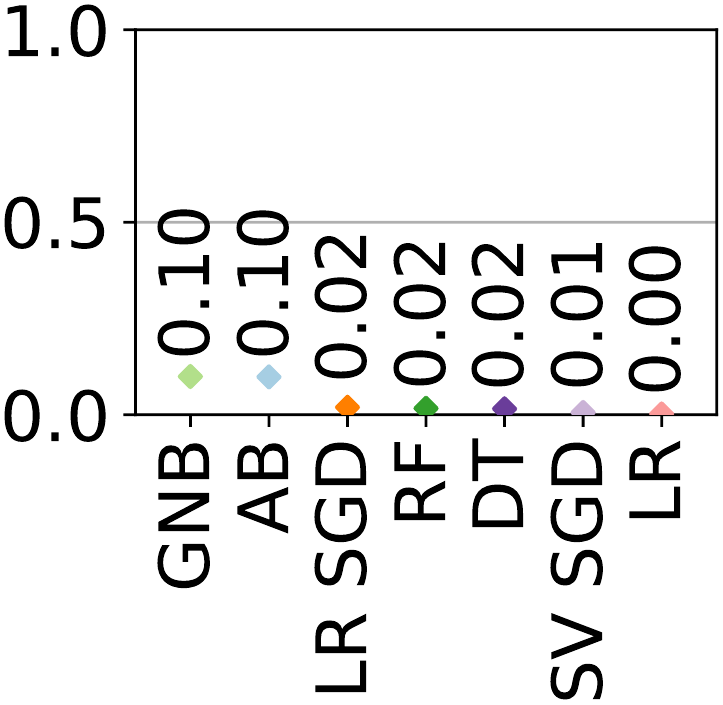} &
    \includegraphics[width=\widthfate\textwidth]{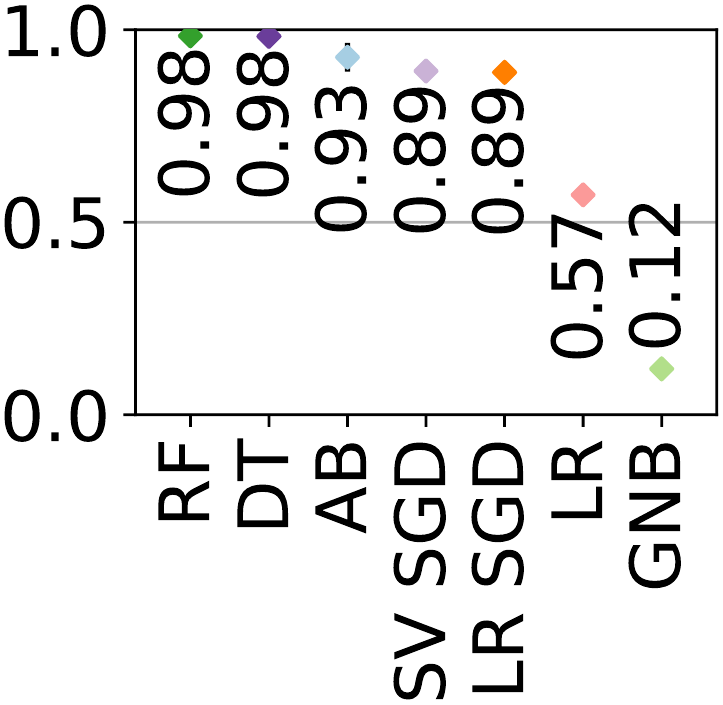} &
    \includegraphics[width=\widthfate\textwidth]{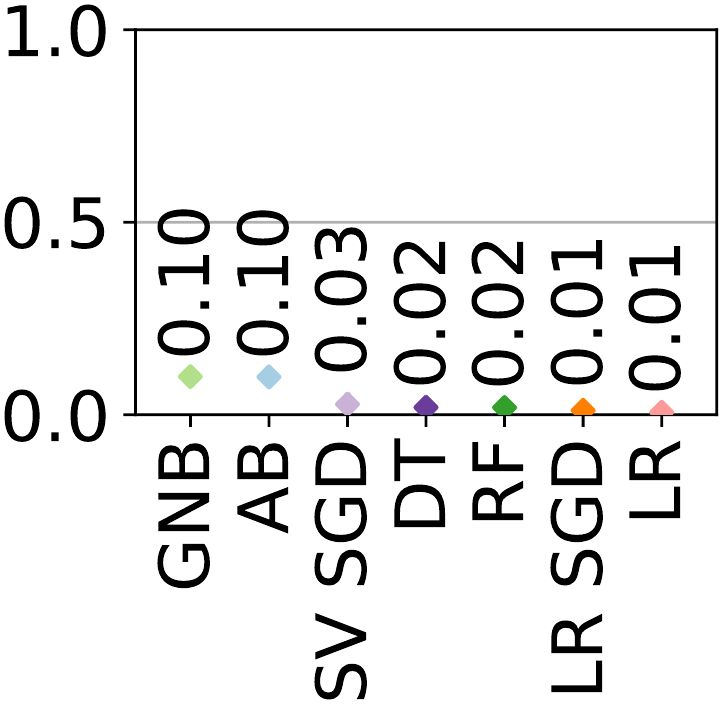} &
    \includegraphics[width=\widthfate\textwidth]{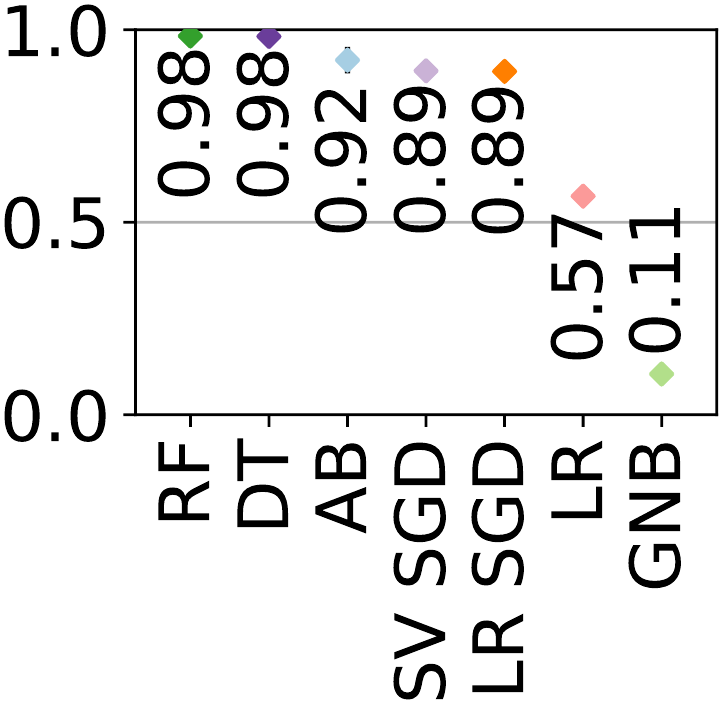} \\
    
    \rotatebox{90}{\hspace{0em} Tunnel classification} &
    \includegraphics[width=\widthfate\textwidth]{fa/tc/te_tc_am1500unaupa_n50_bb_mlad_at_cropped.pdf} &
    \includegraphics[width=\widthfate\textwidth]{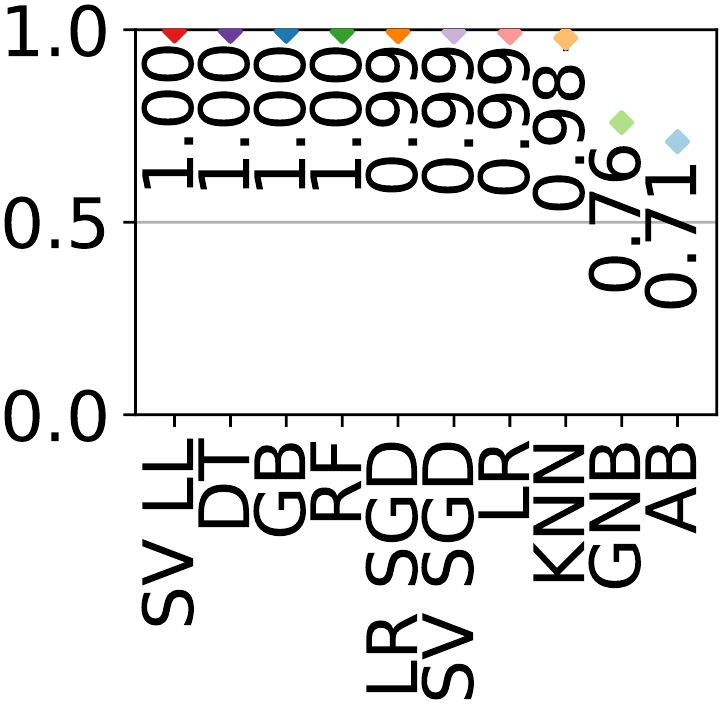} &
    \includegraphics[width=\widthfate\textwidth]{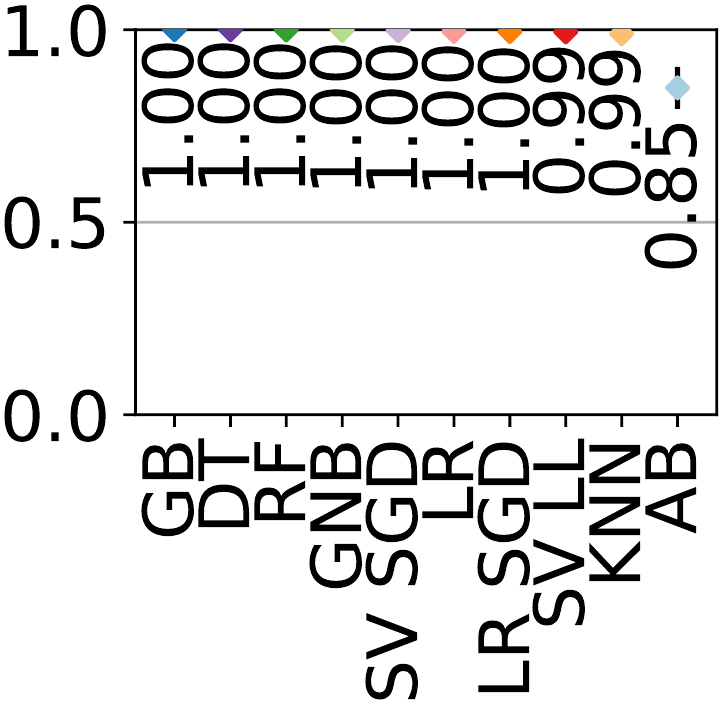} &
    \includegraphics[width=\widthfate\textwidth]{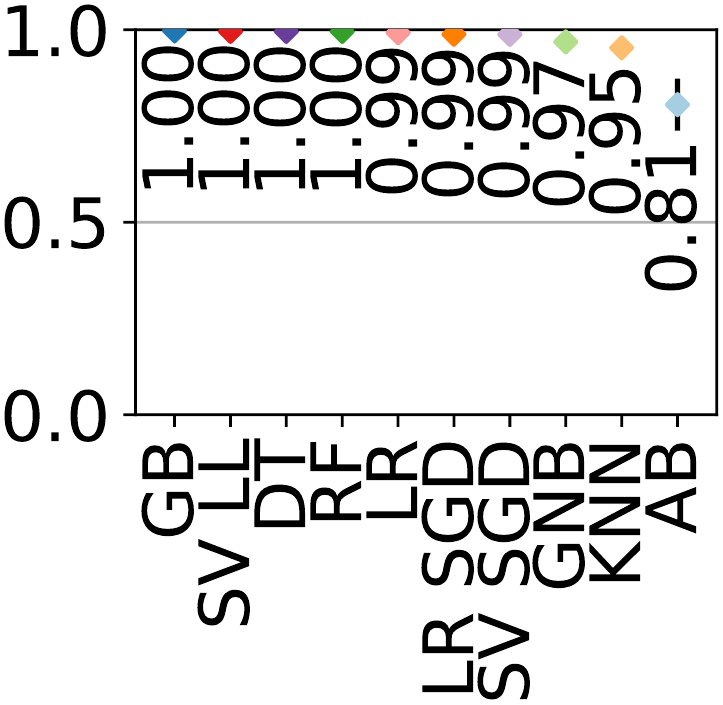} &
    \includegraphics[width=\widthfate\textwidth]{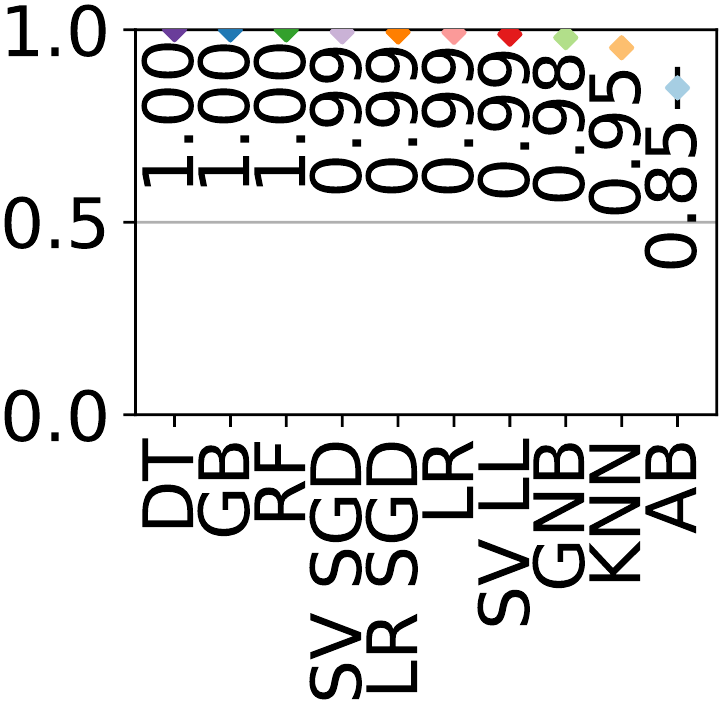} \\

    \rotatebox{90}{\parbox{9em}{\centering Application classification in SSH}} &
    \includegraphics[width=\widthfate\textwidth]{fa/ac/te_ac_am1500unaupa_n150_bb_mlad_ssh_cropped.pdf} &
    \includegraphics[width=\widthfate\textwidth]{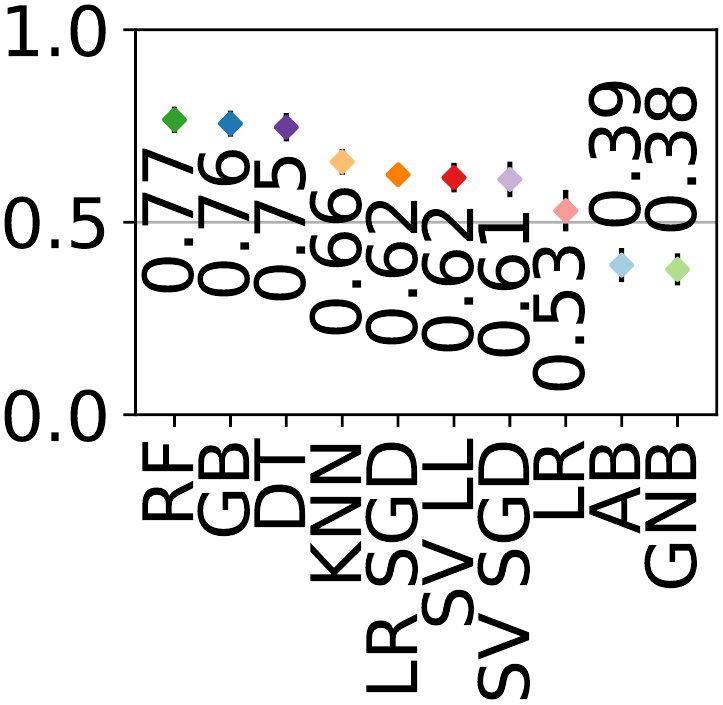} &
    \includegraphics[width=\widthfate\textwidth]{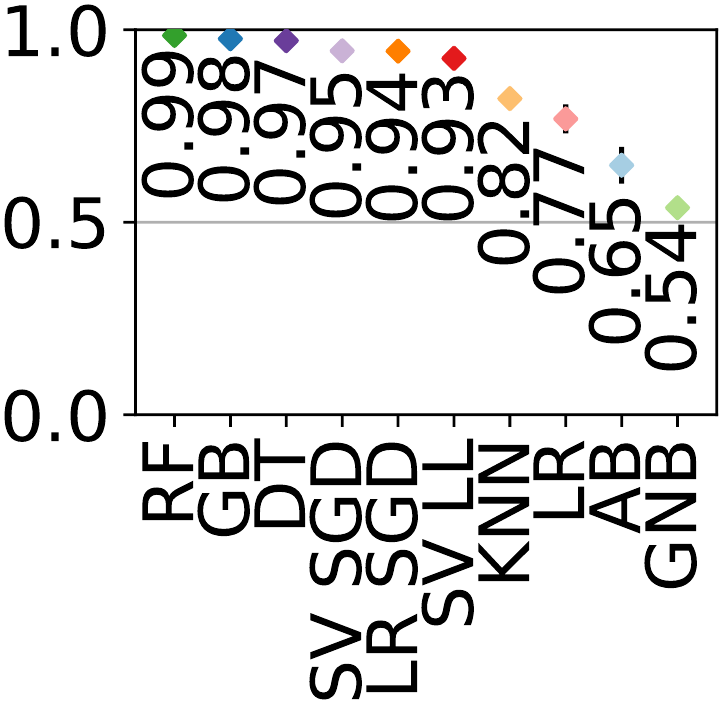} &
    \includegraphics[width=\widthfate\textwidth]{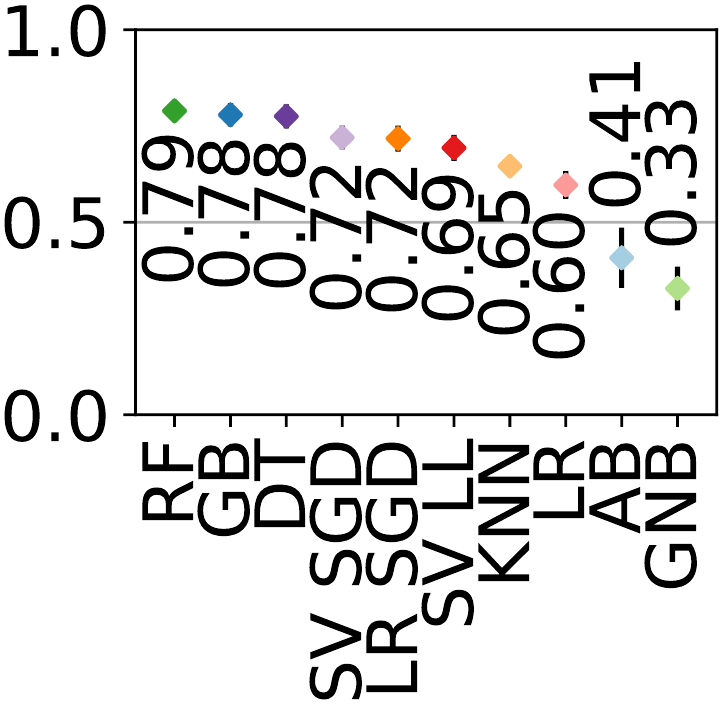} &
    \includegraphics[width=\widthfate\textwidth]{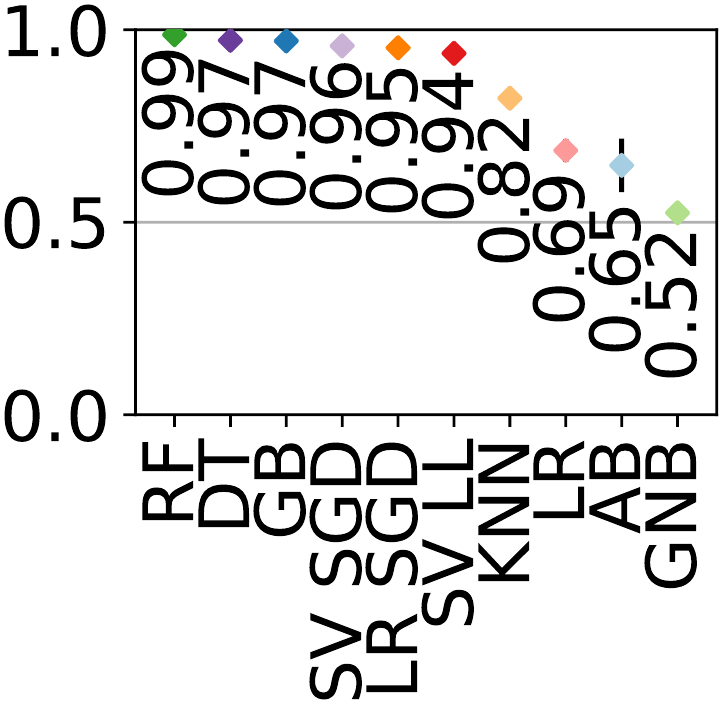} \\
    
  \end{tabular}
  
  \caption{Comparison byte burst, byte burst bigram only, byte burst with 
  bigram, byte burst trigram only, and byte burst with trigram for all pipeline steps.
  Error bars represent a confidence interval with 99\% confidence level.}
  \label{fig:feature_tdtcac_bb2go}
\end{figure*}

\Cref{fig:feature_tdtcac_bb2go} pictures the F1 scores obtained
with ML algorithms for features from the packet burst family.

\bibliographystyle{IEEEtran}
\bibliography{references}

\begin{thebibliography}{10}
\providecommand{\url}[1]{#1}
\csname url@samestyle\endcsname
\providecommand{\newblock}{\relax}
\providecommand{\bibinfo}[2]{#2}
\providecommand{\BIBentrySTDinterwordspacing}{\spaceskip=0pt\relax}
\providecommand{\BIBentryALTinterwordstretchfactor}{4}
\providecommand{\BIBentryALTinterwordspacing}{\spaceskip=\fontdimen2\font plus
\BIBentryALTinterwordstretchfactor\fontdimen3\font minus
  \fontdimen4\font\relax}
\providecommand{\BIBforeignlanguage}[2]{{%
\expandafter\ifx\csname l@#1\endcsname\relax
\typeout{** WARNING: IEEEtran.bst: No hyphenation pattern has been}%
\typeout{** loaded for the language `#1'. Using the pattern for}%
\typeout{** the default language instead.}%
\else
\language=\csname l@#1\endcsname
\fi
#2}}
\providecommand{\BIBdecl}{\relax}
\BIBdecl

\bibitem{Bernaille2006Traffic}
L.~Bernaille, R.~Teixeira, I.~Akodkenou, A.~Soule, and K.~Salamatian, ``Traffic
  classification on the fly,'' \emph{ACM SIGCOMM Computer Communication
  Review}, vol.~36, no.~2, pp. 23--26, 2006.

\bibitem{Bernaille2006Early}
L.~Bernaille, R.~Teixeira, and K.~Salamatian, ``Early application
  identification,'' in \emph{Proceedings of the 2006 ACM CoNEXT
  conference}.\hskip 1em plus 0.5em minus 0.4em\relax ACM, 2006, p.~6.

\bibitem{Kim2008Internet}
H.~Kim, K.~C. Claffy, M.~Fomenkov, D.~Barman, M.~Faloutsos, and K.~Lee,
  ``Internet traffic classification demystified: myths, caveats, and the best
  practices,'' in \emph{Proceedings of the 2008 ACM CoNEXT conference}.\hskip
  1em plus 0.5em minus 0.4em\relax ACM, 2008, p.~11.

\bibitem{Jaber2011Can}
M.~Jaber, R.~G. Cascella, and C.~Barakat, ``Can we trust the inter-packet time
  for traffic classification?'' in \emph{Communications (ICC), 2011 IEEE
  International Conference on}.\hskip 1em plus 0.5em minus 0.4em\relax IEEE,
  2011, pp. 1--5.

\bibitem{Wright2006Inferring}
C.~V. Wright, F.~Monrose, and G.~M. Masson, ``On inferring application protocol
  behaviors in encrypted network traffic,'' \emph{Journal of Machine Learning
  Research}, vol.~7, no. Dec, pp. 2745--2769, 2006.

\bibitem{Bernaille2007Early}
L.~Bernaille and R.~Teixeira, ``Early recognition of encrypted applications,''
  \emph{Passive and Active Network Measurement}, pp. 165--175, 2007.

\bibitem{Maiolini2009Real}
G.~Maiolini, A.~Baiocchi, A.~Iacovazzi, and A.~Rizzi, ``Real time
  identification of ssh encrypted application flows by using cluster analysis
  techniques,'' \emph{NETWORKING 2009}, pp. 182--194, 2009.

\bibitem{Sun2010Novel}
G.-L. Sun, Y.~Xue, Y.~Dong, D.~Wang, and C.~Li, ``An novel hybrid method for
  effectively classifying encrypted traffic,'' in \emph{Global
  Telecommunications Conference (GLOBECOM 2010), 2010 IEEE}.\hskip 1em plus
  0.5em minus 0.4em\relax IEEE, 2010, pp. 1--5.

\bibitem{Okada2011Application}
Y.~Okada, S.~Ata, N.~Nakamura, Y.~Nakahira, and I.~Oka, ``Application
  identification from encrypted traffic based on characteristic changes by
  encryption,'' in \emph{Communications Quality and Reliability (CQR), 2011
  IEEE International Workshop Technical Committee on}.\hskip 1em plus 0.5em
  minus 0.4em\relax IEEE, 2011, pp. 1--6.

\bibitem{Okada2011Comparisons}
------, ``Comparisons of machine learning algorithms for application
  identification of encrypted traffic,'' in \emph{Machine Learning and
  Applications and Workshops (ICMLA), 2011 10th International Conference on},
  vol.~2.\hskip 1em plus 0.5em minus 0.4em\relax IEEE, 2011, pp. 358--361.

\bibitem{Korczynski2014Markov}
M.~Korczy{\'n}ski and A.~Duda, ``Markov chain fingerprinting to classify
  encrypted traffic,'' in \emph{Infocom, 2014 Proceedings IEEE}.\hskip 1em plus
  0.5em minus 0.4em\relax IEEE, 2014, pp. 781--789.

\bibitem{Kumano2014Towards}
Y.~Kumano, S.~Ata, N.~Nakamura, Y.~Nakahira, and I.~Oka, ``Towards real-time
  processing for application identification of encrypted traffic,'' in
  \emph{Computing, Networking and Communications (ICNC), 2014 International
  Conference on}.\hskip 1em plus 0.5em minus 0.4em\relax IEEE, 2014, pp.
  136--140.

\bibitem{Netflow_v1578}
``{Cisco NetFlow v1, v5, v7 and v8},''
  \url{https://www.cisco.com/c/en/us/td/docs/net_mgmt/netflow_collection_engine/3-6/user/guide/format.html},
  accessed: 2017-01-31.

\bibitem{Netflow_v9}
``{Cisco NetFlow v9},''
  \url{https://www.cisco.com/en/US/technologies/tk648/tk362/technologies_white_paper09186a00800a3db9.html},
  accessed: 2017-01-31.

\bibitem{Argus}
``{Audit Record Generation and Utilization System},''
  \url{https://www.qosient.com/argus/index.shtml}, accessed: 2017-01-31.

\bibitem{zeek}
``{Zeek IDS},'' \url{https://www.zeek.org/}, accessed: 2017-01-31.

\bibitem{verizon}
``{Verizon Injecting Perma-Cookies to Track Mobile Customers, Bypassing Privacy
  Controls},'' \url{https://www.eff.org/deeplinks/2014/11/verizon-x-uidh},
  accessed: 2021-09-29.

\bibitem{scikit-learn}
F.~Pedregosa, G.~Varoquaux, A.~Gramfort, V.~Michel, B.~Thirion, O.~Grisel,
  M.~Blondel, P.~Prettenhofer, R.~Weiss, V.~Dubourg, J.~Vanderplas, A.~Passos,
  D.~Cournapeau, M.~Brucher, M.~Perrot, and E.~Duchesnay, ``Scikit-learn:
  Machine learning in {P}ython,'' \emph{Journal of Machine Learning Research},
  vol.~12, pp. 2825--2830, 2011.

\bibitem{wang2014effective}
T.~Wang, X.~Cai, R.~Nithyanand, R.~Johnson, and I.~Goldberg, ``Effective
  attacks and provable defenses for website fingerprinting,'' in \emph{23rd
  $\{$USENIX$\}$ Security Symposium ($\{$USENIX$\}$ Security 14)}, 2014, pp.
  143--157.

\bibitem{unibs}
``{UNIBS: Data sharing},'' \url{http://netweb.ing.unibs.it/~ntw/tools/traces/},
  accessed: 2021-09-29.

\bibitem{bujlow2015independent}
T.~Bujlow, V.~Carela-Espa{\~n}ol, and P.~Barlet-Ros, ``Independent comparison
  of popular dpi tools for traffic classification,'' \emph{Computer Networks},
  vol.~76, pp. 75--89, 2015.

\bibitem{upc}
``{UPC Traffic Classification },''
  \url{https://cba.upc.edu/monitoring/traffic-classification}, accessed:
  2021-09-29.

\bibitem{sklearn_ada_boost}
``{Scikit-learn - AdaBoost classifier},''
  \url{https://scikit-learn.org/stable/modules/generated/sklearn.ensemble.AdaBoostClassifier.html},
  accessed: 2021-09-29.

\bibitem{sklearn_gradient_boosting}
``{Scikit-learn - Gradient boosting classifier},''
  \url{https://scikit-learn.org/stable/modules/generated/sklearn.ensemble.GradientBoostingClassifier.html},
  accessed: 2021-09-29.

\bibitem{sklearn_gaussian_naive_bayes}
``{Scikit-learn - Gausian naive Bayes},''
  \url{https://scikit-learn.org/stable/modules/generated/sklearn.naive_bayes.GaussianNB.html},
  accessed: 2021-09-29.

\bibitem{sklearn_knn}
``{Scikit-learn - K nearest neighbors classifier},''
  \url{https://scikit-learn.org/stable/modules/generated/sklearn.neighbors.KNeighborsClassifier.html},
  accessed: 2021-09-29.

\bibitem{sklearn_linear_svc}
``{Scikit-learn - Linear SVC},''
  \url{https://scikit-learn.org/stable/modules/generated/sklearn.svm.LinearSVC.html},
  accessed: 2021-09-29.

\bibitem{sklearn_logistic_regression}
``{Scikit-learn - Logistic regression classifier},''
  \url{https://scikit-learn.org/stable/modules/generated/sklearn.linear_model.LogisticRegression.html},
  accessed: 2021-09-29.

\bibitem{sklearn_sgd_classifier}
``{Scikit-learn - SGD classifier},''
  \url{https://scikit-learn.org/stable/modules/generated/sklearn.linear_model.SGDClassifier.html},
  accessed: 2021-09-29.

\bibitem{sklearn_random_forest}
``{Scikit-learn - Random forest classifier},''
  \url{https://scikit-learn.org/stable/modules/generated/sklearn.ensemble.RandomForestClassifier.html},
  accessed: 2021-09-29.

\bibitem{sklearn_decision_tree}
``{Scikit-learn - Decision tree classifier},''
  \url{https://scikit-learn.org/stable/modules/generated/sklearn.tree.DecisionTreeClassifier.html},
  accessed: 2021-09-29.

\bibitem{raschka2018model}
S.~Raschka, ``Model evaluation, model selection, and algorithm selection in
  machine learning,'' \emph{arXiv preprint arXiv:1811.12808}, 2018.

\bibitem{novo2020flow}
C.~Novo and R.~Morla, ``Flow-based detection and proxy-based evasion of
  encrypted malware c2 traffic,'' in \emph{Proceedings of the 13th ACM Workshop
  on Artificial Intelligence and Security}, 2020, pp. 83--91.

\bibitem{Pietrzyk2009Challenging}
M.~Pietrzyk, J.-L. Costeux, G.~Urvoy-Keller, and T.~En-Najjary, ``Challenging
  statistical classification for operational usage: the adsl case,'' in
  \emph{Proceedings of the 9th ACM SIGCOMM conference on Internet
  measurement}.\hskip 1em plus 0.5em minus 0.4em\relax ACM, 2009, pp. 122--135.

\bibitem{custura2018exploring}
A.~Custura, G.~Fairhurst, and I.~Learmonth, ``Exploring usable path mtu in the
  internet,'' in \emph{Network Traffic Measurement and Analysis
  Conference}.\hskip 1em plus 0.5em minus 0.4em\relax IFIP Open Digital
  Library, 2018.

\bibitem{mapie}
``{MAPIE - Model Agnostic Prediction Interval Estimator},''
  \url{https://mapie.readthedocs.io/en/latest/index.html}, accessed:
  2021-09-29.

\end{thebibliography}

\end{document}